\documentclass[aps,prc,superscriptaddress,twocolumn]{revtex4-1}
\usepackage{amssymb}
\usepackage{amsmath}
\usepackage{wallpaper}
\usepackage{bm}
\usepackage{multirow}
\usepackage{CJK}
\usepackage[colorlinks,linkcolor=blue, urlcolor=blue, anchorcolor=blue, citecolor=blue]{hyperref}

\begin{document}

\title{Light and heavy $\Lambda$ hyperclusters in nuclear matter with relativistic-mean-field models}

\author{Cheng-Jun Xia}
\email{cjxia@yzu.edu.cn}
\affiliation{Center for Gravitation and Cosmology, College of Physical Science and Technology, Yangzhou University, Yangzhou 225009, China}

\author{Yu-Ting Rong}
\email{rongyuting@gxnu.edu.cn}
\affiliation{Department of Physics, Guangxi Normal University, Guilin, 541004, China}
\affiliation{Guangxi Key Laboratory of Nuclear Physics and Technology, Guangxi Normal University, Guilin, 541004, China}

\author{Ting-Ting Sun}
\email{ttsunphy@zzu.edu.cn}
\affiliation{School of Physics, Zhengzhou University, Zhengzhou 450001, China}

\date{\today}

\begin{abstract}
In the framework of relativistic-mean-field (RMF) models, we investigate the properties of light and heavy $\Lambda$ hyperclusters emersed in nuclear matter at various densities $n_{\mathrm{gas}}$ and proton fractions $Y_p$. In particular, the (hyper)clusters are fixed by solving the Dirac equations imposing the Dirichlet-Neumann boundary condition, while the nuclear matter take constant densities and is treated with Thomas-Fermi approximation. The binding energies of (hyper)clusters decrease with the density of nuclear matter $n_{\mathrm{gas}}$, which eventually become unbound and melt in the presence of nuclear medium, i.e., Mott transition. For light clusters with proton numbers $N_p < 4$, with the addition of $\Lambda$ hyperons, the binding energies per baryon for $\Lambda$ hyperclusters become smaller and decrease faster with $n_{\mathrm{gas}}$ due to the weaker $N$-$\Lambda$ attraction. For heavy clusters with $N_p \geq 4$, on the contrary,  the addition of $\Lambda$ hyperons increases the stability of (hyper)clusters so that the Mott transition density becomes larger as nucleons occupying higher energy states while $\Lambda$ hyperons remain in the $1s_{1/2}$ orbital. The isovector effects on (hyper)clusters in nuclear medium are also identified, where the binding energies for (hyper)clusters with $N_p> N_n$ ($N_p< N_n$) increase (decrease) with $Y_p$. For those predicted by nonlinear relativistic density functionals, light (hyper)clusters are destabilized drastically as $n_{\mathrm{gas}}$ increases, while the binding energies of heavier (hyper)clusters vary smoothly with $n_{\mathrm{gas}}$. The binding energy shifts of various (hyper)clusters due to the impact of nuclear medium are fitted to an analytical formula, which could be employed to examine the evolutions of (hyper)clusters in both heavy-ion collisions and neutron stars.
\end{abstract}

\maketitle


\section{\label{sec:intro}Introduction}

Being the lightest hyperon, $\Lambda$ hyperons are the easiest to synthesize experimentally. In 1953, Danysz and Pniewski observed the first $\Lambda$ hypernucleus (hypertriton) produced by cosmic ray~\cite{Danysz1953_PM44-348}. Since then, extensive investigations on $\Lambda$ hypernuclei were carried out. Particularly, two-body reactions such as $N$($K^-$, $\pi^{-,0}$)$\Lambda$,  $N$($\pi^{+,-}$, $K^{+,0}$)$\Lambda$,  $N$($e$, $e'$$K^{0,+}$)$\Lambda$, and $N$($\gamma$, $K^{0,+}$)$\Lambda$ with $N = n, p$ were employed to synthesize $\Lambda$ hypernuclei after 1972. With these experimental efforts, a large number of single-$\Lambda$ hypernuclei are now produced~\cite{Hashimoto2006_PPNP57-564, Gal2016_RMP88-035004}, where the corresponding $\Lambda$ separation energies $B_\Lambda$ can be obtained with
\begin{equation}
  B_\Lambda =  M({}^{A-1}Z) + M_{\Lambda} - M({}^{A}_{\Lambda}Z). \label{eq:BL}
\end{equation}
Here $M({}^{A-1}Z)$, $M_{\Lambda}$, and $M({}^{A}_{\Lambda}Z)$ are the masses of nucleus ${}^{A-1}Z$, $\Lambda$ hyperon, and $\Lambda$ hypernucleus ${}^{A}_{\Lambda}Z$. Meanwhile, double $\Lambda$ hypernuclei have also been observed, such as ${}^6_{\Lambda\Lambda}$He, ${}^{10, 11, 12}_{\Lambda\Lambda}$Be, and ${}^{12, 13}_{\Lambda\Lambda}$B~\cite{Aoki2009_NPA828-191, Ahn2013_PRC88-014003}. The $\Lambda\Lambda$ binding energies $B_{\Lambda\Lambda}$ can then be obtained, from which the $\Lambda$$\Lambda$ interaction energy $\Delta B_{\Lambda\Lambda}$ is fixed via
\begin{eqnarray}
  \Delta B_{\Lambda\Lambda} ({}^{A}_{\Lambda\Lambda}Z)
  &=& 2 M({}^{A-1}_{\Lambda}Z) - M({}^{A-2}Z)- M({}^{A}_{\Lambda\Lambda}Z) \nonumber  \\
  &=& B_{\Lambda\Lambda}({}^{A}_{\Lambda\Lambda}Z) - 2B_{\Lambda}({}^{A-1}_{\Lambda}Z). \label{eq:DBLL}
\end{eqnarray}
The experimental values for $B_\Lambda$ and $\Delta B_{\Lambda\Lambda}$ provide important constraints on the $N$-$\Lambda$ and $\Lambda$-$\Lambda$ interactions. In practice, one can construct the $N$-$\Lambda$ and $\Lambda$-$\Lambda$ interactions via various nuclear structure models, e.g., the shell model~\cite{Gal1971_AP63-53, Dalitz1978_AP116-167, Millener2008_NPA804-84, Millener2013_NPA914-109}, cluster model~\cite{Motoba1983_PTP70-189, Hiyama2002_PRC66-024007, Hiyama2006_PRC74-054312, Hiyama2009_PRC80-054321, Bando1990_IJMPA05-4021}, antisymmetrized molecular dynamics~\cite{Isaka2013_PRC87-021304}, quark mean field model~\cite{Hu2014_PRC89-025802}, relativistic-mean-field (RMF) models~\cite{Brockmann1977_PLB69-167, Boguta1981_PLB102-93, Mares1989_ZPA333-209, Mares1994_PRC49-2472, Toki1994_PTP92-803, Song2010_IJMPE19-2538, Tanimura2012_PRC85-014306, Wang2013_CTP60-479, Liu2018_PRC98-024316, Rong2021_PRC104-054321, Rong2025}, Skyrme-Hartree-Fock model~\cite{Zhou2007_PRC76-034312}, and quark-meson coupling model~\cite{Tsushima1997_PLB411-9, Tsushima1998_NPA630-691, Guichon2008_NPA814-66}. Alternatively, the $N$-$\Lambda$ and $\Lambda$-$\Lambda$ interactions can be estimated directly with lattice QCD simulations~\cite{Aoki2011_PPNP66-687}.

In heavy-ion collisions (with beam energy approximately larger than 1.58 GeV per nucleon), hyperons and hypernuclei can also be produced under the extreme conditions of high energy densities~\cite{Chen2025}. The existence of $\rm{^3_\Lambda H}$ in heavy-ion reactions were observed~\cite{Adam2020_NP16-409, Acharya2023_PRL131-102302}, while heavier (multi-strangeness) hypernulcei are expected to be produced as well~\cite{Cheng2022_PLB824-136849}. Nevertheless, their detailed production mechanism is still unclear, where two different formalisms were adopted to describe the production of these states, i.e., the statistical thermal method~\cite{Andronic2018_Nature561-321, Botvina2023_JPCS2586-012045, Buyukcizmeci2018_PRC98-064603} and the coalescence model~\cite{She2021_PRC103-014906, Zhou2025, LeFevre2019_PRC100-034904}. Due to the differences in $N$-$\Lambda$ and $N$-$N$ interactions, the production of hypernuclei has distinctive features in heavy-ion collisions compared with the corresponding normal nuclei of the same mass number~\cite{Botta2012_EPJA48-41}. Meanwhile, for heavy-ion collisions with beam energy close to hyperon threshold, higher-density nuclear matter is produced and the $N$-$N$ and $N$-$\Lambda$ interactions as well as the mean field and kaon in-medium potentials are expected to impact the hyperon dynamics, e.g., collective flows, transverse momentum spectra, and rapidity distribution~\cite{Yong2025_PLB866-139549, Yong2024_PLB853-138662, Yong2025_PRC111-054617, Feng2024_PLB851-138580, Wei2024_PLB853-138658, Zhou2025, Hartnack2012_PR510-119}. As hypernuclei are produced in a fast expanding fireball created in heavy-ion collisions~\cite{Reisdorf2010_NPA848-366}, the in-medium binding energy shift of hyperclusters maybe important for understanding hypernuclei production mechanism, particularly at beam energies around the hyperon threshold.

For neutron star matter, according to the $N$-$\Lambda$ interaction constrained by the experimental values of $B_\Lambda$, $\Lambda$ hyperons are expected to emerge at densities around 2--3$n_0$ with $n_0\approx 0.16$ fm${}^{-3}$ being the nuclear saturation density~\cite{Ishizuka2008_JPG35-085201, Shen2011_ApJ197-20, Sun2018_CPC42-25101, Sun2019_PRD99-023004}. Meanwhile, during supernova explosions or binary neutron star mergers, hyperons are also created and compete with the formation of light nuclear clusters in the low density regions, e.g., those indicated in the CompOSE database \footnote{https://compose.obspm.fr/}. The emergence of hyperons and hyperclusters will reduce the free energy of dense stellar matter and thus alter the corresponding equation of state (EOS) and matter content~\cite{Custodio2021_PRC104-035801}, which affects various processes in core-collapse
supernova~\cite{Woosley2002_RMP74-1015}, neutron star properties~\cite{Lattimer2012_ARNPS62-485, Ozel2016_ApJ820-28, Ozel2016_ARAA54-401}, and binary neutron star mergers~\cite{Hotokezaka2011_PRD83-124008}. It is thus necessary to examine the properties of hyperclusters in dense stellar matter, where the in-medium effects may be important.

In such cases, in this work we investigate the clustering phenomenon of $\Lambda$ hypernuclei in nuclear medium. In our previous work we have proposed a unified treatment for investigating the clustering phenomenon of normal nuclei inside a spherical Wigner-Seitz cell employing RMF models~\cite{Xia2025_PRC112-025801}, which will be extended to the hyperonic sector here. The clusters are fixed by solving the Dirac equations imposing the Dirichlet-Neumann boundary condition~\cite{Negele1973_NPA207-298}, while the nuclear medium is treated with Thomas-Fermi approximation and assume constant densities. By calibrating the coupling constants $g_{\sigma\Lambda}$ and $g_{\omega\Lambda}$ according to the experimental separation energies of $\Lambda$ hyperons $B_\Lambda$ for single $\Lambda$ hypernuclei~\cite{Hashimoto2006_PPNP57-564, Gal2016_RMP88-035004}, the properties of $\Lambda$ hypernuclei in vacuum and in nuclear matter can be fixed. 

The paper is organized as follows. In Sec.~\ref{sec:the} we present our theoretical framework, including the Lagrangian density of RMF models including $\Lambda$ hyperons and the formalism for fixing the properties of clusters in nuclear medium. The adopted parameters to reproduce the experimental separation energies of $\Lambda$ hyperons $B_\Lambda$ for single $\Lambda$ hypernuclei as well as the obtained results are presented in Sec.~\ref{sec:res}, and our conclusion is given in Sec.~\ref{sec:con}.

\section{\label{sec:the}Theoretical framework}
The Lagrangian density for RMF models of static systems can be obtained with
\begin{eqnarray}
\mathcal{L}
 &=& \sum_{i=n,p,\Lambda} \bar{\psi}_i
       \left[  i \gamma^\mu \partial_\mu - \gamma^0 \left(g_{\omega i}\omega + g_{\rho i}\rho\tau_i + A q_i\right)- m_i^* \right] \psi_i
\nonumber \\
 &&\mbox{} + \frac{1}{2}\partial_\mu \sigma \partial^\mu \sigma  - \frac{1}{2}m_\sigma^2 \sigma^2
           - \frac{1}{4} \Omega_{\mu\nu}\Omega^{\mu\nu} + \frac{1}{2}m_\omega^2 \omega^2
\nonumber \\
 &&\mbox{} - \frac{1}{4} R_{\mu\nu}R^{\mu\nu} + \frac{1}{2}m_\rho^2 \rho^2 - \frac{1}{4} F_{\mu\nu}F^{\mu\nu} + U(\sigma, \omega),
\label{eq:Lagrange}
\end{eqnarray}
where $\psi_{i}$ is the Dirac spinor, $m_{i}^*\equiv m_{i} + g_{\sigma i} \sigma$ the effective masses for nucleons and hyperons, $\tau_n=-\tau_p=1$ and $\tau_\Lambda=0$ the 3rd components of isospin, $q_{n}=q_{\Lambda}=0$ and $q_{p}=-q_{e}=1$ the charge number. The boson fields $\sigma$, $\omega$, $\rho$, and electromagnetic field $A$ take mean values and are left with only the time components due to time-reversal symmetry, where the field tensors $\Omega_{\mu\nu}$, $R_{\mu\nu}$, and $F_{\mu\nu}$ vanish except for
\begin{equation}
\Omega_{i0} = -\Omega_{0i} = \partial_i \omega,
 R_{i0}  = -R_{0i}   = \partial_i  \rho,
  F_{i0}    = -F_{0i}      = \partial_i A. \nonumber
\end{equation}

For the relativistic density functionals PK1~\cite{Long2004_PRC69-034319} and TM1~\cite{Sugahara1994_NPA579-557}, the mesonic nonlinear self couplings in Eq.~(\ref{eq:Lagrange}) are nonzero and are fixed with
\begin{equation}
U(\sigma, \omega) = -\frac{1}{3}g_2\sigma^3 - \frac{1}{4}g_3\sigma^4 + \frac{1}{4}c_3\omega^4.  \label{eq:U_NL}
\end{equation}
Alternatively, for the functional DD-LZ1~\cite{Wei2020_CPC44-074107}, explicit density-dependent couplings are adopted while taking $U(\sigma, \omega)=0$, i.e.,
\begin{eqnarray}
g_{\xi i}(n_\mathrm{b}) &=& g_{\xi i} a_{\xi} \frac{1+b_{\xi}(n_\mathrm{b}/n_0+d_{\xi})^2}
                          {1+c_{\xi}(n_\mathrm{b}/n_0+d_{\xi})^2}, \label{eq:ddcp_TW} \\
g_{\rho i}(n_\mathrm{b}) &=& g_{\rho i} \exp{\left[-a_\rho(n_\mathrm{b}/n_0 + b_\rho)\right]}. \label{eq:ddcp_rho}
\end{eqnarray}
Here $\xi=\sigma$, $\omega$ and $n_\mathrm{b} = n_p+n_n+n_\Lambda$ the baryon number density.

To fix the properties of (hyper)clusters in nuclear medium inside spherical Wigner-Seitz cells, the Dirac spinors of nucleons and hyperons can be expanded as
\begin{equation}
 \psi_{n\kappa m}({\bm r}) =\frac{1}{r}
 \left(\begin{array}{c}
   iG_{n\kappa}(r) \\
    F_{n\kappa}(r) {\bm\sigma}\cdot{\hat{\bm r}} \\
 \end{array}\right) Y_{jm}^l(\theta,\phi)\:,
\label{EQ:RWF}
\end{equation}
where $G_{n\kappa}(r)/r$ and $F_{n\kappa}(r)/r$ are the upper and lower components for the radial wave functions, $Y_{jm}^l(\theta,\phi)$ is the spinor spherical harmonics, $\kappa$ is connected to the angular momenta $(l,j)$ via $\kappa=(-1)^{j+l+1/2}(j+1/2)$. The Dirac equation is then reduced into
\begin{equation}
 \left(\begin{array}{cc}
  V_i+S_i                             & {\displaystyle -\frac{\mbox{d}}{\mbox{d}r}+\frac{\kappa}{r}}\\
  {\displaystyle \frac{\mbox{d}}{\mbox{d}r}+\frac{\kappa}{r}} & V_i-S_i-2m_i                       \\
 \end{array}\right)
 \left(\begin{array}{c}
  G_{n\kappa} \\
  F_{n\kappa} \\
 \end{array}\right)
 = \varepsilon_{n\kappa}
 \left(\begin{array}{c}
  G_{n\kappa} \\
  F_{n\kappa} \\
 \end{array}\right) \:
\label{EQ:RDirac}
\end{equation}
with $\varepsilon_{n\kappa}$ being the single-particle energy. Note that the Dirichlet-Neumann boundary condition is imposed so that the density approaches to constant values at the edge of Wigner-Seitz cells ($r=R_\mathrm{W}$ with $R_\mathrm{W}$ being the cell size)~\cite{Negele1973_NPA207-298}. The scalar ($S_i$) and vector ($V_i$) potentials are fixed by
\begin{eqnarray}
S_i &=&  g_{\sigma i} \sigma, \label{eq:MF_scalar} \\
V_i &=&  \Sigma^\mathrm{R} + g_{\omega i} \omega + g_{\rho i}\tau_{i} \rho + q_i  A, \label{eq:MF_vector}
\end{eqnarray}
where the ``rearrangement" term $\Sigma^\mathrm{R}$ emerges due to the density-dependency of coupling constants and is obtained with~\cite{Lenske1995_PLB345-355}
\begin{equation}
\Sigma^\mathrm{R}= \sum_{i=n,p,\Lambda} \left(
 \frac{\mbox{d} g_{\sigma i}}{\mbox{d} n_\mathrm{b}} \sigma n^\mathrm{s}_i+
   \frac{\mbox{d} g_{\omega i}}{\mbox{d} n_\mathrm{b}} \omega n_i+
   \frac{\mbox{d} g_{\rho i}}{\mbox{d} n_\mathrm{b}} \rho \tau_i n_i \right).
\label{eq:re_B}
\end{equation}
The equations of motion for bosons are fixed by
\begin{eqnarray}
(-\nabla^2 + m_\sigma^2) \sigma &=& -\sum_{i=n,p,\Lambda} g_{\sigma i} n_i^\mathrm{s} - g_2\sigma^2 - g_3\sigma^3, \label{eq:KG_sigma} \\
(-\nabla^2 + m_\omega^2) \omega &=& \sum_{i=n,p,\Lambda} g_{\omega i} n_i + c_3\omega^3, \label{eq:KG_omega}\\
(-\nabla^2 + m_\rho^2) \rho     &=& \sum_{i=n,p,\Lambda} g_{\rho i}\tau_{i} n_i, \label{eq:KG_rho}\\
                   -\nabla^2 A  &=& e(n_p - n_e), \label{eq:KG_photon}
\end{eqnarray}
where reflective boundary conditions are imposed at $r=0$ and the edge of Wigner-Seitz cell with $r=R_\mathrm{W}$~\cite{Xia2021_PRC103-055812}. Note that the global charge neutrality is imposed by adding negatively charged electron background $n_e$ so that $A(R_\mathrm{W})=\nabla A(R_\mathrm{W})=0$, avoiding the possible divergence in Coulomb potential. Due to the reduction of Coulomb energy~\cite{Xia2025_PRC112-025801}, this will lead to an increment of binding energies $B_0$ as indicated in Table~\ref{table:bind_exp} in comparison with that of vacuum $B$ in Table~\ref{table:Cluster}. For small systems such as heavy-ion collisions, the charge neutrality is not required and the variation of binding energies can be fixed by replacing $B_0$ with $B$. The source currents are comprised of contributions from both the nuclear medium (gas) and cluster, i.e.,
\begin{eqnarray}
  n_{i} &=& \langle \bar{\psi}_{i} \gamma^{0} \psi_{i} \rangle = n_{i,\mathrm{gas}} + n_{i,\mathrm{cluster}}, \label{eq:nv_all} \\
  n_{i}^\mathrm{s} &=& \langle \bar{\psi}_{i} \psi_{i} \rangle=n^\mathrm{s}_{i,\mathrm{gas}} + n^\mathrm{s}_{i,\mathrm{cluster}}. \label{eq:ni_all}
\end{eqnarray}
The scalar ($n_{i}^\mathrm{s}$) and vector ($n_{i}$) densities of clusters (with given particle numbers $N_i$) are fixed according to the radial wave functions, i.e.,
\begin{eqnarray}
 n_{i,\mathrm{cluster}}^{s}(r) &=& \frac{1}{4\pi r^2}\sum_{k=1}^{N_i}  \left[|G_{k i}(r)|^2-|F_{k i}(r)|^2\right], \label{eq:np_cluster} \\
 n_{i,\mathrm{cluster}}(r) &=& \frac{1}{4\pi r^2}\sum_{k=1}^{N_i}  \left[|G_{k i}(r)|^2+|F_{k i}(r)|^2\right],
\end{eqnarray}
while those of nuclear medium are obtained adopting Thomas-Fermi approximation (TFA), i.e.,
\begin{eqnarray}
n_{i,\mathrm{gas}}^\mathrm{s}(r) &=&  \frac{{m_i^*}^3}{2\pi^2} \left[ f(x_i) - f(\bar{x}_i)  \right], \label{eq:nsgas}\\
n_{i,\mathrm{gas}}           &=&  \frac{\nu_i^3-\bar{\nu}_i^3}{3\pi^2} = \mathrm{constant}, \label{eq:ngas}
\end{eqnarray}
with $f(x) = x \sqrt{x^2+1} - \mathrm{arcsh}(x)$, ${x}_i =\nu_i/m_i^*$, and $\bar{x}_i =\bar{\nu}_i/m_i^*$. Here $\nu_i(r)$ represents the Fermi momentum and varies with the space coordinate, while $\bar{\nu}_i(r)$ is the minimum momentum corresponding to the lowest energy state with vanishing momentum in nuclear medium
\begin{equation}
\varepsilon_{i, \mathrm{gas}}^\mathrm{b} = S_{i, \mathrm{gas}} + V_{i, \mathrm{gas}}.
\end{equation}
In practice, the value of $\varepsilon_{i, \mathrm{gas}}^\mathrm{b}$ is fixed for uniform nuclear matter, where $\bar{\nu}_i(r)$ is obtained with
\begin{equation}
\sqrt{\bar{\nu}_i^2+{m_i^*}^2}-m_i + \Sigma^\mathrm{R} + g_{\omega} \omega + g_{\rho}\tau_{i} \rho + q_i  A \equiv \varepsilon_{i, \mathrm{gas}}^\mathrm{b}.  \label{eq:nub_gas}
\end{equation}
The Fermi momentum $\nu_i(r)$ can then be fixed according to Eq.~(\ref{eq:ngas}) with given densities $n_{i,\mathrm{gas}}$ for nuclear medium.

The particle numbers of a cluster $N_i$ are determined by
\begin{eqnarray}
N_i =   \int 4\pi r^2 n_{i,\mathrm{cluster}}(r) \mbox{d}r,
\end{eqnarray}
which gives the total baryon number for the cluster by $A=N_p+N_n+N_\Lambda$. Based on the proton density profiles, the charge radius $R_\mathrm{ch}$ can be obtained with~\cite{Sugahara1994_NPA579-557}
\begin{equation}
  R_\mathrm{ch}^2 = R_p^2+ \delta R_p^2  + (0.862\ \mathrm{fm})^2 - (0.336\ \mathrm{fm})^2\frac{N_n}{N_p}, \label{eq:Rch}
\end{equation}
where the proton radius $R_p$ fixed by
\begin{equation}
  R_p^2 = \frac{1}{N_p}\int 4\pi r^4 n_{p,\mathrm{cluster}}(r) \mbox{d}r. \label{eq:Rp}
\end{equation}
The center-of-mass correction to the proton radius $\delta R_p^2$ is determined by~\cite{Long2004_PRC69-034319}
\begin{equation}
 \delta R_p^2 =  - \frac{2}{A} R_p^2 + \frac{1}{A} R_M^2, \label{eq:Rch1}
\end{equation}
where $R_M$ represents the matter radius of the cluster and is fixed by
\begin{equation}
  R_M^2 = \frac{1}{A}\sum_{i=n,p,\Lambda} \int 4\pi r^4 n_{i,\mathrm{cluster}}(r) \mbox{d}r. \label{eq:Rm}
\end{equation}
Note that the center-of-mass corrections for proton radius $\delta R_p^2$ are small and often neglected, for simplicity, in this work we neglect the mass difference between nucleons and hyperons in estimating $\delta R_p^2$ as the center-of-mass coordinate $\vec{R}_\mathrm{c.m.} = \sum_i^A m_i \vec{r}_i / \sum_i^A m_i$ varies little if we take $m_p=m_n=m_\Lambda$.

The energy of the system is obtained with
\begin{equation}
E_\mathrm{MF}=\int \langle {\cal{T}}_{00} \rangle \mbox{d}^3 r, \label{eq:energy}
\end{equation}
where the energy momentum tensor is determined by
\begin{eqnarray}
\langle {\cal{T}}_{00} \rangle
&=&  \sum_{i=n,p,\Lambda}\left[\sum_{k=1}^{N_i} \varepsilon_{ki} \langle \bar{\psi}_{ki} \gamma^{0} \psi_{ki} \rangle+ (m_i -   V_i) n_{i,\mathrm{cluster}}  \right]  \nonumber \\
&&   +\sum_{i=n,p,\Lambda}\mathcal{E}_{i,\mathrm{gas}} + \frac{1}{2}(\nabla \sigma)^2 + \frac{1}{2}m_\sigma^2 \sigma^2 + \frac{1}{2}(\nabla \omega)^2   \nonumber \\
&&   + \frac{1}{2}m_\omega^2 \omega^2 + \frac{1}{2}(\nabla \rho)^2 + \frac{1}{2}m_\rho^2 \rho^2
     + \frac{1}{2}(\nabla A)^2 \nonumber \\
&&   + c_3\omega^4 - U(\sigma, \omega), \label{eq:ener_dens} \\
 \mathcal{E}_{i,\mathrm{gas}} &=& \frac {{m^*_i}^4}{8\pi^{2}}\left[g(x_i) - g(\bar{x}_i) \right],
\end{eqnarray}
with $g(x) = x(2x^2+1)\sqrt{x^2+1}-\mathrm{arcsh}(x)$. Here $\varepsilon_{ki}$ is the single-particle energy of particle $i$ in a cluster fixed by solving Dirac Eq.~(\ref{EQ:RDirac}). The center-of-mass energy needs to be subtracted from Eq.~(\ref{eq:energy}) and can be fixed by~\cite{Bender2000_EPJA7-467}
\begin{equation}
E_\mathrm{c.m.} = -\frac{\langle P_\mathrm{c.m.}^2 \rangle}{2 \sum_{i=n,p,\Lambda}m_i  N_i}.  \label{eq:ecm}
\end{equation}
Then the total mass of a cluster (with particle numbers $N_i$) emersed in nuclear medium (with densities $n_{i,\mathrm{gas}}$) can be obtained with
\begin{equation}
E_\mathrm{tot} = E_\mathrm{MF} + E_\mathrm{c.m.} - E_\mathrm{gas},  \label{eq:etot}
\end{equation}
where $E_\mathrm{gas}$ represents the energy contribution of nuclear medium in the absence of clusters, i.e., the uniform nuclear matter with densities $n_{i,\mathrm{gas}}$. The binding energy of the cluster is fixed by adopting the formalism of a generalized relativistic density-functional (gRDF) approach~\cite{Typel2010_PRC81-015803}, i.e.,
\begin{equation}
B = \sum_{i=n,p, \Lambda}(m_i + S_{i,\mathrm{gas}} + V_{i,\mathrm{gas}}) N_i - E_\mathrm{tot}, \label{eq:bind}
\end{equation}
where $S_{i,\mathrm{gas}}$ and $V_{i,\mathrm{gas}}$ are the scalar and vector mean-field potentials of baryon $i$ in the uniform nuclear medium. Note that taking the surrounding medium as a nuclear gas is a simplification of the real situation, where there may exist various clusters in the medium. In the gRDF approach with the binding energy shifts treated with a $\sigma$-cluster coupling, this could lead to a reduction of binding energies of clusters as $S_{i,\mathrm{gas}}$ is reduced in comparison with uniform nuclear gas scenarios at same $n_{i,\mathrm{gas}}$~\cite{Custodio2021_PRC104-035801}.

In summary, to fix the properties of a cluster (with fixed $N_i$) emersed in nuclear medium with density $n_{i,\mathrm{gas}}$, we solve the Dirac Eq.~(\ref{EQ:RDirac}), mean-field potential Eqs.~(\ref{eq:MF_scalar}-\ref{eq:re_B}) with the boson fields determined by Eqs.~(\ref{eq:KG_sigma}-\ref{eq:KG_photon}), and densities Eqs.~(\ref{eq:nv_all}-\ref{eq:nub_gas}) iteratively in a box size of $R_\mathrm{W} = 12.8~{\rm fm}$ with a grid distance of $0.1~{\rm fm}$, which is large enough to enclose the (hyper)nuclei considered in this work. Once convergency is reached, the energy and binding energy are obtained with Eqs.~(\ref{eq:etot}) and (\ref{eq:bind}).

\section{\label{sec:res}Results and discussions}

In this work we adopt the relativistic density functionals DD-LZ1~\cite{Wei2020_CPC44-074107}, TM1~\cite{Sugahara1994_NPA579-557}, and PK1~\cite{Long2004_PRC69-034319}, where the corresponding saturation properties of nuclear matter are listed in Table.~\ref{table:NM}. Note that the functionals PK1 and DD-LZ1 were
widely adopted in our previous investigations on the properties of (hyper)nuclei~\cite{Liu2018_PRC98-024316, Ding2023_CPC47-124103}, while that of TM1 was extensively used for supernova simulations, i.e., the Shen EoSs~\cite{Shen2011_ApJ197-20}. Meanwhile, according to the state-of-the-art constraints on nuclear matter properties~\cite{Russotto2016_PRC94-034608, LeFevre2016_NPA945-112, Zhang2020_PRC101-034303, Essick2021_PRL127-192701, Huth2022_Nature606-276}, the functional DD-LZ1~\cite{Wei2020_CPC44-074107} predicts the most realistic equation of state.

\begin{table}
\caption{\label{table:NM} Binding energy $B$, incompressibility $K$, skewness $J$, symmetry energy $S$, the slope and curvature of symmetry energy $L$ and $K_\mathrm{sym}$ of nuclear matter at the saturation density $n_0$, which are predicted by the relativistic density functionals DD-LZ1~\cite{Wei2020_CPC44-074107}, TM1~\cite{Sugahara1994_NPA579-557}, and PK1~\cite{Long2004_PRC69-034319}.}
\begin{tabular}{c|ccccccc} \hline \hline
       & $n_0$        &   $B$    &   $K$  &  $J$   & $S$    &  $L$  & $K_\mathrm{sym}$        \\
       & fm${}^{-3}$  &   MeV    &   MeV  &  MeV   &  MeV   &  MeV  &   MeV             \\ \hline
DD-LZ1 &  0.158       &   16.06  &  230.7 & 1330   & 32.0   &  42.5 &  $-20$            \\
TM1    &  0.145       &   16.26  &  281.2 & $-285$ & 36.9   & 110.8 &   34           \\
PK1    &  0.148       &   16.27  &  282.7 &$-27.8$ & 37.6   & 115.9 &   55           \\
\hline
\end{tabular}
\end{table}

\subsection{\label{sec:res_param} $\Lambda$ hypernuclei}

We then investigate the properties of $\Lambda$ hypernuclei based on these effective $N$-$N$ interactions in the absence of pairing interactions. The negligence of pairing interactions is justified for double magic nuclei since the formation of pairs is hindered by the large shell gap, while for open-shell nuclei this may lead to slight deviations of binding energies ($\sim 6/\sqrt{A}$ MeV) from the experimental values~\cite{Zubov2009_PPN40-847}. For the $N$-$\Lambda$ interactions, we take the depth of $\Lambda$ mean field potential in symmetric nuclear matter (SNM) at saturation density to be
\begin{equation}
  U_\Lambda (n_0) = g_{\sigma\Lambda}\sigma + g_{\omega\Lambda}\omega \equiv -30\ \mathrm{MeV},  \label{eq:Vlam}
\end{equation}
which is expected to reproduce the single-$\Lambda$ binding energies irrespective of the particular values of $g_{\sigma\Lambda}$ and $g_{\omega\Lambda}$~\cite{Bouyssy1976_PLB64-276, Millener1988_PRC38-2700, Wang2013_CTP60-479, Sun2018_CPC42-25101, Rong2021_PRC104-054321}. Consider the two-solar-mass constraint on neutron stars' masses~\cite{Sun2018_CPC42-25101, Sun2019_PRD99-023004}, we take $g_{\sigma\Lambda} = g_{\sigma N}$ and then fix $g_{\omega\Lambda}$ according to Eq.~(\ref{eq:Vlam}), which gives $g_{\omega\Lambda} \approx 1.14 g_{\omega N}$. Note that slightly smaller values for $g_{\sigma\Lambda}$ and $g_{\omega\Lambda}$ were adopted in our previous study~\cite{Sun2018_CPC42-25101, Sun2019_PRD99-023004}, which barely affect our predictions on single-$\Lambda$ binding energies in $\Lambda$ hypernuclei as long as Eq.~(\ref{eq:Vlam}) is satisfied.

\begin{figure}
\includegraphics[width=\linewidth]{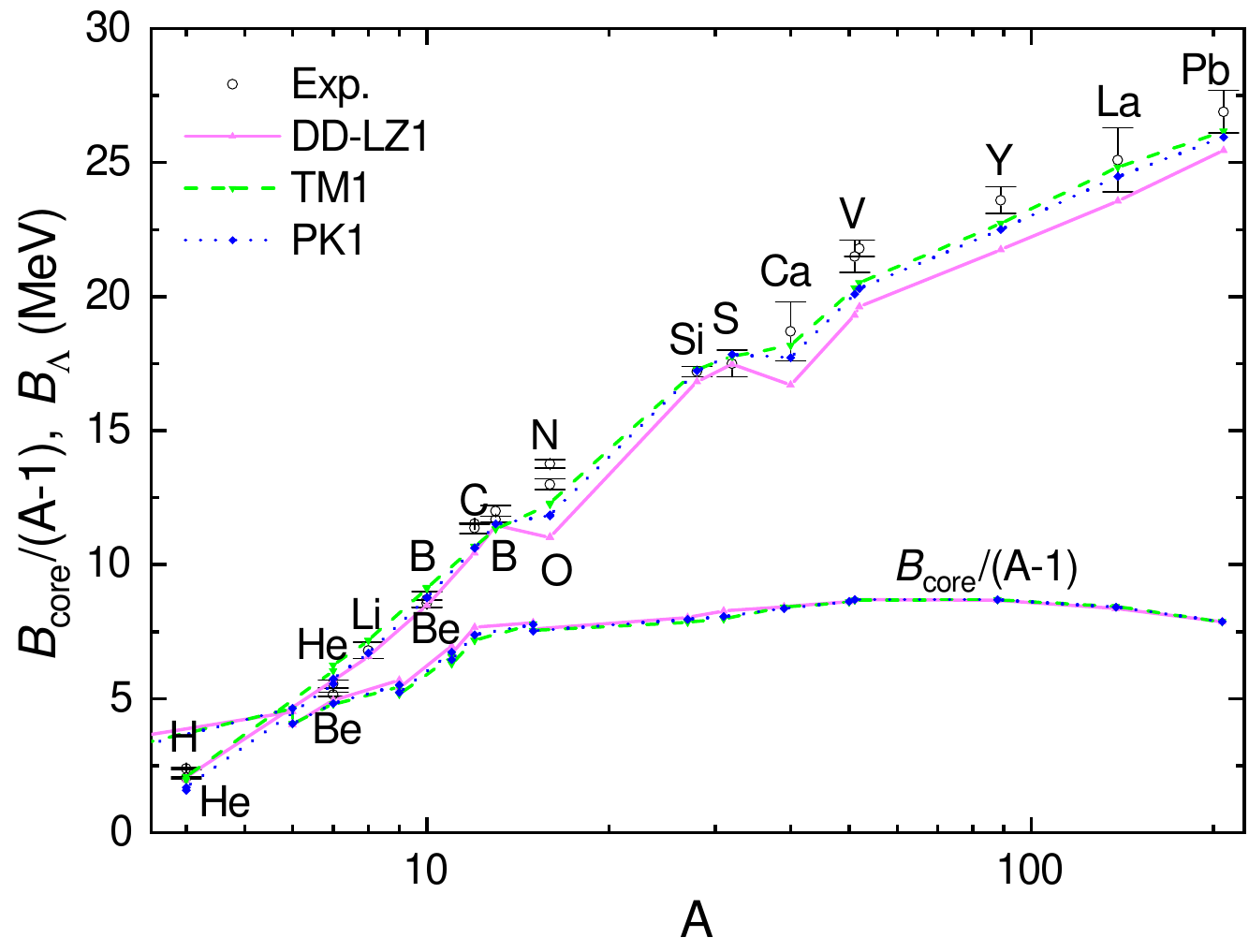}
\caption{\label{Fig:B_Lam} Binding energies of $\Lambda$ hyperon $B_\Lambda$ in the $1s_{1/2}$ state of hypernuclei predicted by the relativistic density functionals DD-LZ1~\cite{Wei2020_CPC44-074107}, TM1~\cite{Sugahara1994_NPA579-557}, and PK1~\cite{Long2004_PRC69-034319}. The corresponding experimental values are indicated by the open circles for comparison~\cite{Hashimoto2006_PPNP57-564, Gal2016_RMP88-035004}. The binding energy per nucleon $B_\mathrm{core}/(A-1)$ for the nuclear cores are presented as well. The element symbols (proton numbers $N_p$) are indicated for both $\Lambda$ hypernuclei and their core nuclei with $A\rightarrow A-1$, where each nucleus is presented independently. }
\end{figure}

The binding energies of $\Lambda$ hyperon in the $1s_{1/2}$ states of $\Lambda$ hypernuclei (${}^{4}_\Lambda$H, ${}^{4}_\Lambda$He, ${}^{7}_\Lambda$He, ${}^{7}_\Lambda$Be, ${}^{8}_\Lambda$Li, ${}^{10}_\Lambda$Be, ${}^{10}_\Lambda$B, ${}^{12}_\Lambda$B, ${}^{12}_\Lambda$C, ${}^{13}_\Lambda$C, ${}^{13}_\Lambda$C, ${}^{16}_\Lambda$N, ${}^{16}_\Lambda$O, ${}^{28}_\Lambda$Si, ${}^{32}_\Lambda$S, ${}^{40}_\Lambda$Ca, ${}^{51}_\Lambda$V, ${}^{52}_\Lambda$V, ${}^{89}_\Lambda$Y, ${}^{139}_\Lambda$La, and ${}^{208}_\Lambda$Pb) are then estimated with Eq.~(\ref{eq:BL}). The obtained results are presented in Fig.~\ref{Fig:B_Lam}, while the corresponding experimental values are indicated as well for comparison~\cite{Hashimoto2006_PPNP57-564, Gal2016_RMP88-035004}. Evidently, the relativistic density functionals adopted in this work generally well reproduce the experimental binding energies of $\Lambda$ hyperon in $\Lambda$ hypernuclei in a large mass range of $A=4\sim 208$, where the corresponding root-mean-square deviations of $B_\Lambda$ from the experimental values are $\Delta = 1.30$, 0.74, and 0.86 MeV when the functionals DD-LZ1~\cite{Wei2020_CPC44-074107}, TM1~\cite{Sugahara1994_NPA579-557}, and PK1~\cite{Long2004_PRC69-034319} are adopted for the effective $NN$ interactions. The corresponding mean squared error are $\delta = 8.54\%$, 9.85\%, and 7.34\%, so that the single $\Lambda$ binding energies are sufficiently well reproduced considering the uncertainties in the experimental data. Note that the deviations may be further reduced if we readjust the coupling constants and consider more realistic scenarios, e.g., pairing, deformation, and charge symmetry breaking~\cite{Rong2025, Sun2025_PLB865-139460}. The binding energy per nucleon $B_\mathrm{core}/(A-1)$ for the nuclear cores are presented as well. For light nuclei such as ${}^{3}$H and ${}^{3}$He, their binding energies per nucleon are larger than the binding energies of $\Lambda$ hyperon for ${}^{4}_\Lambda$H and ${}^{4}_\Lambda$He, which is mainly due to the weaker $N$-$\Lambda$ interaction than that of $N$-$N$ interaction. As mass number ($A\gtrsim 8$) increases, nucleons eventually occupy single-particle states of larger energies, while $\Lambda$ hyperons still occupy the $1s_{1/2}$ state, so that the binding energies of $\Lambda$ hyperon eventually surpass that of nucleons.

\begin{table}
  \centering
  \caption{\label{table:Cluster}Binding energies $B$ and charge radii $R_\mathrm{ch}$ for finite nuclei predicted by the relativistic density functional DD-LZ1~\cite{Wei2020_CPC44-074107}. By adding one (two) $\Lambda$ hyperon(s) to these nuclei, the hyperon separation energies $B_\Lambda$ ($B_{\Lambda\Lambda}$) of $\Lambda$ ($\Lambda$$\Lambda$) hypernuclei as well as their charge radii $R_\mathrm{ch}^\Lambda$ ($R_\mathrm{ch}^{\Lambda\Lambda}$) are obtained and listed here. The experimental binding energies $B^\mathrm{exp}$ and charge radii $R_\mathrm{ch}^\mathrm{exp}$ for the core nuclei are indicated as well~\cite{Angeli2013_ADNDT99-69, Huang2021_CPC45-30002, Wang2021_CPC45-030003}.}
  \begin{tabular}{c|cccc|cc|cc}
    \hline \hline
      Core &  $B/A$ & $B^\mathrm{exp}/A$ &  $R_\mathrm{ch}$ &   $R_\mathrm{ch}^\mathrm{exp}$  &  $B_\Lambda$ & $R_\mathrm{ch}^\Lambda$  &  $B_{\Lambda\Lambda}/2$ &  $R_\mathrm{ch}^{\Lambda\Lambda}$  \\
             &     MeV & MeV    &  fm       &fm     &  MeV  & fm    &  MeV  &  fm     \\\hline
  $^3$H      &   3.747 & 2.827  &  2.090  &1.759    & 1.996 & 2.164 & 1.923 &  2.187  \\
  $^3$He     &   3.419 & 2.573  &  2.337  &1.966    & 2.064 & 2.407 & 1.972 &  2.414  \\
  $^4$He     &   6.832 & 7.074  &  1.988  &1.675    & 2.710 & 2.030 & 2.644 &  2.020  \\
  $^5$Li     &   5.107 & 5.266  &  3.529  & --      & 4.383 & 2.667 & 4.070 &  2.442  \\
  $^8$Be     &   5.220 & 7.062  &  2.464  & --      & 7.503 & 2.357 & 6.969 &  2.292  \\
  $^{12}$C   &   7.663 & 7.680  &  2.354  &2.470    & 11.48 & 2.313 & 11.05 &  2.292  \\
  $^{16}$O   &   8.028 & 7.976  &  2.671  &2.699    & 10.98 & 2.637 & 10.68 &  2.616  \\
  $^{40}$Ca  &   8.610 & 8.551  &  3.440  &3.478    & 16.64 & 3.415 & 16.38 &  3.397  \\
  $^{48}$Ca  &   8.659 & 8.667  &  3.440  &3.477    & 18.93 & 3.416 & 18.67 &  3.399  \\
  $^{208}$Pb &   7.871 & 7.867  &  5.485  &5.501    & 25.49 & 5.468 & 25.35 &  5.452  \\   \hline
  \end{tabular}
\end{table}

We then investigate the properties of various $\Lambda$ hypernuclei and $\Lambda$$\Lambda$ hypernuclei that may be produced in heavy-ion collisions as well as in violent astrophysical processes such as supernova explosions and binary neutron star mergers. As an example, we employ the relativistic density functional DD-LZ1~\cite{Wei2020_CPC44-074107} for the effective $N$-$N$ interactions and $g_{\omega\Lambda} = 1.143 g_{\omega N}$ ($g_{\sigma\Lambda} = g_{\sigma N}$) for the $\Lambda$-$N$ interactions. The obtained binding energy $B$ and charge radii $R_\mathrm{ch}$ of finite nuclei are then summarized in Table~\ref{table:Cluster}, which are generally consistent with the experimental data~\cite{Angeli2013_ADNDT99-69, Huang2021_CPC45-30002, Wang2021_CPC45-030003} except for the lightest nuclei at  $A=3$ with $\Delta B/A\approx 1$ MeV and $\Delta R_\mathrm{ch}\approx 0.3$ fm~\cite{Xia2025_PRC112-025801}. This should be improved in our future works by properly incorporating various corrections to the mean field approximation employed here, e.g., the inclusion of proton-neutron pairing~\cite{Bally2025_EPJA61-140}. The binding energy of $^8$Be also deviates from experimental values, which can be fixed if we consider deformations and introduce rotational corrections~\cite{Rong2023_PRC108-054314}. By adding one or two $\Lambda$s to these nuclei, we can further obtain the binding energies $B_\Lambda$ and $B_{\Lambda\Lambda}$ for single $\Lambda$ and double $\Lambda$s in $\Lambda$ hypernuclei and $\Lambda$$\Lambda$ hypernuclei, while the corresponding charge radii $R_\mathrm{ch}^\Lambda$ and $R_\mathrm{ch}^{\Lambda\Lambda}$ are fixed as well. As indicated in Table~\ref{table:Cluster}, due to the attractive $\Lambda$-$N$ interaction, the binding energy of $\Lambda$ hyperon is always positive and increases with the mass number $A$, which are expected to reach $B_\Lambda=30$ MeV as $A\rightarrow \infty$ according to Eq.~(\ref{eq:Vlam}). Except for the light nuclei with $A=3$ or 4, the charge radii decrease with the inclusion of $\Lambda$ hyperons, i.e., the core becomes more compact due to the impurity effects of $\Lambda$ hyperons with $R_\mathrm{ch}^{\Lambda\Lambda} < R_\mathrm{ch}^{\Lambda} < R_\mathrm{ch}$. Based on the binding energies $B_\Lambda$ and $B_{\Lambda\Lambda}$, the $\Lambda$-$\Lambda$ interaction energy $\Delta B_{\Lambda\Lambda}$ can be obtained with Eq.~(\ref{eq:DBLL}), which are negative and correspond to a repulsive $\Lambda$-$\Lambda$ interaction. This contradicts the experimental constraints as $\Lambda$-$\Lambda$ interaction was found to be attractive in ${}^6_{\Lambda\Lambda}$He~\cite{Takahashi2001_PRL87-212502, Ahn2013_PRC88-014003}.

\begin{figure}
  \centering
  \includegraphics[width=\linewidth]{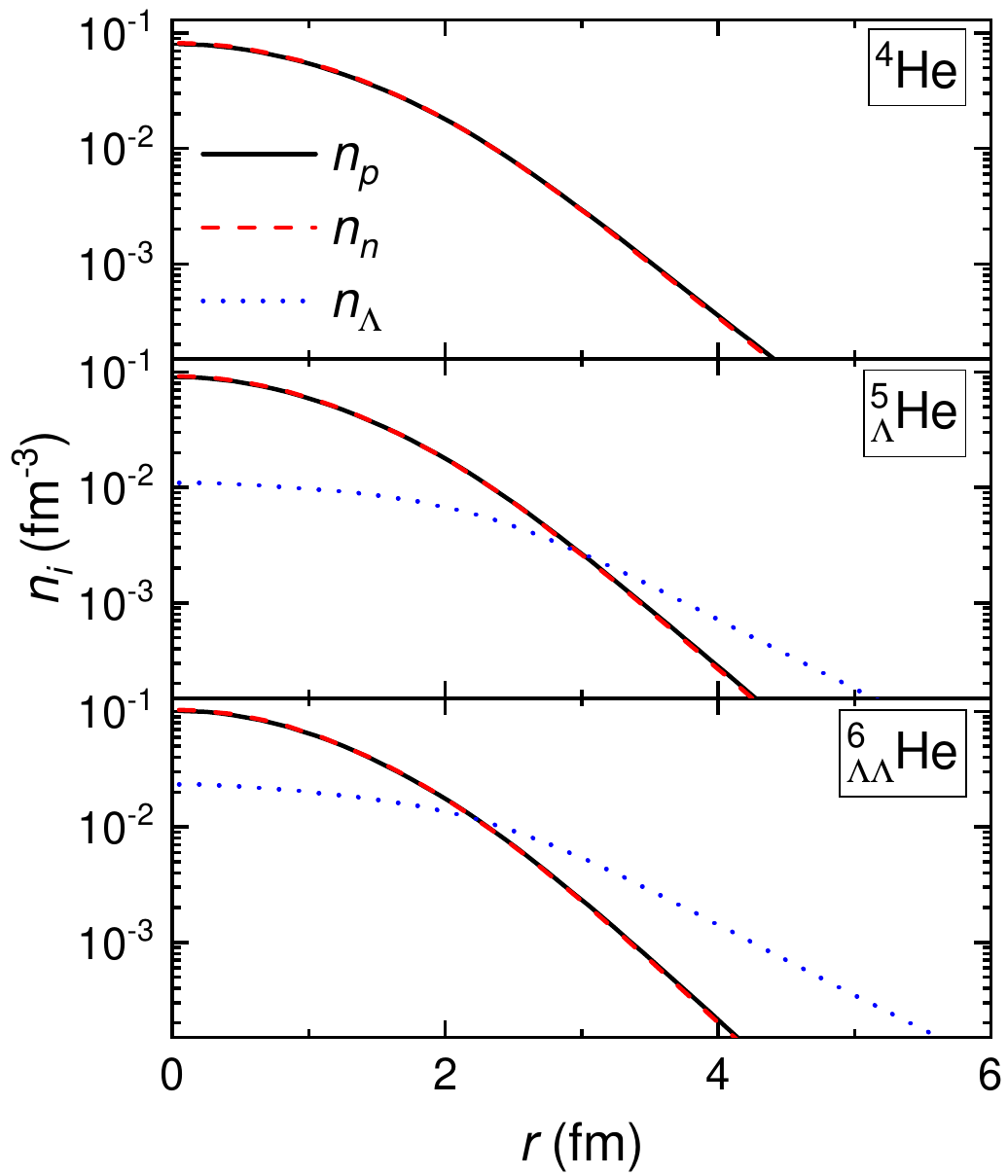}
  \caption{\label{Fig:Dense}Density profiles of $^4$He, ${}^{5}_\Lambda$He, and ${}^{6}_{\Lambda\Lambda}$He in vacuum, where the relativistic density functional DD-LZ1~\cite{Wei2020_CPC44-074107} is employed for the effective $N$-$N$ interactions and $g_{\omega\Lambda} = 1.143 g_{\omega N}$ ($g_{\sigma\Lambda} = g_{\sigma N}$) for the $\Lambda$-$N$ interactions.}
\end{figure}

To show these more explicitly, in Fig.~\ref{Fig:Dense} we present the obtained density profiles of $^{4}$He, ${}^{5}_\Lambda$He, and ${}^{6}_{\Lambda\Lambda}$He in vacuum with $n_\mathrm{gas}=0$ and the Coulomb potential $A=e N_p/4\pi r$ outside of the nucleus. The density profile of $\Lambda$ hyperons for ${}^{6}_{\Lambda\Lambda}$He is about twice higher than that of ${}^{5}_\Lambda$He since they both occupy the same orbital $1s_{1/2}$, while the density of each $\Lambda$ hyperon at the center becomes slightly larger for ${}^{6}_{\Lambda\Lambda}$He. Due to the presence of $\Lambda$ hyperons, the core $^4$He becomes more compact with its increasing central density, where $n_p(0) + n_n(0) = 0.161$, 0.182, and 0.204 fm${}^{-3}$ for $^4$He, ${}^{5}_\Lambda$He, and ${}^{6}_{\Lambda\Lambda}$He. This destabilizes the core and leads to a reduction of $B_\Lambda \approx 2.7$ MeV by approximately 1 MeV from that of the single-particle energy for $\Lambda$ hyperon ($\varepsilon_{1s_{1/2}}\approx -4$ MeV) in ${}^{5}_\Lambda$He. The corresponding $\Lambda$-$\Lambda$ interaction energy for ${}^{6}_{\Lambda\Lambda}$He is $\Delta B_{\Lambda\Lambda} = -0.133$ MeV and indicates a repulsive $\Lambda$-$\Lambda$ interaction, while in fact the $\Lambda$-$\Lambda$ interaction is attractive with $\Delta B_{\Lambda\Lambda} = 0.67 \pm 0.17$ MeV according to the KEK hybrid-emulsion experiment E373~\cite{Takahashi2001_PRL87-212502, Ahn2013_PRC88-014003}. This can be fixed by introducing explicitly $\Lambda$-$\Lambda$ attractive interaction via $\sigma^*$ meson exchange~\cite{Rong2021_PRC104-054321}, which is nonetheless neglected here since $\Delta B_{\Lambda\Lambda}$ is relatively small compared with $B_{\Lambda\Lambda}$ ($= 6.91 \pm 0.16$ MeV) and we only consider systems with $N_\Lambda\leq 2$.

\subsection{\label{sec:res_DDLZ1} $\Lambda$ hyperclusters in nuclear medium}

\begin{figure*}
  \centering
  \includegraphics[width=0.38\linewidth]{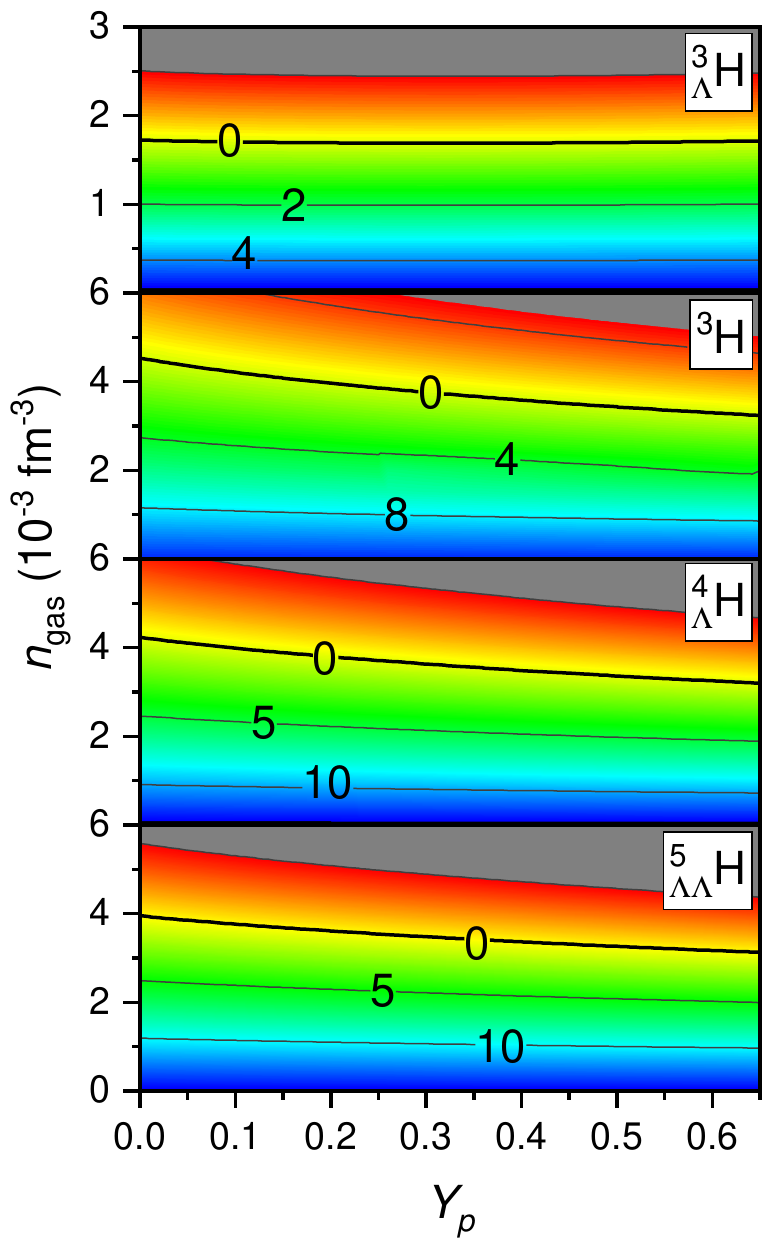} \ \ \ \ \ \ \ \
  \includegraphics[width=0.38\linewidth]{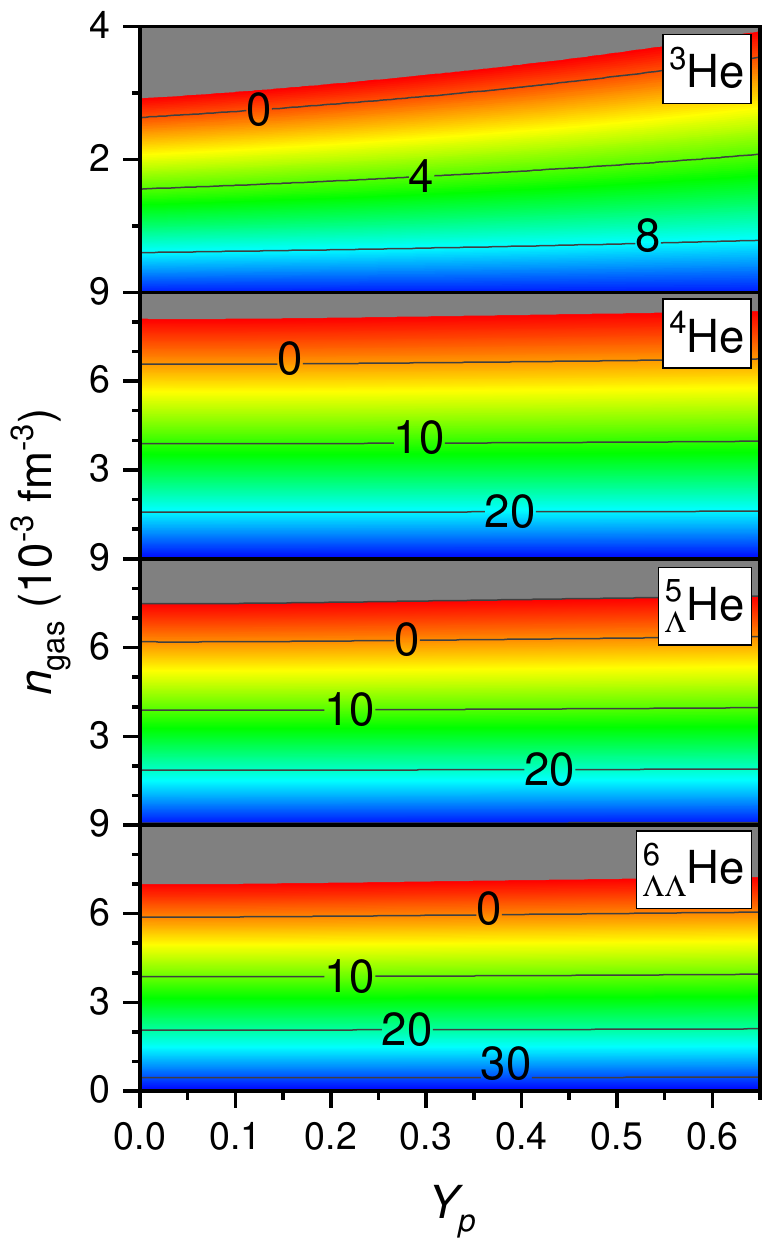}
  \caption{\label{Fig:DBall_DDLZ1} Binding energies $B$ (in MeV) of H and He clusters emersed in nuclear medium with density $n_{\mathrm{gas}}=n_{p,\mathrm{gas}}+n_{n,\mathrm{gas}}$ and proton fraction $Y_p = n_{p,\mathrm{gas}}/n_{\mathrm{gas}}$, which are obtained by employing the relativistic density functional DD-LZ1~\cite{Wei2020_CPC44-074107}.}
\end{figure*}

We then investigate the properties of various (hyper)clusters in nuclear medium. In Fig.~\ref{Fig:DBall_DDLZ1} the binding energy of H and He clusters emersed in nuclear medium with density $n_{\mathrm{gas}}=n_{p,\mathrm{gas}}+n_{n,\mathrm{gas}}$ and proton fraction $Y_p = n_{p,\mathrm{gas}}/n_{\mathrm{gas}}$ are presented, which are obtained by employing the relativistic density functional DD-LZ1~\cite{Wei2020_CPC44-074107}. The proton fraction $Y_p$ is thus related to the asymmetry parameter of the nuclear matter by $\delta = (n_{n, \mathrm{gas}}-n_{p, \mathrm{gas}})/n_\mathrm{gas}=1-2Y_p$. At vanishing densities $n_{\mathrm{gas}}\rightarrow 0$, as indicated by $B_0$ in Table~\ref{table:bind_exp}, in-medium clusters become more bound than those in vacuum due to the reduction of Coulomb energy~\cite{Xia2025_PRC112-025801}. The inclusion of $\Lambda$ hyperons also increase the binding energy by $B_\Lambda$ and $B_{\Lambda\Lambda}$ for single and double $\Lambda$ hyperclusters. As $n_{\mathrm{gas}}$ increases, the binding energies of clusters decrease, which would eventually become unstable and melt at sufficient large densities, i.e., Mott transition. Despite the increased binding energy for $\Lambda$ hyperclusters at small $n_{\mathrm{gas}}$, the critical densities for vanishing binding energy $B=0$ are smaller for $\Lambda$ hyperclusters than that of normal clusters without $\Lambda$ hyperons. This is attributed to the reduction of binding energy per baryon for light nuclei with addition of $\Lambda$ hyperons, i.e., $B_\mathrm{core}/(A-1)>B_\Lambda$ as indicated in Fig.~\ref{Fig:B_Lam}. Meanwhile, the variation of binding energies with respect to $Y_p$ are found to be sensitive to the differences between neutron and proton numbers in the cluster. For (hyper)clusters $^3$H, $^4_\Lambda$H, and $^5_{\Lambda\Lambda}$H, their binding energies decrease with the proton fraction of nuclear medium $Y_p$. If we replace a neutron in $^3$H by a $\Lambda$ hyperon, as indicated in Fig.~\ref{Fig:DBall_DDLZ1}, the binding energy for $^3_\Lambda$H is generally insensitive to $Y_p$ since $N_p = N_n=1$. Similarly, the binding energy for $^4$He, $^5_\Lambda$He, and $^6_{\Lambda\Lambda}$He also vary little with respect to the variation of $Y_p$  since $N_p=N_n=2$.

\begin{table}
  \centering
  \caption{\label{table:bind_exp} Best-fit coefficients for the expansion formula (\ref{eq:bind_exp}) of binding energy for various clusters presented in Fig.~\ref{Fig:DBall_DDLZ1} and Fig.~\ref{Fig:Bind-DDLZ1}, where the relativistic density functional DD-LZ1 is employed~\cite{Wei2020_CPC44-074107}.}
  \begin{tabular}{c|cccccc}
    \hline \hline
               &  $B_0$       & $a$       & $b$   &   $c$          &  $d$                 &  $f$    \\
             &  MeV         & GeV$\cdot$fm$^3$ &  GeV$\cdot$fm$^6$ &  GeV$\cdot$fm$^3$ & MeV &   MeV   \\ \hline
 $^3_\Lambda$H          &  5.208  &  $-$3.380 &   $195$ & $-$0.0131 & $-$0.31 & $0.49$  \\
    $^3$H               &  11.53  &  $-$3.089 &   $113$ & $-$1.11   & $-$1.2  & $1.1$  \\
 $^4_\Lambda$H          &  13.48  &  $-$3.822 &   $144$ & $-$1.19   & $-$1.1  & $1.2$   \\
$^5_{\Lambda\Lambda}$H  &  15.31  &  $-$4.580 &   $170$ & $-$1.19   & $-$1.1  & $1.2$   \\ \hline
             $^3$He     &  10.57  &  $-$4.388 &   $149$ & $1.197$   & $-$0.26 & $1.0$   \\
   $^4_\Lambda$He       &  12.70  &  $-$5.208 &   $182$ & $1.377$   & $-$0.34 & $0.81$  \\
    $^4$He              &  27.70  &  $-$5.029 &   $122$ & $0.166$   & $-$0.60 & $0.77$  \\
$^5_\Lambda$He          &  30.39  &  $-$5.800 &   $141$ & $0.233$   & $-$0.53 & $0.51$  \\
$^6_{\Lambda\Lambda}$He &  32.97  &  $-$6.577 &   $162$ & $0.285$   & $-$0.47 & $0.30$  \\ \hline
    $^5$Li              &  26.12  &  $-$6.076 &   $219$ & $0.374$   & $-$0.11 & $1.3$  \\
  $^6_\Lambda$Li        &  30.47  &  $-$7.383 &   $275$ & $0.512$   &    0.16 & $1.6$  \\
$^7_{\Lambda\Lambda}$Li &  34.35  &  $-$8.371 &   $266$ & $0.915$   &   0.024 & $1.6$  \\  \hline
             $^8$Be     &  43.34  &  $-$10.06 &   $262$ & $0.268$   & $0.10$  &   0     \\
     $^9_\Lambda$Be     &  50.80  &  $-$10.66 &   $230$ & $0.438$   & $0.049$ &   0     \\ \hline
             $^{12}$C   &  95.40  &  $-$14.55 &   $221$ & $1.03$    & $-$0.37 &   0     \\
     $^{13}_\Lambda$C   &  106.9  &  $-$15.44 &   $249$ & $1.03$    & $-$0.34 &   0     \\ \hline
             $^{16}$O   &  134.6  &  $-$19.32 &   $337$ & $1.41$    & $-$0.44 &   0    \\
     $^{17}_\Lambda$O   &  145.6  &  $-$20.19 &   $337$ & $1.47$    & $-$0.47 &   0    \\ \hline
             $^{40}$Ca  &  383.0  &  $-$48.47 &   $714$ & $4.53$    & $-$1.3  &   0     \\
     $^{41}_\Lambda$Ca  &  399.7  &  $-$49.27 &   $723$ & $4.69$    & $-$1.5  &   0     \\ \hline
             $^{48}$Ca  &  454.2  &  $-$52.32 &   $712$ & $-$4.81   & $-$3.4  &   0     \\
     $^{49}_\Lambda$Ca  &  473.1  &  $-$53.08 &   $715$ & $-$4.61   & $-$3.5  &   0     \\ \hline
  \end{tabular}
\end{table}

Similar to our previous investigation~\cite{Xia2025_PRC112-025801}, we fit the variation of binding energy with respect to the density $n_{\mathrm{gas}}$ and proton fraction $Y_p$ of nuclear medium with a polynomial formula, i.e.,
\begin{equation}
  B = B_0  + a n_{\mathrm{gas}}  + b n_{\mathrm{gas}}^2 + c Y_p n_{\mathrm{gas}} + d Y_p + f Y_p^2. \label{eq:bind_exp}
\end{equation}
The best-fit coefficients are then indicated in Table~\ref{table:bind_exp}. The coefficient $B_0$ represent the binding energy of (hyper)clusters in nuclear medium at vanishing $n_{\mathrm{gas}}$. The coefficients $a$ and $b$ indicate the density dependent behavior for the binding energy of clusters in nuclear medium. Despite the increased binding energy with the inclusion of $\Lambda$ hyperons, $a$ becomes more negative and leads to a faster decline for the binding energy as $n_{\mathrm{gas}}$ increases. The coefficients $c$, $d$, and $f$ are connected to the variation of binding energy with respect to the isospin asymmetry of nuclear medium, which are dominated by the differences between neutron and proton numbers in the cluster. For example, the binding energies for $^3_\Lambda$H, $^4$He, $^5_\Lambda$He, and $^6_{\Lambda\Lambda}$He ($N_p = N_n$) are roughly insensitive to $Y_p$. The binding energies for $^3$He,  $^5$Li, $^6_\Lambda$Li, and $^7_{\Lambda\Lambda}$Li ($N_p> N_n$) increase with $Y_p$, while those of $^3$H, $^4_\Lambda$H, $^5_{\Lambda\Lambda}$H, $^{48}$Ca, and $^{49}_\Lambda$Ca ($N_p< N_n$) decrease with $Y_p$ as $c$ takes negative values. This is mainly attributed to the isovector effects in nuclear medium with the differences in the mean-field potentials of protons and neutrons change as we vary $Y_p$.

\begin{figure}
  \centering
  \includegraphics[width=0.85\linewidth]{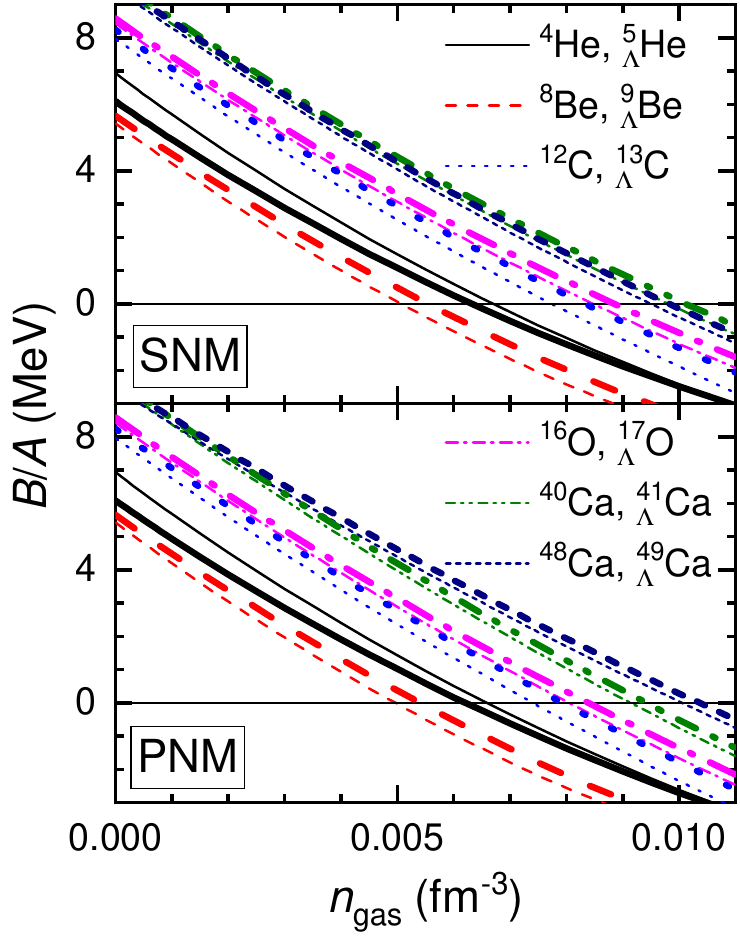}
  \caption{\label{Fig:Bind-DDLZ1} Binding energy per nucleon for various (hyper)clusters emersed in SNM and PNM as functions of the density of nuclear medium $n_{\mathrm{gas}}$, which are obtained by employing the relativistic density functional DD-LZ1~\cite{Wei2020_CPC44-074107}. The thin curves indicate normal clusters and thick curves correspond to those with one additional $\Lambda$ hyperon.}
\end{figure}

In Fig.~\ref{Fig:Bind-DDLZ1} we present the binding energy per baryon for heavier (hyper)clusters emersed in SNM and pure neutron matter (PNM, $Y_p = 0$) as functions of $n_{\mathrm{gas}}$. Similar to light clusters, the binding energies of (hyper)clusters decrease with $n_{\mathrm{gas}}$. Nevertheless, since the binding energy per baryon increase for $\Lambda$ hyperclusters with $N_p\geq 4$, they become more bound with the Mott transition densities increased. By comparing the upper and lower panels in Fig.~\ref{Fig:Bind-DDLZ1}, the variations of binding energies with respect to $Y_p$ can be identified, where ($\Lambda$ hyper)clusters with $N_p=N_n\lesssim 6$ are generally insensitive to the variations in $Y_p$. By fitting the binding energies indicated in Fig.~\ref{Fig:Bind-DDLZ1} with Eq.~(\ref{eq:bind_exp}), we can fix the corresponding coefficients, which are indicated in Table~\ref{table:bind_exp}. The variations of binding energies with respect to $n_{\mathrm{gas}}$ and $Y_p$ can then be reproduced.

\begin{figure}
  \centering
  \includegraphics[width=0.8\linewidth]{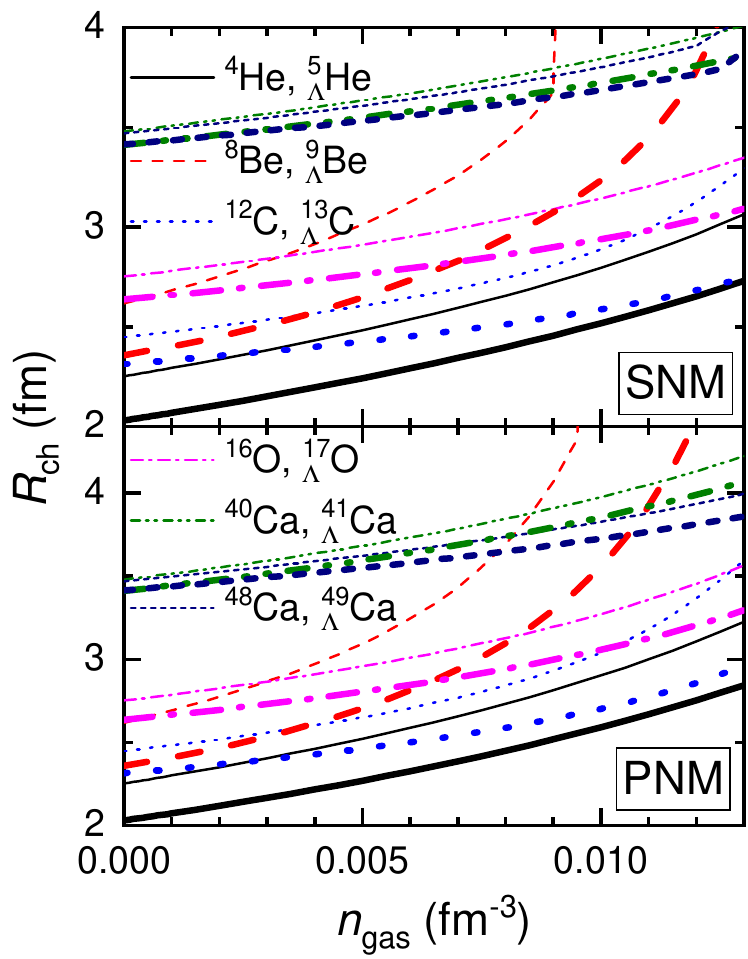}
  \caption{\label{Fig:Rch-DDLZ1} Similar as Fig.~\ref{Fig:Bind-DDLZ1} but for charge radii $R_\mathrm{ch}$ of in-medium (hyper)clusters.}
\end{figure}

The charge radii $R_\mathrm{ch}$ for the (hyper)clusters as functions of the density of nuclear medium $n_{\mathrm{gas}}$ are presented in Fig.~\ref{Fig:Rch-DDLZ1}, which correspond to those in Fig.~\ref{Fig:Bind-DDLZ1} and are obtained with the relativistic density functional DD-LZ1~\cite{Wei2020_CPC44-074107}. It is found that $R_\mathrm{ch}$ are generally increasing with $n_{\mathrm{gas}}$ while the binding energy is decreasing. Similar to the cases in vacuum as indicated in Table~\ref{table:Cluster}, the addition of a $\Lambda$ hyperon reduces the charge radii. The slope of $R_\mathrm{ch}$ as a function of $n_{\mathrm{gas}}$ also becomes smaller with the addition of a $\Lambda$ hyperon. By comparing the upper and lower panels in Fig.~\ref{Fig:Rch-DDLZ1}, the isovector effects in nuclear medium can be identified, e.g., $R_\mathrm{ch}$ increases slightly faster for (hyper)clusters with in PNM than those in SNM as we increase $n_{\mathrm{gas}}$. For heavy (hyper)clusters such as $^{48}$Ca and $^{48}_\Lambda$Ca, the variations of $R_\mathrm{ch}$ with respect to $Y_p$ are less significant.

\begin{figure*}
\includegraphics[width=0.38\linewidth]{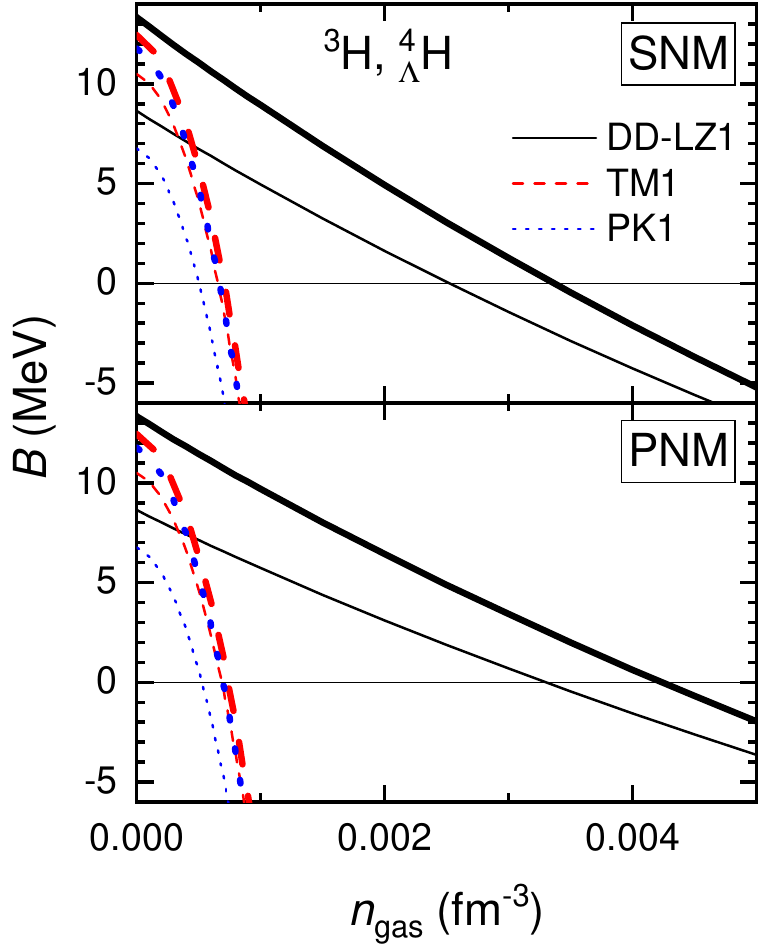} \ \ \  \ \ \
\includegraphics[width=0.38\linewidth]{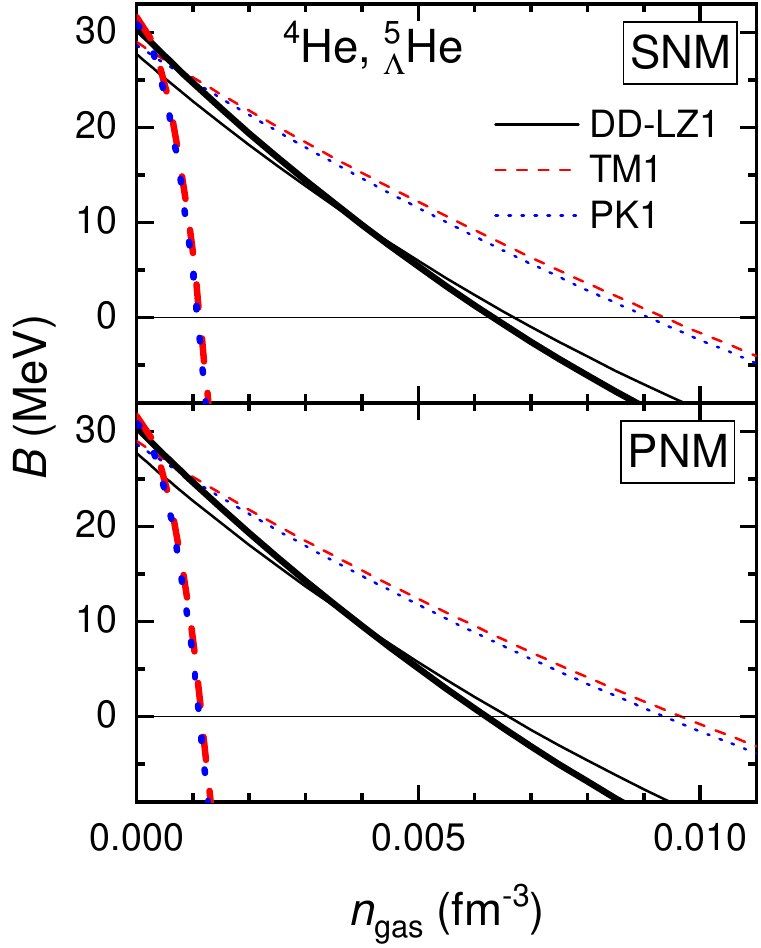}
  \caption{\label{Fig:Bind_HeH} Binding energies of $^3$H (thin curves), $^4$He (thin curves), $^4_\Lambda$H (thick curves) and $^5_\Lambda$He (thick curves) emersed in SNM and PNM as functions of $n_{\mathrm{gas}}$, which are obtained adopting the relativistic density functionals PK1~\cite{Long2004_PRC69-034319}, TM1~\cite{Sugahara1994_NPA579-557}, and DD-LZ1~\cite{Wei2020_CPC44-074107}. The results are fitted with Eq.~(\ref{eq:bind_exp}) with the coefficients indicated in Table~\ref{table:bind_exp_L}.   }
\end{figure*}

Finally, to show the variations due to the uncertainties in relativistic density functionals, we examine the binding energies of $^3$H, $^4$He, $^4_\Lambda$H, and $^5_\Lambda$He emersed in SNM and PNM adopting the functionals PK1~\cite{Long2004_PRC69-034319}, TM1~\cite{Sugahara1994_NPA579-557}, and DD-LZ1~\cite{Wei2020_CPC44-074107}, which are presented in Fig.~\ref{Fig:Bind_HeH}. As expected, the binding energy increases with the inclusion of $\Lambda$ hyperon, while the binding energy per baryon decreases since $B_\mathrm{core}/(A-1)>B_\Lambda$ for light hyperclusters as indicated in Fig.~\ref{Fig:B_Lam}. The binding energies of the (hyper)clusters are generally decreasing with the density of nuclear matter $n_{\mathrm{gas}}$. For the cluster $^4$He, the binding energy decline slowly with $n_{\mathrm{gas}}$, while slight variations on the in-medium binding energy shift are observed when we adopt different functionals. Due to the non-linear self-couplings of mesons, the binding energies of $^3$H, $^4_\Lambda$H, and $^5_\Lambda$He decrease drastically at $n_{\mathrm{gas}}\approx 0.0005$-0.0011 $\mathrm{fm}^{-3}$ when we adopt non-linear relativistic density functionals PK1~\cite{Long2004_PRC69-034319} and TM1~\cite{Sugahara1994_NPA579-557}, while the density-dependent functional DD-LZ1~\cite{Wei2020_CPC44-074107} predicts relatively slow decline of the binding energies. It is worth mentioning that the binding energy of $^5_\Lambda$He decreases drastically when non-linear relativistic density functionals are adopted, even though that of $^4$He decreases slowly with $n_{\mathrm{gas}}$. In such cases, the addition of $\Lambda$ hyperons destabilize the light (hyper)clusters which is particular the case when non-linear relativistic density functionals are adopted.

\begin{table}
  \centering
  \caption{\label{table:bind_exp_L} Coefficients for the expansion formula (\ref{eq:bind_exp}) of binding energies for
  in-medium $^3$H, $^4$He, $^4_\Lambda$H, and $^5_\Lambda$He presented in Fig.~\ref{Fig:Bind_HeH}, where $f=0$ is assumed.}
  \begin{tabular}{c|c|ccccc}
    \hline \hline
  &           &  $B_0$       & $a$       & $b$   &   $c$          &  $d$                     \\
  &           &  MeV         & GeV$\cdot$fm$^3$ &  GeV$\cdot$fm$^6$ &  GeV$\cdot$fm$^3$ & MeV    \\ \hline
\multirow{2}*{$^3$H}
  &  PK1      &  $6.746$    & $-$1.86  &  $-$20752   &   $-$2.48     & $0.076$      \\
  &  TM1      &  $10.47$    & $-$1.70  &  $-$19761   &   $-$30.1     & $1.4$      \\ \hline
\multirow{2}*{$^4_\Lambda$H}
  &  PK1      &  $11.75$    & $-$2.35  &  $-$20913   &   $-$2.76     & $0.10$      \\
  &  TM1      &  $12.45$    & $-$2.28  &  $-$19747   &   $-$3.36     & $0.19$      \\ \hline
\multirow{2}*{$^4$He}
 &  PK1      &  $28.46$    & $-$3.58   &     57.0    &   $-$0.213    & $0.30$      \\
 &  TM1      &  $28.78$    & $-$3.51   &     53.8    &   $-$0.205    & $0.31$      \\ \hline
\multirow{2}*{$^5_\Lambda$He}
 &  PK1      &  $30.85$    & $-$2.80   &   $-$21831  &   $-$3.41     & $0.42$      \\
 &  TM1      &  $31.58$    & $-$2.98   &   $-$20408  &   $-$3.39     & $0.41$      \\ \hline
 \end{tabular}
\end{table}

\section{\label{sec:con}Conclusion}

Based on a hybrid treatment for clusters in nuclear matter~\cite{Xia2025_PRC112-025801}, we investigate the properties of light and heavy $\Lambda$ hyperclusters emersed in nuclear medium at various densities $n_{\mathrm{gas}}$ and proton fractions $Y_p$. Our findings are summarized as follows.
\begin{itemize}
  \item As the density of nuclear medium increases, the (hyper)clusters eventually become unbound with an increasing root-mean-square radius, which will cause the (hyper)clusters to melt, i.e., a Mott transition.
  \item The inclusion of $\Lambda$ hyperons generally increases the binding energies of ($\Lambda$ hyper)clusters. For light $\Lambda$ hyperclusters with $N_p < 4$, the binding energies per baryon are smaller and decreasing faster with $n_{\mathrm{gas}}$.
  \item The binding energies of in-medium (hyper)clusters with $N_p> N_n$ ($N_p< N_n$) increase (decrease) with $Y_p$, while those with $N_p = N_n\lesssim 6$ are generally insensitive to $Y_p$.
  \item Light (hyper)clusters are destabilized drastically as $n_{\mathrm{gas}}$ increases when nonlinear relativistic density functionals are adopted.
  \item The variation of binding energies for various (hyper)clusters are well described by Eq.~(\ref{eq:bind_exp}) with the coefficients indicated in Table~\ref{table:bind_exp} and \ref{table:bind_exp_L}.
\end{itemize}

\begin{acknowledgments}
This work was supported by the National Natural Science Foundation of China (Grant Nos. 12275234 and 12205057), the National SKA Program of China (Grant No. 2020SKA0120300), the Science and Technology Plan Project of Guangxi (Guike AD23026250), the Natural Science Foundation  of Guangxi (2025GXNSFBA069076), the Central Government Guides Local Scientific and Technological Development Fund Projects (Guike ZY22096024), the Natural Science Foundation of Henan Province (242300421156), and the National  Natural Science Foundation of China (U2032141).
\end{acknowledgments}


\begin{thebibliography}{80}%
\makeatletter
\providecommand \@ifxundefined [1]{%
 \@ifx{#1\undefined}
}%
\providecommand \@ifnum [1]{%
 \ifnum #1\expandafter \@firstoftwo
 \else \expandafter \@secondoftwo
 \fi
}%
\providecommand \@ifx [1]{%
 \ifx #1\expandafter \@firstoftwo
 \else \expandafter \@secondoftwo
 \fi
}%
\providecommand \natexlab [1]{#1}%
\providecommand \enquote  [1]{``#1''}%
\providecommand \bibnamefont  [1]{#1}%
\providecommand \bibfnamefont [1]{#1}%
\providecommand \citenamefont [1]{#1}%
\providecommand \href@noop [0]{\@secondoftwo}%
\providecommand \href [0]{\begingroup \@sanitize@url \@href}%
\providecommand \@href[1]{\@@startlink{#1}\@@href}%
\providecommand \@@href[1]{\endgroup#1\@@endlink}%
\providecommand \@sanitize@url [0]{\catcode `\\12\catcode `\$12\catcode
  `\&12\catcode `\#12\catcode `\^12\catcode `\_12\catcode `\%12\relax}%
\providecommand \@@startlink[1]{}%
\providecommand \@@endlink[0]{}%
\providecommand \url  [0]{\begingroup\@sanitize@url \@url }%
\providecommand \@url [1]{\endgroup\@href {#1}{\urlprefix }}%
\providecommand \urlprefix  [0]{URL }%
\providecommand \Eprint [0]{\href }%
\providecommand \doibase [0]{http://dx.doi.org/}%
\providecommand \selectlanguage [0]{\@gobble}%
\providecommand \bibinfo  [0]{\@secondoftwo}%
\providecommand \bibfield  [0]{\@secondoftwo}%
\providecommand \translation [1]{[#1]}%
\providecommand \BibitemOpen [0]{}%
\providecommand \bibitemStop [0]{}%
\providecommand \bibitemNoStop [0]{.\EOS\space}%
\providecommand \EOS [0]{\spacefactor3000\relax}%
\providecommand \BibitemShut  [1]{\csname bibitem#1\endcsname}%
\let\auto@bib@innerbib\@empty
\bibitem [{\citenamefont {Danysz}\ and\ \citenamefont
  {Pniewski}(1953)}]{Danysz1953_PM44-348}%
  \BibitemOpen
  \bibfield  {author} {\bibinfo {author} {\bibfnamefont {M.}~\bibnamefont
  {Danysz}}\ and\ \bibinfo {author} {\bibfnamefont {J.}~\bibnamefont
  {Pniewski}},\ }\href@noop {} {\bibfield  {journal} {\bibinfo  {journal}
  {Philos. Mag.}\ }\textbf {\bibinfo {volume} {44}},\ \bibinfo {pages} {348}
  (\bibinfo {year} {1953})}\BibitemShut {NoStop}%
\bibitem [{\citenamefont {Hashimoto}\ and\ \citenamefont
  {Tamura}(2006)}]{Hashimoto2006_PPNP57-564}%
  \BibitemOpen
  \bibfield  {author} {\bibinfo {author} {\bibfnamefont {O.}~\bibnamefont
  {Hashimoto}}\ and\ \bibinfo {author} {\bibfnamefont {H.}~\bibnamefont
  {Tamura}},\ }\href {\doibase http://dx.doi.org/10.1016/j.ppnp.2005.07.001}
  {\bibfield  {journal} {\bibinfo  {journal} {Prog. Part. Nucl. Phys.}\
  }\textbf {\bibinfo {volume} {57}},\ \bibinfo {pages} {564 } (\bibinfo {year}
  {2006})}\BibitemShut {NoStop}%
\bibitem [{\citenamefont {Gal}\ \emph {et~al.}(2016)\citenamefont {Gal},
  \citenamefont {Hungerford},\ and\ \citenamefont
  {Millener}}]{Gal2016_RMP88-035004}%
  \BibitemOpen
  \bibfield  {author} {\bibinfo {author} {\bibfnamefont {A.}~\bibnamefont
  {Gal}}, \bibinfo {author} {\bibfnamefont {E.~V.}\ \bibnamefont {Hungerford}},
  \ and\ \bibinfo {author} {\bibfnamefont {D.~J.}\ \bibnamefont {Millener}},\
  }\href {\doibase 10.1103/RevModPhys.88.035004} {\bibfield  {journal}
  {\bibinfo  {journal} {Rev. Mod. Phys.}\ }\textbf {\bibinfo {volume} {88}},\
  \bibinfo {pages} {035004} (\bibinfo {year} {2016})}\BibitemShut {NoStop}%
\bibitem [{\citenamefont {Aoki}\ \emph {et~al.}(2009)\citenamefont {Aoki},
  \citenamefont {Bahk}, \citenamefont {Chung}, \citenamefont {Funahashi},
  \citenamefont {Hahn}, \citenamefont {Hanabata}, \citenamefont {Hara},
  \citenamefont {Hirata}, \citenamefont {Hoshino}, \citenamefont {Ieiri},
  \citenamefont {Iijima}, \citenamefont {Imai}, \citenamefont {Itow},
  \citenamefont {Jin-ya}, \citenamefont {Kazuno}, \citenamefont {Kim},
  \citenamefont {Kim}, \citenamefont {Kim}, \citenamefont {Kodama},
  \citenamefont {Kuze}, \citenamefont {Maeda}, \citenamefont {Masaike},
  \citenamefont {Masuoka}, \citenamefont {Matsuda}, \citenamefont {Matsui},
  \citenamefont {Nagase}, \citenamefont {Nagoshi}, \citenamefont {Nakamura},
  \citenamefont {Nakanishi}, \citenamefont {Nakano}, \citenamefont {Nakazawa},
  \citenamefont {Niwa}, \citenamefont {Oda}, \citenamefont {Okabe},
  \citenamefont {Ono}, \citenamefont {Ozaki}, \citenamefont {Park},
  \citenamefont {Park}, \citenamefont {Sakai}, \citenamefont {Sasaki},
  \citenamefont {Sato}, \citenamefont {Shibuya}, \citenamefont {Shimizu},
  \citenamefont {Song}, \citenamefont {Sugimoto}, \citenamefont {Tajima},
  \citenamefont {Takahashi}, \citenamefont {Takashima}, \citenamefont
  {Takeutchi}, \citenamefont {Tanaka}, \citenamefont {Teranaka}, \citenamefont
  {Tezuka}, \citenamefont {Togawa}, \citenamefont {Tsunemi}, \citenamefont
  {Ukai}, \citenamefont {Ushida}, \citenamefont {Watanabe}, \citenamefont
  {Yasuda}, \citenamefont {Yokota},\ and\ \citenamefont
  {Yoon}}]{Aoki2009_NPA828-191}%
  \BibitemOpen
  \bibfield  {author} {\bibinfo {author} {\bibfnamefont {S.}~\bibnamefont
  {Aoki}}, \bibinfo {author} {\bibfnamefont {S.}~\bibnamefont {Bahk}}, \bibinfo
  {author} {\bibfnamefont {S.}~\bibnamefont {Chung}}, \bibinfo {author}
  {\bibfnamefont {H.}~\bibnamefont {Funahashi}}, \bibinfo {author}
  {\bibfnamefont {C.}~\bibnamefont {Hahn}}, \bibinfo {author} {\bibfnamefont
  {M.}~\bibnamefont {Hanabata}}, \bibinfo {author} {\bibfnamefont
  {T.}~\bibnamefont {Hara}}, \bibinfo {author} {\bibfnamefont {S.}~\bibnamefont
  {Hirata}}, \bibinfo {author} {\bibfnamefont {K.}~\bibnamefont {Hoshino}},
  \bibinfo {author} {\bibfnamefont {M.}~\bibnamefont {Ieiri}}, \bibinfo
  {author} {\bibfnamefont {T.}~\bibnamefont {Iijima}}, \bibinfo {author}
  {\bibfnamefont {K.}~\bibnamefont {Imai}}, \bibinfo {author} {\bibfnamefont
  {Y.}~\bibnamefont {Itow}}, \bibinfo {author} {\bibfnamefont {T.}~\bibnamefont
  {Jin-ya}}, \bibinfo {author} {\bibfnamefont {M.}~\bibnamefont {Kazuno}},
  \bibinfo {author} {\bibfnamefont {C.}~\bibnamefont {Kim}}, \bibinfo {author}
  {\bibfnamefont {J.}~\bibnamefont {Kim}}, \bibinfo {author} {\bibfnamefont
  {S.}~\bibnamefont {Kim}}, \bibinfo {author} {\bibfnamefont {K.}~\bibnamefont
  {Kodama}}, \bibinfo {author} {\bibfnamefont {T.}~\bibnamefont {Kuze}},
  \bibinfo {author} {\bibfnamefont {Y.}~\bibnamefont {Maeda}}, \bibinfo
  {author} {\bibfnamefont {A.}~\bibnamefont {Masaike}}, \bibinfo {author}
  {\bibfnamefont {A.}~\bibnamefont {Masuoka}}, \bibinfo {author} {\bibfnamefont
  {Y.}~\bibnamefont {Matsuda}}, \bibinfo {author} {\bibfnamefont
  {A.}~\bibnamefont {Matsui}}, \bibinfo {author} {\bibfnamefont
  {Y.}~\bibnamefont {Nagase}}, \bibinfo {author} {\bibfnamefont
  {C.}~\bibnamefont {Nagoshi}}, \bibinfo {author} {\bibfnamefont
  {M.}~\bibnamefont {Nakamura}}, \bibinfo {author} {\bibfnamefont
  {S.}~\bibnamefont {Nakanishi}}, \bibinfo {author} {\bibfnamefont
  {T.}~\bibnamefont {Nakano}}, \bibinfo {author} {\bibfnamefont
  {K.}~\bibnamefont {Nakazawa}}, \bibinfo {author} {\bibfnamefont
  {K.}~\bibnamefont {Niwa}}, \bibinfo {author} {\bibfnamefont {H.}~\bibnamefont
  {Oda}}, \bibinfo {author} {\bibfnamefont {H.}~\bibnamefont {Okabe}}, \bibinfo
  {author} {\bibfnamefont {S.}~\bibnamefont {Ono}}, \bibinfo {author}
  {\bibfnamefont {R.}~\bibnamefont {Ozaki}}, \bibinfo {author} {\bibfnamefont
  {B.}~\bibnamefont {Park}}, \bibinfo {author} {\bibfnamefont {I.}~\bibnamefont
  {Park}}, \bibinfo {author} {\bibfnamefont {K.}~\bibnamefont {Sakai}},
  \bibinfo {author} {\bibfnamefont {T.}~\bibnamefont {Sasaki}}, \bibinfo
  {author} {\bibfnamefont {Y.}~\bibnamefont {Sato}}, \bibinfo {author}
  {\bibfnamefont {H.}~\bibnamefont {Shibuya}}, \bibinfo {author} {\bibfnamefont
  {H.}~\bibnamefont {Shimizu}}, \bibinfo {author} {\bibfnamefont
  {J.}~\bibnamefont {Song}}, \bibinfo {author} {\bibfnamefont {M.}~\bibnamefont
  {Sugimoto}}, \bibinfo {author} {\bibfnamefont {H.}~\bibnamefont {Tajima}},
  \bibinfo {author} {\bibfnamefont {H.}~\bibnamefont {Takahashi}}, \bibinfo
  {author} {\bibfnamefont {R.}~\bibnamefont {Takashima}}, \bibinfo {author}
  {\bibfnamefont {F.}~\bibnamefont {Takeutchi}}, \bibinfo {author}
  {\bibfnamefont {K.}~\bibnamefont {Tanaka}}, \bibinfo {author} {\bibfnamefont
  {M.}~\bibnamefont {Teranaka}}, \bibinfo {author} {\bibfnamefont
  {I.}~\bibnamefont {Tezuka}}, \bibinfo {author} {\bibfnamefont
  {H.}~\bibnamefont {Togawa}}, \bibinfo {author} {\bibfnamefont
  {T.}~\bibnamefont {Tsunemi}}, \bibinfo {author} {\bibfnamefont
  {M.}~\bibnamefont {Ukai}}, \bibinfo {author} {\bibfnamefont {N.}~\bibnamefont
  {Ushida}}, \bibinfo {author} {\bibfnamefont {T.}~\bibnamefont {Watanabe}},
  \bibinfo {author} {\bibfnamefont {N.}~\bibnamefont {Yasuda}}, \bibinfo
  {author} {\bibfnamefont {J.}~\bibnamefont {Yokota}}, \ and\ \bibinfo {author}
  {\bibfnamefont {C.}~\bibnamefont {Yoon}},\ }\href {\doibase
  https://doi.org/10.1016/j.nuclphysa.2009.07 .005} {\bibfield  {journal}
  {\bibinfo  {journal} {Nucl. Phys. A}\ }\textbf {\bibinfo {volume} {828}},\
  \bibinfo {pages} {191} (\bibinfo {year} {2009})}\BibitemShut {NoStop}%
\bibitem [{\citenamefont {Ahn}\ \emph {et~al.}(2013)\citenamefont {Ahn},
  \citenamefont {Akikawa}, \citenamefont {Aoki}, \citenamefont {Arai},
  \citenamefont {Bahk}, \citenamefont {Baik}, \citenamefont {Bassalleck},
  \citenamefont {Chung}, \citenamefont {Chung}, \citenamefont {Davis},
  \citenamefont {Fukuda}, \citenamefont {Hoshino}, \citenamefont {Ichikawa},
  \citenamefont {Ieiri}, \citenamefont {Imai}, \citenamefont {Itonaga},
  \citenamefont {Iwata}, \citenamefont {Iwata}, \citenamefont {Kanda},
  \citenamefont {Kaneko}, \citenamefont {Kawai}, \citenamefont {Kawasaki},
  \citenamefont {Kim}, \citenamefont {Kim}, \citenamefont {Kim}, \citenamefont
  {Kim}, \citenamefont {Kondo}, \citenamefont {Kouketsu}, \citenamefont {Kyaw},
  \citenamefont {Lee}, \citenamefont {McNabb}, \citenamefont {Min},
  \citenamefont {Mitsuhara}, \citenamefont {Miwa}, \citenamefont {Nakazawa},
  \citenamefont {Nagase}, \citenamefont {Nagoshi}, \citenamefont {Nakanishi},
  \citenamefont {Noumi}, \citenamefont {Ogawa}, \citenamefont {Okabe},
  \citenamefont {Oyama}, \citenamefont {Park}, \citenamefont {Park},
  \citenamefont {Park}, \citenamefont {Parker}, \citenamefont {Ra},
  \citenamefont {Rhee}, \citenamefont {Rusek}, \citenamefont {Sawa},
  \citenamefont {Shibuya}, \citenamefont {Sim}, \citenamefont {Saha},
  \citenamefont {Seki}, \citenamefont {Sekimoto}, \citenamefont {Song},
  \citenamefont {Takahashi}, \citenamefont {Takahashi}, \citenamefont
  {Takeutchi}, \citenamefont {Tanaka}, \citenamefont {Tanida}, \citenamefont
  {Tint}, \citenamefont {Tojo}, \citenamefont {Torii}, \citenamefont {Torikai},
  \citenamefont {Tovee}, \citenamefont {Tsunemi}, \citenamefont {Ukai},
  \citenamefont {Ushida}, \citenamefont {Wint}, \citenamefont {Yamamoto},
  \citenamefont {Yasuda}, \citenamefont {Yang}, \citenamefont {Yoon},
  \citenamefont {Yoon}, \citenamefont {Yosoi}, \citenamefont {Yoshida},\ and\
  \citenamefont {Zhu}}]{Ahn2013_PRC88-014003}%
  \BibitemOpen
  \bibfield  {author} {\bibinfo {author} {\bibfnamefont {J.~K.}\ \bibnamefont
  {Ahn}}, \bibinfo {author} {\bibfnamefont {H.}~\bibnamefont {Akikawa}},
  \bibinfo {author} {\bibfnamefont {S.}~\bibnamefont {Aoki}}, \bibinfo {author}
  {\bibfnamefont {K.}~\bibnamefont {Arai}}, \bibinfo {author} {\bibfnamefont
  {S.~Y.}\ \bibnamefont {Bahk}}, \bibinfo {author} {\bibfnamefont {K.~M.}\
  \bibnamefont {Baik}}, \bibinfo {author} {\bibfnamefont {B.}~\bibnamefont
  {Bassalleck}}, \bibinfo {author} {\bibfnamefont {J.~H.}\ \bibnamefont
  {Chung}}, \bibinfo {author} {\bibfnamefont {M.~S.}\ \bibnamefont {Chung}},
  \bibinfo {author} {\bibfnamefont {D.~H.}\ \bibnamefont {Davis}}, \bibinfo
  {author} {\bibfnamefont {T.}~\bibnamefont {Fukuda}}, \bibinfo {author}
  {\bibfnamefont {K.}~\bibnamefont {Hoshino}}, \bibinfo {author} {\bibfnamefont
  {A.}~\bibnamefont {Ichikawa}}, \bibinfo {author} {\bibfnamefont
  {M.}~\bibnamefont {Ieiri}}, \bibinfo {author} {\bibfnamefont
  {K.}~\bibnamefont {Imai}}, \bibinfo {author} {\bibfnamefont {K.}~\bibnamefont
  {Itonaga}}, \bibinfo {author} {\bibfnamefont {Y.~H.}\ \bibnamefont {Iwata}},
  \bibinfo {author} {\bibfnamefont {Y.~S.}\ \bibnamefont {Iwata}}, \bibinfo
  {author} {\bibfnamefont {H.}~\bibnamefont {Kanda}}, \bibinfo {author}
  {\bibfnamefont {M.}~\bibnamefont {Kaneko}}, \bibinfo {author} {\bibfnamefont
  {T.}~\bibnamefont {Kawai}}, \bibinfo {author} {\bibfnamefont
  {M.}~\bibnamefont {Kawasaki}}, \bibinfo {author} {\bibfnamefont {C.~O.}\
  \bibnamefont {Kim}}, \bibinfo {author} {\bibfnamefont {J.~Y.}\ \bibnamefont
  {Kim}}, \bibinfo {author} {\bibfnamefont {S.~H.}\ \bibnamefont {Kim}},
  \bibinfo {author} {\bibfnamefont {S.~J.}\ \bibnamefont {Kim}}, \bibinfo
  {author} {\bibfnamefont {Y.}~\bibnamefont {Kondo}}, \bibinfo {author}
  {\bibfnamefont {T.}~\bibnamefont {Kouketsu}}, \bibinfo {author}
  {\bibfnamefont {H.~N.}\ \bibnamefont {Kyaw}}, \bibinfo {author}
  {\bibfnamefont {Y.~L.}\ \bibnamefont {Lee}}, \bibinfo {author} {\bibfnamefont
  {J.~W.~C.}\ \bibnamefont {McNabb}}, \bibinfo {author} {\bibfnamefont {A.~A.}\
  \bibnamefont {Min}}, \bibinfo {author} {\bibfnamefont {M.}~\bibnamefont
  {Mitsuhara}}, \bibinfo {author} {\bibfnamefont {K.}~\bibnamefont {Miwa}},
  \bibinfo {author} {\bibfnamefont {K.}~\bibnamefont {Nakazawa}}, \bibinfo
  {author} {\bibfnamefont {Y.}~\bibnamefont {Nagase}}, \bibinfo {author}
  {\bibfnamefont {C.}~\bibnamefont {Nagoshi}}, \bibinfo {author} {\bibfnamefont
  {Y.}~\bibnamefont {Nakanishi}}, \bibinfo {author} {\bibfnamefont
  {H.}~\bibnamefont {Noumi}}, \bibinfo {author} {\bibfnamefont
  {S.}~\bibnamefont {Ogawa}}, \bibinfo {author} {\bibfnamefont
  {H.}~\bibnamefont {Okabe}}, \bibinfo {author} {\bibfnamefont
  {K.}~\bibnamefont {Oyama}}, \bibinfo {author} {\bibfnamefont {B.~D.}\
  \bibnamefont {Park}}, \bibinfo {author} {\bibfnamefont {H.~M.}\ \bibnamefont
  {Park}}, \bibinfo {author} {\bibfnamefont {I.~G.}\ \bibnamefont {Park}},
  \bibinfo {author} {\bibfnamefont {J.}~\bibnamefont {Parker}}, \bibinfo
  {author} {\bibfnamefont {Y.~S.}\ \bibnamefont {Ra}}, \bibinfo {author}
  {\bibfnamefont {J.~T.}\ \bibnamefont {Rhee}}, \bibinfo {author}
  {\bibfnamefont {A.}~\bibnamefont {Rusek}}, \bibinfo {author} {\bibfnamefont
  {A.}~\bibnamefont {Sawa}}, \bibinfo {author} {\bibfnamefont {H.}~\bibnamefont
  {Shibuya}}, \bibinfo {author} {\bibfnamefont {K.~S.}\ \bibnamefont {Sim}},
  \bibinfo {author} {\bibfnamefont {P.~K.}\ \bibnamefont {Saha}}, \bibinfo
  {author} {\bibfnamefont {D.}~\bibnamefont {Seki}}, \bibinfo {author}
  {\bibfnamefont {M.}~\bibnamefont {Sekimoto}}, \bibinfo {author}
  {\bibfnamefont {J.~S.}\ \bibnamefont {Song}}, \bibinfo {author}
  {\bibfnamefont {H.}~\bibnamefont {Takahashi}}, \bibinfo {author}
  {\bibfnamefont {T.}~\bibnamefont {Takahashi}}, \bibinfo {author}
  {\bibfnamefont {F.}~\bibnamefont {Takeutchi}}, \bibinfo {author}
  {\bibfnamefont {H.}~\bibnamefont {Tanaka}}, \bibinfo {author} {\bibfnamefont
  {K.}~\bibnamefont {Tanida}}, \bibinfo {author} {\bibfnamefont {K.~T.}\
  \bibnamefont {Tint}}, \bibinfo {author} {\bibfnamefont {J.}~\bibnamefont
  {Tojo}}, \bibinfo {author} {\bibfnamefont {H.}~\bibnamefont {Torii}},
  \bibinfo {author} {\bibfnamefont {S.}~\bibnamefont {Torikai}}, \bibinfo
  {author} {\bibfnamefont {D.~N.}\ \bibnamefont {Tovee}}, \bibinfo {author}
  {\bibfnamefont {T.}~\bibnamefont {Tsunemi}}, \bibinfo {author} {\bibfnamefont
  {M.}~\bibnamefont {Ukai}}, \bibinfo {author} {\bibfnamefont {N.}~\bibnamefont
  {Ushida}}, \bibinfo {author} {\bibfnamefont {T.}~\bibnamefont {Wint}},
  \bibinfo {author} {\bibfnamefont {K.}~\bibnamefont {Yamamoto}}, \bibinfo
  {author} {\bibfnamefont {N.}~\bibnamefont {Yasuda}}, \bibinfo {author}
  {\bibfnamefont {J.~T.}\ \bibnamefont {Yang}}, \bibinfo {author}
  {\bibfnamefont {C.~J.}\ \bibnamefont {Yoon}}, \bibinfo {author}
  {\bibfnamefont {C.~S.}\ \bibnamefont {Yoon}}, \bibinfo {author}
  {\bibfnamefont {M.}~\bibnamefont {Yosoi}}, \bibinfo {author} {\bibfnamefont
  {T.}~\bibnamefont {Yoshida}}, \ and\ \bibinfo {author} {\bibfnamefont
  {L.}~\bibnamefont {Zhu}} (\bibinfo {collaboration} {E373 (KEK-PS)
  Collaboration}),\ }\href {\doibase 10.1103/PhysRevC.88.014003} {\bibfield
  {journal} {\bibinfo  {journal} {Phys. Rev. C}\ }\textbf {\bibinfo {volume}
  {88}},\ \bibinfo {pages} {014003} (\bibinfo {year} {2013})}\BibitemShut
  {NoStop}%
\bibitem [{\citenamefont {Gal}\ \emph {et~al.}(1971)\citenamefont {Gal},
  \citenamefont {Soper},\ and\ \citenamefont {Dalitz}}]{Gal1971_AP63-53}%
  \BibitemOpen
  \bibfield  {author} {\bibinfo {author} {\bibfnamefont {A.}~\bibnamefont
  {Gal}}, \bibinfo {author} {\bibfnamefont {J.}~\bibnamefont {Soper}}, \ and\
  \bibinfo {author} {\bibfnamefont {R.}~\bibnamefont {Dalitz}},\ }\href
  {\doibase http://dx.doi.org/10.1016/0003-4916(71)90297-1} {\bibfield
  {journal} {\bibinfo  {journal} {Ann. Phys.}\ }\textbf {\bibinfo {volume}
  {63}},\ \bibinfo {pages} {53 } (\bibinfo {year} {1971})}\BibitemShut
  {NoStop}%
\bibitem [{\citenamefont {Dalitz}\ and\ \citenamefont
  {Gal}(1978)}]{Dalitz1978_AP116-167}%
  \BibitemOpen
  \bibfield  {author} {\bibinfo {author} {\bibfnamefont {R.}~\bibnamefont
  {Dalitz}}\ and\ \bibinfo {author} {\bibfnamefont {A.}~\bibnamefont {Gal}},\
  }\href {\doibase http://dx.doi.org/10.1016/0003-4916(78)90008-8} {\bibfield
  {journal} {\bibinfo  {journal} {Ann. Phys.}\ }\textbf {\bibinfo {volume}
  {116}},\ \bibinfo {pages} {167 } (\bibinfo {year} {1978})}\BibitemShut
  {NoStop}%
\bibitem [{\citenamefont {Millener}(2008)}]{Millener2008_NPA804-84}%
  \BibitemOpen
  \bibfield  {author} {\bibinfo {author} {\bibfnamefont {D.}~\bibnamefont
  {Millener}},\ }\href {\doibase
  http://dx.doi.org/10.1016/j.nuclphysa.2008.02.252} {\bibfield  {journal}
  {\bibinfo  {journal} {Nucl. Phys. A}\ }\textbf {\bibinfo {volume} {804}},\
  \bibinfo {pages} {84} (\bibinfo {year} {2008})}\BibitemShut {NoStop}%
\bibitem [{\citenamefont {Millener}(2013)}]{Millener2013_NPA914-109}%
  \BibitemOpen
  \bibfield  {author} {\bibinfo {author} {\bibfnamefont {D.}~\bibnamefont
  {Millener}},\ }\href {\doibase
  http://dx.doi.org/10.1016/j.nuclphysa.2013.01.023} {\bibfield  {journal}
  {\bibinfo  {journal} {Nucl. Phys. A}\ }\textbf {\bibinfo {volume} {914}},\
  \bibinfo {pages} {109} (\bibinfo {year} {2013})}\BibitemShut {NoStop}%
\bibitem [{\citenamefont {Motoba}\ \emph {et~al.}(1983)\citenamefont {Motoba},
  \citenamefont {Band\={o}},\ and\ \citenamefont
  {Ikeda}}]{Motoba1983_PTP70-189}%
  \BibitemOpen
  \bibfield  {author} {\bibinfo {author} {\bibfnamefont {T.}~\bibnamefont
  {Motoba}}, \bibinfo {author} {\bibfnamefont {H.}~\bibnamefont {Band\={o}}}, \
  and\ \bibinfo {author} {\bibfnamefont {K.}~\bibnamefont {Ikeda}},\ }\href
  {\doibase 10.1143/PTP.70.189} {\bibfield  {journal} {\bibinfo  {journal}
  {Prog. Theor. Phys.}\ }\textbf {\bibinfo {volume} {70}},\ \bibinfo {pages}
  {189} (\bibinfo {year} {1983})}\BibitemShut {NoStop}%
\bibitem [{\citenamefont {Hiyama}\ \emph {et~al.}(2002)\citenamefont {Hiyama},
  \citenamefont {Kamimura}, \citenamefont {Motoba}, \citenamefont {Yamada},\
  and\ \citenamefont {Yamamoto}}]{Hiyama2002_PRC66-024007}%
  \BibitemOpen
  \bibfield  {author} {\bibinfo {author} {\bibfnamefont {E.}~\bibnamefont
  {Hiyama}}, \bibinfo {author} {\bibfnamefont {M.}~\bibnamefont {Kamimura}},
  \bibinfo {author} {\bibfnamefont {T.}~\bibnamefont {Motoba}}, \bibinfo
  {author} {\bibfnamefont {T.}~\bibnamefont {Yamada}}, \ and\ \bibinfo {author}
  {\bibfnamefont {Y.}~\bibnamefont {Yamamoto}},\ }\href {\doibase
  10.1103/PhysRevC.66.024007} {\bibfield  {journal} {\bibinfo  {journal} {Phys.
  Rev. C}\ }\textbf {\bibinfo {volume} {66}},\ \bibinfo {pages} {024007}
  (\bibinfo {year} {2002})}\BibitemShut {NoStop}%
\bibitem [{\citenamefont {Hiyama}\ \emph {et~al.}(2006)\citenamefont {Hiyama},
  \citenamefont {Yamamoto}, \citenamefont {Rijken},\ and\ \citenamefont
  {Motoba}}]{Hiyama2006_PRC74-054312}%
  \BibitemOpen
  \bibfield  {author} {\bibinfo {author} {\bibfnamefont {E.}~\bibnamefont
  {Hiyama}}, \bibinfo {author} {\bibfnamefont {Y.}~\bibnamefont {Yamamoto}},
  \bibinfo {author} {\bibfnamefont {T.~A.}\ \bibnamefont {Rijken}}, \ and\
  \bibinfo {author} {\bibfnamefont {T.}~\bibnamefont {Motoba}},\ }\href
  {\doibase 10.1103/PhysRevC.74.054312} {\bibfield  {journal} {\bibinfo
  {journal} {Phys. Rev. C}\ }\textbf {\bibinfo {volume} {74}},\ \bibinfo
  {pages} {054312} (\bibinfo {year} {2006})}\BibitemShut {NoStop}%
\bibitem [{\citenamefont {Hiyama}\ \emph {et~al.}(2009)\citenamefont {Hiyama},
  \citenamefont {Yamamoto}, \citenamefont {Motoba},\ and\ \citenamefont
  {Kamimura}}]{Hiyama2009_PRC80-054321}%
  \BibitemOpen
  \bibfield  {author} {\bibinfo {author} {\bibfnamefont {E.}~\bibnamefont
  {Hiyama}}, \bibinfo {author} {\bibfnamefont {Y.}~\bibnamefont {Yamamoto}},
  \bibinfo {author} {\bibfnamefont {T.}~\bibnamefont {Motoba}}, \ and\ \bibinfo
  {author} {\bibfnamefont {M.}~\bibnamefont {Kamimura}},\ }\href {\doibase
  10.1103/PhysRevC.80.054321} {\bibfield  {journal} {\bibinfo  {journal} {Phys.
  Rev. C}\ }\textbf {\bibinfo {volume} {80}},\ \bibinfo {pages} {054321}
  (\bibinfo {year} {2009})}\BibitemShut {NoStop}%
\bibitem [{\citenamefont {Band\={o}}\ \emph {et~al.}(1990)\citenamefont
  {Band\={o}}, \citenamefont {Motoba},\ and\ \citenamefont
  {\u{Z}ofka}}]{Bando1990_IJMPA05-4021}%
  \BibitemOpen
  \bibfield  {author} {\bibinfo {author} {\bibfnamefont {H.}~\bibnamefont
  {Band\={o}}}, \bibinfo {author} {\bibfnamefont {T.}~\bibnamefont {Motoba}}, \
  and\ \bibinfo {author} {\bibfnamefont {J.}~\bibnamefont {\u{Z}ofka}},\ }\href
  {\doibase 10.1142/S0217751X90001732} {\bibfield  {journal} {\bibinfo
  {journal} {Int. J. Mod. Phys. A}\ }\textbf {\bibinfo {volume} {05}},\
  \bibinfo {pages} {4021} (\bibinfo {year} {1990})}\BibitemShut {NoStop}%
\bibitem [{\citenamefont {Isaka}\ \emph {et~al.}(2013)\citenamefont {Isaka},
  \citenamefont {Kimura}, \citenamefont {Dot\'e},\ and\ \citenamefont
  {Ohnishi}}]{Isaka2013_PRC87-021304}%
  \BibitemOpen
  \bibfield  {author} {\bibinfo {author} {\bibfnamefont {M.}~\bibnamefont
  {Isaka}}, \bibinfo {author} {\bibfnamefont {M.}~\bibnamefont {Kimura}},
  \bibinfo {author} {\bibfnamefont {A.}~\bibnamefont {Dot\'e}}, \ and\ \bibinfo
  {author} {\bibfnamefont {A.}~\bibnamefont {Ohnishi}},\ }\href {\doibase
  10.1103/PhysRevC.87.021304} {\bibfield  {journal} {\bibinfo  {journal} {Phys.
  Rev. C}\ }\textbf {\bibinfo {volume} {87}},\ \bibinfo {pages} {021304}
  (\bibinfo {year} {2013})}\BibitemShut {NoStop}%
\bibitem [{\citenamefont {Hu}\ \emph {et~al.}(2014)\citenamefont {Hu},
  \citenamefont {Li}, \citenamefont {Toki},\ and\ \citenamefont
  {Zuo}}]{Hu2014_PRC89-025802}%
  \BibitemOpen
  \bibfield  {author} {\bibinfo {author} {\bibfnamefont {J.~N.}\ \bibnamefont
  {Hu}}, \bibinfo {author} {\bibfnamefont {A.}~\bibnamefont {Li}}, \bibinfo
  {author} {\bibfnamefont {H.}~\bibnamefont {Toki}}, \ and\ \bibinfo {author}
  {\bibfnamefont {W.}~\bibnamefont {Zuo}},\ }\href {\doibase
  10.1103/PhysRevC.89.025802} {\bibfield  {journal} {\bibinfo  {journal} {Phys.
  Rev. C}\ }\textbf {\bibinfo {volume} {89}},\ \bibinfo {pages} {025802}
  (\bibinfo {year} {2014})}\BibitemShut {NoStop}%
\bibitem [{\citenamefont {Brockmann}\ and\ \citenamefont
  {Weise}(1977)}]{Brockmann1977_PLB69-167}%
  \BibitemOpen
  \bibfield  {author} {\bibinfo {author} {\bibfnamefont {R.}~\bibnamefont
  {Brockmann}}\ and\ \bibinfo {author} {\bibfnamefont {W.}~\bibnamefont
  {Weise}},\ }\href {\doibase http://dx.doi.org/10.1016/0370-2693(77)90635-9}
  {\bibfield  {journal} {\bibinfo  {journal} {Phys. Lett. B}\ }\textbf
  {\bibinfo {volume} {69}},\ \bibinfo {pages} {167} (\bibinfo {year}
  {1977})}\BibitemShut {NoStop}%
\bibitem [{\citenamefont {Boguta}\ and\ \citenamefont
  {Bohrmann}(1981)}]{Boguta1981_PLB102-93}%
  \BibitemOpen
  \bibfield  {author} {\bibinfo {author} {\bibfnamefont {J.}~\bibnamefont
  {Boguta}}\ and\ \bibinfo {author} {\bibfnamefont {S.}~\bibnamefont
  {Bohrmann}},\ }\href {\doibase
  http://dx.doi.org/10.1016/0370-2693(81)91037-6} {\bibfield  {journal}
  {\bibinfo  {journal} {Phys. Lett. B}\ }\textbf {\bibinfo {volume} {102}},\
  \bibinfo {pages} {93} (\bibinfo {year} {1981})}\BibitemShut {NoStop}%
\bibitem [{\citenamefont {Mare{\v{s}}}\ and\ \citenamefont
  {{\v{Z}}ofka}(1989)}]{Mares1989_ZPA333-209}%
  \BibitemOpen
  \bibfield  {author} {\bibinfo {author} {\bibfnamefont {J.}~\bibnamefont
  {Mare{\v{s}}}}\ and\ \bibinfo {author} {\bibfnamefont {J.}~\bibnamefont
  {{\v{Z}}ofka}},\ }\href {\doibase 10.1007/BF01565152} {\bibfield  {journal}
  {\bibinfo  {journal} {Z. Phys. A}\ }\textbf {\bibinfo {volume} {333}},\
  \bibinfo {pages} {209} (\bibinfo {year} {1989})}\BibitemShut {NoStop}%
\bibitem [{\citenamefont {Mare\v{s}}\ and\ \citenamefont
  {Jennings}(1994)}]{Mares1994_PRC49-2472}%
  \BibitemOpen
  \bibfield  {author} {\bibinfo {author} {\bibfnamefont {J.}~\bibnamefont
  {Mare\v{s}}}\ and\ \bibinfo {author} {\bibfnamefont {B.~K.}\ \bibnamefont
  {Jennings}},\ }\href {\doibase 10.1103/PhysRevC.49.2472} {\bibfield
  {journal} {\bibinfo  {journal} {Phys. Rev. C}\ }\textbf {\bibinfo {volume}
  {49}},\ \bibinfo {pages} {2472} (\bibinfo {year} {1994})}\BibitemShut
  {NoStop}%
\bibitem [{\citenamefont {Sugahara}\ and\ \citenamefont
  {Toki}(1994{\natexlab{a}})}]{Toki1994_PTP92-803}%
  \BibitemOpen
  \bibfield  {author} {\bibinfo {author} {\bibfnamefont {Y.}~\bibnamefont
  {Sugahara}}\ and\ \bibinfo {author} {\bibfnamefont {H.}~\bibnamefont
  {Toki}},\ }\href {\doibase 10.1143/ptp/92.4.803} {\bibfield  {journal}
  {\bibinfo  {journal} {Prog. Theor. Phys.}\ }\textbf {\bibinfo {volume}
  {92}},\ \bibinfo {pages} {803} (\bibinfo {year}
  {1994}{\natexlab{a}})}\BibitemShut {NoStop}%
\bibitem [{\citenamefont {Song}\ \emph {et~al.}(2010)\citenamefont {Song},
  \citenamefont {Yao}, \citenamefont {LV},\ and\ \citenamefont
  {Meng}}]{Song2010_IJMPE19-2538}%
  \BibitemOpen
  \bibfield  {author} {\bibinfo {author} {\bibfnamefont {C.~Y.}\ \bibnamefont
  {Song}}, \bibinfo {author} {\bibfnamefont {J.~M.}\ \bibnamefont {Yao}},
  \bibinfo {author} {\bibfnamefont {H.~F.}\ \bibnamefont {LV}}, \ and\ \bibinfo
  {author} {\bibfnamefont {J.}~\bibnamefont {Meng}},\ }\href {\doibase
  10.1142/S0218301310017058} {\bibfield  {journal} {\bibinfo  {journal} {Int.
  J. Mod. Phys. E}\ }\textbf {\bibinfo {volume} {19}},\ \bibinfo {pages} {2538}
  (\bibinfo {year} {2010})}\BibitemShut {NoStop}%
\bibitem [{\citenamefont {Tanimura}\ and\ \citenamefont
  {Hagino}(2012)}]{Tanimura2012_PRC85-014306}%
  \BibitemOpen
  \bibfield  {author} {\bibinfo {author} {\bibfnamefont {Y.}~\bibnamefont
  {Tanimura}}\ and\ \bibinfo {author} {\bibfnamefont {K.}~\bibnamefont
  {Hagino}},\ }\href {\doibase 10.1103/PhysRevC.85.014306} {\bibfield
  {journal} {\bibinfo  {journal} {Phys. Rev. C}\ }\textbf {\bibinfo {volume}
  {85}},\ \bibinfo {pages} {014306} (\bibinfo {year} {2012})}\BibitemShut
  {NoStop}%
\bibitem [{\citenamefont {Wang}\ \emph {et~al.}(2013)\citenamefont {Wang},
  \citenamefont {Sang}, \citenamefont {Wang},\ and\ \citenamefont
  {Lv}}]{Wang2013_CTP60-479}%
  \BibitemOpen
  \bibfield  {author} {\bibinfo {author} {\bibfnamefont {X.-S.}\ \bibnamefont
  {Wang}}, \bibinfo {author} {\bibfnamefont {H.-Y.}\ \bibnamefont {Sang}},
  \bibinfo {author} {\bibfnamefont {J.-H.}\ \bibnamefont {Wang}}, \ and\
  \bibinfo {author} {\bibfnamefont {H.-F.}\ \bibnamefont {Lv}},\ }\href
  {http://ctp.itp.ac.cn/EN/abstract/abstract16149.shtml#} {\bibfield  {journal}
  {\bibinfo  {journal} {Commun. Theor. Phys.}\ }\textbf {\bibinfo {volume}
  {60}},\ \bibinfo {pages} {479} (\bibinfo {year} {2013})}\BibitemShut
  {NoStop}%
\bibitem [{\citenamefont {Liu}\ \emph {et~al.}(2018)\citenamefont {Liu},
  \citenamefont {Xia}, \citenamefont {Lu}, \citenamefont {Li}, \citenamefont
  {Hu},\ and\ \citenamefont {Sun}}]{Liu2018_PRC98-024316}%
  \BibitemOpen
  \bibfield  {author} {\bibinfo {author} {\bibfnamefont {Z.-X.}\ \bibnamefont
  {Liu}}, \bibinfo {author} {\bibfnamefont {C.-J.}\ \bibnamefont {Xia}},
  \bibinfo {author} {\bibfnamefont {W.-L.}\ \bibnamefont {Lu}}, \bibinfo
  {author} {\bibfnamefont {Y.-X.}\ \bibnamefont {Li}}, \bibinfo {author}
  {\bibfnamefont {J.~N.}\ \bibnamefont {Hu}}, \ and\ \bibinfo {author}
  {\bibfnamefont {T.-T.}\ \bibnamefont {Sun}},\ }\href {\doibase
  10.1103/PhysRevC.98.024316} {\bibfield  {journal} {\bibinfo  {journal} {Phys.
  Rev. C}\ }\textbf {\bibinfo {volume} {98}},\ \bibinfo {pages} {024316}
  (\bibinfo {year} {2018})}\BibitemShut {NoStop}%
\bibitem [{\citenamefont {Rong}\ \emph {et~al.}(2021)\citenamefont {Rong},
  \citenamefont {Tu},\ and\ \citenamefont {Zhou}}]{Rong2021_PRC104-054321}%
  \BibitemOpen
  \bibfield  {author} {\bibinfo {author} {\bibfnamefont {Y.-T.}\ \bibnamefont
  {Rong}}, \bibinfo {author} {\bibfnamefont {Z.-H.}\ \bibnamefont {Tu}}, \ and\
  \bibinfo {author} {\bibfnamefont {S.-G.}\ \bibnamefont {Zhou}},\ }\href
  {\doibase 10.1103/PhysRevC.104.054321} {\bibfield  {journal} {\bibinfo
  {journal} {Phys. Rev. C}\ }\textbf {\bibinfo {volume} {104}},\ \bibinfo
  {pages} {054321} (\bibinfo {year} {2021})}\BibitemShut {NoStop}%
\bibitem [{\citenamefont {Rong}\ \emph {et~al.}(2025)\citenamefont {Rong},
  \citenamefont {Yang}, \citenamefont {Xia},\ and\ \citenamefont
  {Sun}}]{Rong2025}%
  \BibitemOpen
  \bibfield  {author} {\bibinfo {author} {\bibfnamefont {Y.-T.}\ \bibnamefont
  {Rong}}, \bibinfo {author} {\bibfnamefont {D.}~\bibnamefont {Yang}}, \bibinfo
  {author} {\bibfnamefont {C.-J.}\ \bibnamefont {Xia}}, \ and\ \bibinfo
  {author} {\bibfnamefont {T.-T.}\ \bibnamefont {Sun}},\ }\href@noop {} {\
  (\bibinfo {year} {2025})},\ \Eprint {http://arxiv.org/abs/2506.13499}
  {arXiv:2506.13499 [nucl-th]} \BibitemShut {NoStop}%
\bibitem [{\citenamefont {Zhou}\ \emph {et~al.}(2007)\citenamefont {Zhou},
  \citenamefont {Schulze}, \citenamefont {Sagawa}, \citenamefont {Wu},\ and\
  \citenamefont {Zhao}}]{Zhou2007_PRC76-034312}%
  \BibitemOpen
  \bibfield  {author} {\bibinfo {author} {\bibfnamefont {X.-R.}\ \bibnamefont
  {Zhou}}, \bibinfo {author} {\bibfnamefont {H.-J.}\ \bibnamefont {Schulze}},
  \bibinfo {author} {\bibfnamefont {H.}~\bibnamefont {Sagawa}}, \bibinfo
  {author} {\bibfnamefont {C.-X.}\ \bibnamefont {Wu}}, \ and\ \bibinfo {author}
  {\bibfnamefont {E.-G.}\ \bibnamefont {Zhao}},\ }\href {\doibase
  10.1103/PhysRevC.76.034312} {\bibfield  {journal} {\bibinfo  {journal} {Phys.
  Rev. C}\ }\textbf {\bibinfo {volume} {76}},\ \bibinfo {pages} {034312}
  (\bibinfo {year} {2007})}\BibitemShut {NoStop}%
\bibitem [{\citenamefont {Tsushima}\ \emph {et~al.}(1997)\citenamefont
  {Tsushima}, \citenamefont {Saito},\ and\ \citenamefont
  {Thomas}}]{Tsushima1997_PLB411-9}%
  \BibitemOpen
  \bibfield  {author} {\bibinfo {author} {\bibfnamefont {K.}~\bibnamefont
  {Tsushima}}, \bibinfo {author} {\bibfnamefont {K.}~\bibnamefont {Saito}}, \
  and\ \bibinfo {author} {\bibfnamefont {A.}~\bibnamefont {Thomas}},\ }\href
  {\doibase http://dx.doi.org/10.1016/S0370-2693(97)00944-1} {\bibfield
  {journal} {\bibinfo  {journal} {Phys. Lett. B}\ }\textbf {\bibinfo {volume}
  {411}},\ \bibinfo {pages} {9} (\bibinfo {year} {1997})}\BibitemShut {NoStop}%
\bibitem [{\citenamefont {Tsushima}\ \emph {et~al.}(1998)\citenamefont
  {Tsushima}, \citenamefont {Saito}, \citenamefont {Haidenbauer},\ and\
  \citenamefont {Thomas}}]{Tsushima1998_NPA630-691}%
  \BibitemOpen
  \bibfield  {author} {\bibinfo {author} {\bibfnamefont {K.}~\bibnamefont
  {Tsushima}}, \bibinfo {author} {\bibfnamefont {K.}~\bibnamefont {Saito}},
  \bibinfo {author} {\bibfnamefont {J.}~\bibnamefont {Haidenbauer}}, \ and\
  \bibinfo {author} {\bibfnamefont {A.}~\bibnamefont {Thomas}},\ }\href
  {\doibase http://dx.doi.org/10.1016/S0375-9474(98)00806-9} {\bibfield
  {journal} {\bibinfo  {journal} {Nucl. Phys. A}\ }\textbf {\bibinfo {volume}
  {630}},\ \bibinfo {pages} {691} (\bibinfo {year} {1998})}\BibitemShut
  {NoStop}%
\bibitem [{\citenamefont {Guichon}\ \emph {et~al.}(2008)\citenamefont
  {Guichon}, \citenamefont {Thomas},\ and\ \citenamefont
  {Tsushima}}]{Guichon2008_NPA814-66}%
  \BibitemOpen
  \bibfield  {author} {\bibinfo {author} {\bibfnamefont {P.~A.}\ \bibnamefont
  {Guichon}}, \bibinfo {author} {\bibfnamefont {A.~W.}\ \bibnamefont {Thomas}},
  \ and\ \bibinfo {author} {\bibfnamefont {K.}~\bibnamefont {Tsushima}},\
  }\href {\doibase http://dx.doi.org/10.1016/j.nuclphysa.2008.10.001}
  {\bibfield  {journal} {\bibinfo  {journal} {Nucl. Phys. A}\ }\textbf
  {\bibinfo {volume} {814}},\ \bibinfo {pages} {66} (\bibinfo {year}
  {2008})}\BibitemShut {NoStop}%
\bibitem [{\citenamefont {Aoki}(2011)}]{Aoki2011_PPNP66-687}%
  \BibitemOpen
  \bibfield  {author} {\bibinfo {author} {\bibfnamefont {S.}~\bibnamefont
  {Aoki}},\ }\href {\doibase http://dx.doi.org/10.1016/j.ppnp.2011.07.001}
  {\bibfield  {journal} {\bibinfo  {journal} {Prog. Part. Nucl. Phys.}\
  }\textbf {\bibinfo {volume} {66}},\ \bibinfo {pages} {687 } (\bibinfo {year}
  {2011})}\BibitemShut {NoStop}%
\bibitem [{\citenamefont {Chen}\ \emph {et~al.}(2025)\citenamefont {Chen},
  \citenamefont {Geng}, \citenamefont {Hiyama}, \citenamefont {Liu},\ and\
  \citenamefont {Pochodzalla}}]{Chen2025}%
  \BibitemOpen
  \bibfield  {author} {\bibinfo {author} {\bibfnamefont {J.-H.}\ \bibnamefont
  {Chen}}, \bibinfo {author} {\bibfnamefont {L.-S.}\ \bibnamefont {Geng}},
  \bibinfo {author} {\bibfnamefont {E.}~\bibnamefont {Hiyama}}, \bibinfo
  {author} {\bibfnamefont {Z.-W.}\ \bibnamefont {Liu}}, \ and\ \bibinfo
  {author} {\bibfnamefont {J.}~\bibnamefont {Pochodzalla}},\ }\href@noop {} {\
  (\bibinfo {year} {2025})},\ \Eprint {http://arxiv.org/abs/2506.00864}
  {arXiv:2506.00864 [nucl-th]} \BibitemShut {NoStop}%
\bibitem [{\citenamefont {{The STAR Collaboration}}(2020)}]{Adam2020_NP16-409}%
  \BibitemOpen
\bibfield  {journal} {  }\bibfield  {author} {\bibinfo {author} {\bibnamefont
  {{The STAR Collaboration}}},\ }\href {\doibase 10.1038/s41567-020-0799-7}
  {\bibfield  {journal} {\bibinfo  {journal} {Nat. Phys.}\ }\textbf {\bibinfo
  {volume} {16}},\ \bibinfo {pages} {409} (\bibinfo {year} {2020})}\BibitemShut
  {NoStop}%
\bibitem [{\citenamefont {{ALICE
  Collaboration}}(2023)}]{Acharya2023_PRL131-102302}%
  \BibitemOpen
  \bibfield  {author} {\bibinfo {author} {\bibnamefont {{ALICE Collaboration}}},\ }\href {\doibase
  10.1103/PhysRevLett.131.102302} {\bibfield  {journal} {\bibinfo  {journal}
  {Phys. Rev. Lett.}\ }\textbf {\bibinfo {volume} {131}},\ \bibinfo {pages}
  {102302} (\bibinfo {year} {2023})}\BibitemShut {NoStop}%
\bibitem [{\citenamefont {Cheng}\ and\ \citenamefont
  {Feng}(2022)}]{Cheng2022_PLB824-136849}%
  \BibitemOpen
  \bibfield  {author} {\bibinfo {author} {\bibfnamefont {H.-G.}\ \bibnamefont
  {Cheng}}\ and\ \bibinfo {author} {\bibfnamefont {Z.-Q.}\ \bibnamefont
  {Feng}},\ }\href {\doibase https://doi.org/10.1016/j.physletb.2021.136849}
  {\bibfield  {journal} {\bibinfo  {journal} {Phys. Lett. B}\ }\textbf
  {\bibinfo {volume} {824}},\ \bibinfo {pages} {136849} (\bibinfo {year}
  {2022})}\BibitemShut {NoStop}%
\bibitem [{\citenamefont {Andronic}\ \emph {et~al.}(2018)\citenamefont
  {Andronic}, \citenamefont {Braun-Munzinger}, \citenamefont {Redlich},\ and\
  \citenamefont {Stachel}}]{Andronic2018_Nature561-321}%
  \BibitemOpen
  \bibfield  {author} {\bibinfo {author} {\bibfnamefont {A.}~\bibnamefont
  {Andronic}}, \bibinfo {author} {\bibfnamefont {P.}~\bibnamefont
  {Braun-Munzinger}}, \bibinfo {author} {\bibfnamefont {K.}~\bibnamefont
  {Redlich}}, \ and\ \bibinfo {author} {\bibfnamefont {J.}~\bibnamefont
  {Stachel}},\ }\href {\doibase 10.1038/s41586-018-0491-6} {\bibfield
  {journal} {\bibinfo  {journal} {Nature}\ }\textbf {\bibinfo {volume} {561}},\
  \bibinfo {pages} {321} (\bibinfo {year} {2018})}\BibitemShut {NoStop}%
\bibitem [{\citenamefont {Botvina}\ \emph {et~al.}(2023)\citenamefont
  {Botvina}, \citenamefont {Bleicher},\ and\ \citenamefont
  {Buyukcizmeci}}]{Botvina2023_JPCS2586-012045}%
  \BibitemOpen
  \bibfield  {author} {\bibinfo {author} {\bibfnamefont {A.~S.}\ \bibnamefont
  {Botvina}}, \bibinfo {author} {\bibfnamefont {M.}~\bibnamefont {Bleicher}}, \
  and\ \bibinfo {author} {\bibfnamefont {N.}~\bibnamefont {Buyukcizmeci}},\
  }\href {\doibase 10.1088/1742-6596/2586/1/012045} {\bibfield  {journal}
  {\bibinfo  {journal} {J. Phys: Conf. Ser.}\ }\textbf {\bibinfo {volume}
  {2586}},\ \bibinfo {pages} {012045} (\bibinfo {year} {2023})}\BibitemShut
  {NoStop}%
\bibitem [{\citenamefont {Buyukcizmeci}\ \emph {et~al.}(2018)\citenamefont
  {Buyukcizmeci}, \citenamefont {Botvina}, \citenamefont {Ergun}, \citenamefont
  {Ogul},\ and\ \citenamefont {Bleicher}}]{Buyukcizmeci2018_PRC98-064603}%
  \BibitemOpen
  \bibfield  {author} {\bibinfo {author} {\bibfnamefont {N.}~\bibnamefont
  {Buyukcizmeci}}, \bibinfo {author} {\bibfnamefont {A.~S.}\ \bibnamefont
  {Botvina}}, \bibinfo {author} {\bibfnamefont {A.}~\bibnamefont {Ergun}},
  \bibinfo {author} {\bibfnamefont {R.}~\bibnamefont {Ogul}}, \ and\ \bibinfo
  {author} {\bibfnamefont {M.}~\bibnamefont {Bleicher}},\ }\href {\doibase
  10.1103/PhysRevC.98.064603} {\bibfield  {journal} {\bibinfo  {journal} {Phys.
  Rev. C}\ }\textbf {\bibinfo {volume} {98}},\ \bibinfo {pages} {064603}
  (\bibinfo {year} {2018})}\BibitemShut {NoStop}%
\bibitem [{\citenamefont {She}\ \emph {et~al.}(2021)\citenamefont {She},
  \citenamefont {Chen}, \citenamefont {Zhou}, \citenamefont {Zheng},
  \citenamefont {Xie},\ and\ \citenamefont {Xu}}]{She2021_PRC103-014906}%
  \BibitemOpen
  \bibfield  {author} {\bibinfo {author} {\bibfnamefont {Z.-L.}\ \bibnamefont
  {She}}, \bibinfo {author} {\bibfnamefont {G.}~\bibnamefont {Chen}}, \bibinfo
  {author} {\bibfnamefont {D.-M.}\ \bibnamefont {Zhou}}, \bibinfo {author}
  {\bibfnamefont {L.}~\bibnamefont {Zheng}}, \bibinfo {author} {\bibfnamefont
  {Y.-L.}\ \bibnamefont {Xie}}, \ and\ \bibinfo {author} {\bibfnamefont
  {H.-G.}\ \bibnamefont {Xu}},\ }\href {\doibase 10.1103/PhysRevC.103.014906}
  {\bibfield  {journal} {\bibinfo  {journal} {Phys. Rev. C}\ }\textbf {\bibinfo
  {volume} {103}},\ \bibinfo {pages} {014906} (\bibinfo {year}
  {2021})}\BibitemShut {NoStop}%
\bibitem [{\citenamefont {Zhou}\ \emph {et~al.}(2025)\citenamefont {Zhou} \emph
  {et~al.}}]{Zhou2025}%
  \BibitemOpen
  \bibfield  {author} {\bibinfo {author} {\bibfnamefont {Y.}~\bibnamefont
  {Zhou}} \emph {et~al.},\ }\href@noop {} {\  (\bibinfo {year} {2025})},\
  \Eprint {http://arxiv.org/abs/2507.14255} {arXiv:2507.14255 [nucl-th]}
  \BibitemShut {NoStop}%
\bibitem [{\citenamefont {Le~F\`evre}\ \emph {et~al.}(2019)\citenamefont
  {Le~F\`evre}, \citenamefont {Aichelin}, \citenamefont {Hartnack},\ and\
  \citenamefont {Leifels}}]{LeFevre2019_PRC100-034904}%
  \BibitemOpen
  \bibfield  {author} {\bibinfo {author} {\bibfnamefont {A.}~\bibnamefont
  {Le~F\`evre}}, \bibinfo {author} {\bibfnamefont {J.}~\bibnamefont
  {Aichelin}}, \bibinfo {author} {\bibfnamefont {C.}~\bibnamefont {Hartnack}},
  \ and\ \bibinfo {author} {\bibfnamefont {Y.}~\bibnamefont {Leifels}},\ }\href
  {\doibase 10.1103/PhysRevC.100.034904} {\bibfield  {journal} {\bibinfo
  {journal} {Phys. Rev. C}\ }\textbf {\bibinfo {volume} {100}},\ \bibinfo
  {pages} {034904} (\bibinfo {year} {2019})}\BibitemShut {NoStop}%
\bibitem [{\citenamefont {Botta}\ \emph {et~al.}(2012)\citenamefont {Botta},
  \citenamefont {Bressani},\ and\ \citenamefont
  {Garbarino}}]{Botta2012_EPJA48-41}%
  \BibitemOpen
  \bibfield  {author} {\bibinfo {author} {\bibfnamefont {E.}~\bibnamefont
  {Botta}}, \bibinfo {author} {\bibfnamefont {T.}~\bibnamefont {Bressani}}, \
  and\ \bibinfo {author} {\bibfnamefont {G.}~\bibnamefont {Garbarino}},\ }\href
  {\doibase 10.1140/epja/i2012-12041-6} {\bibfield  {journal} {\bibinfo
  {journal} {Eur. Phys. J. A}\ }\textbf {\bibinfo {volume} {48}},\ \bibinfo
  {pages} {41} (\bibinfo {year} {2012})}\BibitemShut {NoStop}%
\bibitem [{\citenamefont {Yong}(2025)}]{Yong2025_PLB866-139549}%
  \BibitemOpen
  \bibfield  {author} {\bibinfo {author} {\bibfnamefont {G.-C.}\ \bibnamefont
  {Yong}},\ }\href {\doibase https://doi.org/10.1016/j.physletb.2025.139549}
  {\bibfield  {journal} {\bibinfo  {journal} {Phys. Lett. B}\ }\textbf
  {\bibinfo {volume} {866}},\ \bibinfo {pages} {139549} (\bibinfo {year}
  {2025})}\BibitemShut {NoStop}%
\bibitem [{\citenamefont {Yong}(2024)}]{Yong2024_PLB853-138662}%
  \BibitemOpen
  \bibfield  {author} {\bibinfo {author} {\bibfnamefont {G.-C.}\ \bibnamefont
  {Yong}},\ }\href {\doibase https://doi.org/10.1016/j.physletb.2024.138662}
  {\bibfield  {journal} {\bibinfo  {journal} {Phys. Lett. B}\ }\textbf
  {\bibinfo {volume} {853}},\ \bibinfo {pages} {138662} (\bibinfo {year}
  {2024})}\BibitemShut {NoStop}%
\bibitem [{\citenamefont {Yong}\ \emph {et~al.}(2025)\citenamefont {Yong},
  \citenamefont {Rodr\'{\i}guez-S\'anchez},\ and\ \citenamefont
  {Zhang}}]{Yong2025_PRC111-054617}%
  \BibitemOpen
  \bibfield  {author} {\bibinfo {author} {\bibfnamefont {G.-C.}\ \bibnamefont
  {Yong}}, \bibinfo {author} {\bibfnamefont {J.~L.}\ \bibnamefont
  {Rodr\'{\i}guez-S\'anchez}}, \ and\ \bibinfo {author} {\bibfnamefont
  {Y.}~\bibnamefont {Zhang}},\ }\href {\doibase 10.1103/PhysRevC.111.054617}
  {\bibfield  {journal} {\bibinfo  {journal} {Phys. Rev. C}\ }\textbf {\bibinfo
  {volume} {111}},\ \bibinfo {pages} {054617} (\bibinfo {year}
  {2025})}\BibitemShut {NoStop}%
\bibitem [{\citenamefont {Feng}(2024)}]{Feng2024_PLB851-138580}%
  \BibitemOpen
  \bibfield  {author} {\bibinfo {author} {\bibfnamefont {Z.-Q.}\ \bibnamefont
  {Feng}},\ }\href {\doibase https://doi.org/10.1016/j.physletb.2024.138580}
  {\bibfield  {journal} {\bibinfo  {journal} {Phys. Lett. B}\ }\textbf
  {\bibinfo {volume} {851}},\ \bibinfo {pages} {138580} (\bibinfo {year}
  {2024})}\BibitemShut {NoStop}%
\bibitem [{\citenamefont {Wei}\ \emph {et~al.}(2024)\citenamefont {Wei},
  \citenamefont {Feng},\ and\ \citenamefont {Jiang}}]{Wei2024_PLB853-138658}%
  \BibitemOpen
  \bibfield  {author} {\bibinfo {author} {\bibfnamefont {S.-N.}\ \bibnamefont
  {Wei}}, \bibinfo {author} {\bibfnamefont {Z.-Q.}\ \bibnamefont {Feng}}, \
  and\ \bibinfo {author} {\bibfnamefont {W.-Z.}\ \bibnamefont {Jiang}},\ }\href
  {\doibase https://doi.org/10.1016/j.physletb.2024.138658} {\bibfield
  {journal} {\bibinfo  {journal} {Phys. Lett. B}\ }\textbf {\bibinfo {volume}
  {853}},\ \bibinfo {pages} {138658} (\bibinfo {year} {2024})}\BibitemShut
  {NoStop}%
\bibitem [{\citenamefont {Hartnack}\ \emph {et~al.}(2012)\citenamefont
  {Hartnack}, \citenamefont {Oeschler}, \citenamefont {Leifels}, \citenamefont
  {Bratkovskaya},\ and\ \citenamefont {Aichelin}}]{Hartnack2012_PR510-119}%
  \BibitemOpen
  \bibfield  {author} {\bibinfo {author} {\bibfnamefont {C.}~\bibnamefont
  {Hartnack}}, \bibinfo {author} {\bibfnamefont {H.}~\bibnamefont {Oeschler}},
  \bibinfo {author} {\bibfnamefont {Y.}~\bibnamefont {Leifels}}, \bibinfo
  {author} {\bibfnamefont {E.~L.}\ \bibnamefont {Bratkovskaya}}, \ and\
  \bibinfo {author} {\bibfnamefont {J.}~\bibnamefont {Aichelin}},\ }\href
  {\doibase https://doi.org/10.1016/j.physrep.2011.08.004} {\bibfield
  {journal} {\bibinfo  {journal} {Phys. Rep.}\ }\textbf {\bibinfo {volume}
  {510}},\ \bibinfo {pages} {119} (\bibinfo {year} {2012})}\BibitemShut
  {NoStop}%
\bibitem [{\citenamefont {Reisdorf}\ \emph {et~al.}(2010)\citenamefont
  {Reisdorf}, \citenamefont {Andronic}, \citenamefont {Averbeck},\ and\
  \citenamefont {et~al.}}]{Reisdorf2010_NPA848-366}%
  \BibitemOpen
  \bibfield  {author} {\bibinfo {author} {\bibfnamefont {W.}~\bibnamefont
  {Reisdorf}}, \bibinfo {author} {\bibfnamefont {A.}~\bibnamefont {Andronic}},
  \bibinfo {author} {\bibfnamefont {R.}~\bibnamefont {Averbeck}}, \ and\
  \bibinfo {author} {\bibnamefont {et~al.}},\ }\href {\doibase
  https://doi.org/10.1016/j.nuclphysa.2010.09.008} {\bibfield  {journal}
  {\bibinfo  {journal} {Nucl. Phys. A}\ }\textbf {\bibinfo {volume} {848}},\
  \bibinfo {pages} {366} (\bibinfo {year} {2010})}\BibitemShut {NoStop}%
\bibitem [{\citenamefont {Ishizuka}\ \emph {et~al.}(2008)\citenamefont
  {Ishizuka}, \citenamefont {Ohnishi}, \citenamefont {Tsubakihara},
  \citenamefont {Sumiyoshi},\ and\ \citenamefont
  {Yamada}}]{Ishizuka2008_JPG35-085201}%
  \BibitemOpen
  \bibfield  {author} {\bibinfo {author} {\bibfnamefont {C.}~\bibnamefont
  {Ishizuka}}, \bibinfo {author} {\bibfnamefont {A.}~\bibnamefont {Ohnishi}},
  \bibinfo {author} {\bibfnamefont {K.}~\bibnamefont {Tsubakihara}}, \bibinfo
  {author} {\bibfnamefont {K.}~\bibnamefont {Sumiyoshi}}, \ and\ \bibinfo
  {author} {\bibfnamefont {S.}~\bibnamefont {Yamada}},\ }\href
  {http://stacks.iop.org/0954-3899/35/i=8/a=085201} {\bibfield  {journal}
  {\bibinfo  {journal} {J. Phys. G: Nucl. Part. Phys.}\ }\textbf {\bibinfo
  {volume} {35}},\ \bibinfo {pages} {085201} (\bibinfo {year}
  {2008})}\BibitemShut {NoStop}%
\bibitem [{\citenamefont {Shen}\ \emph {et~al.}(2011)\citenamefont {Shen},
  \citenamefont {Toki}, \citenamefont {Oyamatsu},\ and\ \citenamefont
  {Sumiyoshi}}]{Shen2011_ApJ197-20}%
  \BibitemOpen
  \bibfield  {author} {\bibinfo {author} {\bibfnamefont {H.}~\bibnamefont
  {Shen}}, \bibinfo {author} {\bibfnamefont {H.}~\bibnamefont {Toki}}, \bibinfo
  {author} {\bibfnamefont {K.}~\bibnamefont {Oyamatsu}}, \ and\ \bibinfo
  {author} {\bibfnamefont {K.}~\bibnamefont {Sumiyoshi}},\ }\href
  {http://stacks.iop.org/0067-0049/197/i=2/a=20} {\bibfield  {journal}
  {\bibinfo  {journal} {Astrophys. J.}\ }\textbf {\bibinfo {volume} {197}},\
  \bibinfo {pages} {20} (\bibinfo {year} {2011})}\BibitemShut {NoStop}%
\bibitem [{\citenamefont {Sun}\ \emph {et~al.}(2018)\citenamefont {Sun},
  \citenamefont {Xia}, \citenamefont {Zhang},\ and\ \citenamefont
  {Smith}}]{Sun2018_CPC42-25101}%
  \BibitemOpen
  \bibfield  {author} {\bibinfo {author} {\bibfnamefont {T.-T.}\ \bibnamefont
  {Sun}}, \bibinfo {author} {\bibfnamefont {C.-J.}\ \bibnamefont {Xia}},
  \bibinfo {author} {\bibfnamefont {S.-S.}\ \bibnamefont {Zhang}}, \ and\
  \bibinfo {author} {\bibfnamefont {M.~S.}\ \bibnamefont {Smith}},\ }\href
  {\doibase 10.1088/1674-1137/42/2/025101} {\bibfield  {journal} {\bibinfo
  {journal} {Chin. Phys. C}\ }\textbf {\bibinfo {volume} {42}},\ \bibinfo {eid}
  {25101} (\bibinfo {year} {2018})}\BibitemShut {NoStop}%
\bibitem [{\citenamefont {Sun}\ \emph {et~al.}(2019)\citenamefont {Sun},
  \citenamefont {Zhang}, \citenamefont {Zhang},\ and\ \citenamefont
  {Xia}}]{Sun2019_PRD99-023004}%
  \BibitemOpen
  \bibfield  {author} {\bibinfo {author} {\bibfnamefont {T.-T.}\ \bibnamefont
  {Sun}}, \bibinfo {author} {\bibfnamefont {S.-S.}\ \bibnamefont {Zhang}},
  \bibinfo {author} {\bibfnamefont {Q.-L.}\ \bibnamefont {Zhang}}, \ and\
  \bibinfo {author} {\bibfnamefont {C.-J.}\ \bibnamefont {Xia}},\ }\href
  {\doibase 10.1103/PhysRevD.99.023004} {\bibfield  {journal} {\bibinfo
  {journal} {Phys. Rev. D}\ }\textbf {\bibinfo {volume} {99}},\ \bibinfo
  {pages} {023004} (\bibinfo {year} {2019})}\BibitemShut {NoStop}%
\bibitem [{Note1()}]{Note1}%
  \BibitemOpen
  \href{https://compose.obspm.fr/}{\bibinfo {note} {https://compose.obspm.fr/}}\BibitemShut {NoStop}%
\bibitem [{\citenamefont {Cust\'odio}\ \emph {et~al.}(2021)\citenamefont
  {Cust\'odio}, \citenamefont {Pais},\ and\ \citenamefont
  {Provid\^encia}}]{Custodio2021_PRC104-035801}%
  \BibitemOpen
  \bibfield  {author} {\bibinfo {author} {\bibfnamefont {T.}~\bibnamefont
  {Cust\'odio}}, \bibinfo {author} {\bibfnamefont {H.}~\bibnamefont {Pais}}, \
  and\ \bibinfo {author} {\bibfnamefont {C.~m.~c.}\ \bibnamefont
  {Provid\^encia}},\ }\href {\doibase 10.1103/PhysRevC.104.035801} {\bibfield
  {journal} {\bibinfo  {journal} {Phys. Rev. C}\ }\textbf {\bibinfo {volume}
  {104}},\ \bibinfo {pages} {035801} (\bibinfo {year} {2021})}\BibitemShut
  {NoStop}%
\bibitem [{\citenamefont {Woosley}\ \emph {et~al.}(2002)\citenamefont
  {Woosley}, \citenamefont {Heger},\ and\ \citenamefont
  {Weaver}}]{Woosley2002_RMP74-1015}%
  \BibitemOpen
  \bibfield  {author} {\bibinfo {author} {\bibfnamefont {S.~E.}\ \bibnamefont
  {Woosley}}, \bibinfo {author} {\bibfnamefont {A.}~\bibnamefont {Heger}}, \
  and\ \bibinfo {author} {\bibfnamefont {T.~A.}\ \bibnamefont {Weaver}},\
  }\href {\doibase 10.1103/RevModPhys.74.1015} {\bibfield  {journal} {\bibinfo
  {journal} {Rev. Mod. Phys.}\ }\textbf {\bibinfo {volume} {74}},\ \bibinfo
  {pages} {1015} (\bibinfo {year} {2002})}\BibitemShut {NoStop}%
\bibitem [{\citenamefont {Lattimer}(2012)}]{Lattimer2012_ARNPS62-485}%
  \BibitemOpen
  \bibfield  {author} {\bibinfo {author} {\bibfnamefont {J.~M.}\ \bibnamefont
  {Lattimer}},\ }\href
  {http://www.annualreviews.org/doi/abs/10.1146/annurev-nucl-102711-095018}
  {\bibfield  {journal} {\bibinfo  {journal} {Annu. Rev. Nucl. Part. Sci.}\
  }\textbf {\bibinfo {volume} {62}},\ \bibinfo {pages} {485} (\bibinfo {year}
  {2012})}\BibitemShut {NoStop}%
\bibitem [{\citenamefont {\"Ozel}\ \emph {et~al.}(2016)\citenamefont {\"Ozel},
  \citenamefont {Psaltis}, \citenamefont {Guver}, \citenamefont {Baym},
  \citenamefont {Heinke},\ and\ \citenamefont {Guillot}}]{Ozel2016_ApJ820-28}%
  \BibitemOpen
  \bibfield  {author} {\bibinfo {author} {\bibfnamefont {F.}~\bibnamefont
  {\"Ozel}}, \bibinfo {author} {\bibfnamefont {D.}~\bibnamefont {Psaltis}},
  \bibinfo {author} {\bibfnamefont {T.}~\bibnamefont {Guver}}, \bibinfo
  {author} {\bibfnamefont {G.}~\bibnamefont {Baym}}, \bibinfo {author}
  {\bibfnamefont {C.}~\bibnamefont {Heinke}}, \ and\ \bibinfo {author}
  {\bibfnamefont {S.}~\bibnamefont {Guillot}},\ }\href
  {http://stacks.iop.org/0004-637X/820/i=1/a=28} {\bibfield  {journal}
  {\bibinfo  {journal} {Astrophys. J.}\ }\textbf {\bibinfo {volume} {820}},\
  \bibinfo {pages} {28} (\bibinfo {year} {2016})}\BibitemShut {NoStop}%
\bibitem [{\citenamefont {\"Ozel}\ and\ \citenamefont
  {Freire}(2016)}]{Ozel2016_ARAA54-401}%
  \BibitemOpen
  \bibfield  {author} {\bibinfo {author} {\bibfnamefont {F.}~\bibnamefont
  {\"Ozel}}\ and\ \bibinfo {author} {\bibfnamefont {P.}~\bibnamefont
  {Freire}},\ }\href {\doibase 10.1146/annurev-astro-081915-023322} {\bibfield
  {journal} {\bibinfo  {journal} {Annu. Rev. Astron. Astrophys.}\ }\textbf
  {\bibinfo {volume} {54}},\ \bibinfo {pages} {401} (\bibinfo {year}
  {2016})}\BibitemShut {NoStop}%
\bibitem [{\citenamefont {Hotokezaka}\ \emph {et~al.}(2011)\citenamefont
  {Hotokezaka}, \citenamefont {Kyutoku}, \citenamefont {Okawa}, \citenamefont
  {Shibata},\ and\ \citenamefont {Kiuchi}}]{Hotokezaka2011_PRD83-124008}%
  \BibitemOpen
  \bibfield  {author} {\bibinfo {author} {\bibfnamefont {K.}~\bibnamefont
  {Hotokezaka}}, \bibinfo {author} {\bibfnamefont {K.}~\bibnamefont {Kyutoku}},
  \bibinfo {author} {\bibfnamefont {H.}~\bibnamefont {Okawa}}, \bibinfo
  {author} {\bibfnamefont {M.}~\bibnamefont {Shibata}}, \ and\ \bibinfo
  {author} {\bibfnamefont {K.}~\bibnamefont {Kiuchi}},\ }\href {\doibase
  10.1103/PhysRevD.83.124008} {\bibfield  {journal} {\bibinfo  {journal} {Phys.
  Rev. D}\ }\textbf {\bibinfo {volume} {83}},\ \bibinfo {pages} {124008}
  (\bibinfo {year} {2011})}\BibitemShut {NoStop}%
\bibitem [{\citenamefont {Xia}(2025)}]{Xia2025_PRC112-025801}%
  \BibitemOpen
  \bibfield  {author} {\bibinfo {author} {\bibfnamefont {C.-J.}\ \bibnamefont
  {Xia}},\ }\href {\doibase 10.1103/ggh3-f6vj} {\bibfield  {journal} {\bibinfo
  {journal} {Phys. Rev. C}\ }\textbf {\bibinfo {volume} {112}},\ \bibinfo
  {pages} {025801} (\bibinfo {year} {2025})}\BibitemShut {NoStop}%
\bibitem [{\citenamefont {{Negele}}\ and\ \citenamefont
  {{Vautherin}}(1973)}]{Negele1973_NPA207-298}%
  \BibitemOpen
  \bibfield  {author} {\bibinfo {author} {\bibfnamefont {J.~W.}\ \bibnamefont
  {{Negele}}}\ and\ \bibinfo {author} {\bibfnamefont {D.}~\bibnamefont
  {{Vautherin}}},\ }\href {\doibase 10.1016/0375-9474(73)90349-7} {\bibfield
  {journal} {\bibinfo  {journal} {Nucl. Phys. A}\ }\textbf {\bibinfo {volume}
  {207}},\ \bibinfo {pages} {298} (\bibinfo {year} {1973})}\BibitemShut
  {NoStop}%
\bibitem [{\citenamefont {Long}\ \emph {et~al.}(2004)\citenamefont {Long},
  \citenamefont {Meng}, \citenamefont {Giai},\ and\ \citenamefont
  {Zhou}}]{Long2004_PRC69-034319}%
  \BibitemOpen
  \bibfield  {author} {\bibinfo {author} {\bibfnamefont {W.-H.}\ \bibnamefont
  {Long}}, \bibinfo {author} {\bibfnamefont {J.}~\bibnamefont {Meng}}, \bibinfo
  {author} {\bibfnamefont {N.~V.}\ \bibnamefont {Giai}}, \ and\ \bibinfo
  {author} {\bibfnamefont {S.-G.}\ \bibnamefont {Zhou}},\ }\href {\doibase
  10.1103/PhysRevC.69.034319} {\bibfield  {journal} {\bibinfo  {journal} {Phys.
  Rev. C}\ }\textbf {\bibinfo {volume} {69}},\ \bibinfo {pages} {034319}
  (\bibinfo {year} {2004})}\BibitemShut {NoStop}%
\bibitem [{\citenamefont {Sugahara}\ and\ \citenamefont
  {Toki}(1994{\natexlab{b}})}]{Sugahara1994_NPA579-557}%
  \BibitemOpen
  \bibfield  {author} {\bibinfo {author} {\bibfnamefont {Y.}~\bibnamefont
  {Sugahara}}\ and\ \bibinfo {author} {\bibfnamefont {H.}~\bibnamefont
  {Toki}},\ }\href {\doibase http://dx.doi.org/10.1016/0375-9474(94)90923-7}
  {\bibfield  {journal} {\bibinfo  {journal} {Nucl. Phys. A}\ }\textbf
  {\bibinfo {volume} {579}},\ \bibinfo {pages} {557} (\bibinfo {year}
  {1994}{\natexlab{b}})}\BibitemShut {NoStop}%
\bibitem [{\citenamefont {Wei}\ \emph {et~al.}(2020)\citenamefont {Wei},
  \citenamefont {Zhao}, \citenamefont {Wang}, \citenamefont {Geng},
  \citenamefont {Sun}, \citenamefont {Niu},\ and\ \citenamefont
  {Long}}]{Wei2020_CPC44-074107}%
  \BibitemOpen
  \bibfield  {author} {\bibinfo {author} {\bibfnamefont {B.}~\bibnamefont
  {Wei}}, \bibinfo {author} {\bibfnamefont {Q.}~\bibnamefont {Zhao}}, \bibinfo
  {author} {\bibfnamefont {Z.-H.}\ \bibnamefont {Wang}}, \bibinfo {author}
  {\bibfnamefont {J.}~\bibnamefont {Geng}}, \bibinfo {author} {\bibfnamefont
  {B.-Y.}\ \bibnamefont {Sun}}, \bibinfo {author} {\bibfnamefont {Y.-F.}\
  \bibnamefont {Niu}}, \ and\ \bibinfo {author} {\bibfnamefont {W.-H.}\
  \bibnamefont {Long}},\ }\href {\doibase 10.1088/1674-1137/44/7/074107}
  {\bibfield  {journal} {\bibinfo  {journal} {Chin. Phys. C}\ }\textbf
  {\bibinfo {volume} {44}},\ \bibinfo {pages} {074107} (\bibinfo {year}
  {2020})}\BibitemShut {NoStop}%
\bibitem [{\citenamefont {Lenske}\ and\ \citenamefont
  {Fuchs}(1995)}]{Lenske1995_PLB345-355}%
  \BibitemOpen
  \bibfield  {author} {\bibinfo {author} {\bibfnamefont {H.}~\bibnamefont
  {Lenske}}\ and\ \bibinfo {author} {\bibfnamefont {C.}~\bibnamefont {Fuchs}},\
  }\href {\doibase https://doi.org/10.1016/0370-2693(94)01664-X} {\bibfield
  {journal} {\bibinfo  {journal} {Phys. Lett. B}\ }\textbf {\bibinfo {volume}
  {345}},\ \bibinfo {pages} {355 } (\bibinfo {year} {1995})}\BibitemShut
  {NoStop}%
\bibitem [{\citenamefont {Xia}\ \emph {et~al.}(2021)\citenamefont {Xia},
  \citenamefont {Maruyama}, \citenamefont {Yasutake}, \citenamefont {Tatsumi},\
  and\ \citenamefont {Zhang}}]{Xia2021_PRC103-055812}%
  \BibitemOpen
  \bibfield  {author} {\bibinfo {author} {\bibfnamefont {C.-J.}\ \bibnamefont
  {Xia}}, \bibinfo {author} {\bibfnamefont {T.}~\bibnamefont {Maruyama}},
  \bibinfo {author} {\bibfnamefont {N.}~\bibnamefont {Yasutake}}, \bibinfo
  {author} {\bibfnamefont {T.}~\bibnamefont {Tatsumi}}, \ and\ \bibinfo
  {author} {\bibfnamefont {Y.-X.}\ \bibnamefont {Zhang}},\ }\href {\doibase
  10.1103/PhysRevC.103.055812} {\bibfield  {journal} {\bibinfo  {journal}
  {Phys. Rev. C}\ }\textbf {\bibinfo {volume} {103}},\ \bibinfo {pages}
  {055812} (\bibinfo {year} {2021})}\BibitemShut {NoStop}%
\bibitem [{\citenamefont {Bender}\ \emph {et~al.}(2000)\citenamefont {Bender},
  \citenamefont {Rutz}, \citenamefont {Reinhard},\ and\ \citenamefont
  {Maruhn}}]{Bender2000_EPJA7-467}%
  \BibitemOpen
  \bibfield  {author} {\bibinfo {author} {\bibfnamefont {M.}~\bibnamefont
  {Bender}}, \bibinfo {author} {\bibfnamefont {K.}~\bibnamefont {Rutz}},
  \bibinfo {author} {\bibfnamefont {P.-G.}\ \bibnamefont {Reinhard}}, \ and\
  \bibinfo {author} {\bibfnamefont {J.}~\bibnamefont {Maruhn}},\ }\href
  {\doibase 10.1007/PL00013645} {\bibfield  {journal} {\bibinfo  {journal}
  {Eur. Phys. J. A}\ }\textbf {\bibinfo {volume} {7}},\ \bibinfo {pages} {467}
  (\bibinfo {year} {2000})}\BibitemShut {NoStop}%
\bibitem [{\citenamefont {Typel}\ \emph {et~al.}(2010)\citenamefont {Typel},
  \citenamefont {R\"opke}, \citenamefont {Kl\"ahn}, \citenamefont {Blaschke},\
  and\ \citenamefont {Wolter}}]{Typel2010_PRC81-015803}%
  \BibitemOpen
  \bibfield  {author} {\bibinfo {author} {\bibfnamefont {S.}~\bibnamefont
  {Typel}}, \bibinfo {author} {\bibfnamefont {G.}~\bibnamefont {R\"opke}},
  \bibinfo {author} {\bibfnamefont {T.}~\bibnamefont {Kl\"ahn}}, \bibinfo
  {author} {\bibfnamefont {D.}~\bibnamefont {Blaschke}}, \ and\ \bibinfo
  {author} {\bibfnamefont {H.~H.}\ \bibnamefont {Wolter}},\ }\href {\doibase
  10.1103/PhysRevC.81.015803} {\bibfield  {journal} {\bibinfo  {journal} {Phys.
  Rev. C}\ }\textbf {\bibinfo {volume} {81}},\ \bibinfo {pages} {015803}
  (\bibinfo {year} {2010})}\BibitemShut {NoStop}%
\bibitem [{\citenamefont {Ding}\ \emph {et~al.}(2023)\citenamefont {Ding},
  \citenamefont {Yang},\ and\ \citenamefont {Sun}}]{Ding2023_CPC47-124103}%
  \BibitemOpen
  \bibfield  {author} {\bibinfo {author} {\bibfnamefont {S.-Y.}\ \bibnamefont
  {Ding}}, \bibinfo {author} {\bibfnamefont {W.}~\bibnamefont {Yang}}, \ and\
  \bibinfo {author} {\bibfnamefont {B.-Y.}\ \bibnamefont {Sun}},\ }\href
  {\doibase 10.1088/1674-1137/acf91e} {\bibfield  {journal} {\bibinfo
  {journal} {Chin. Phys. C}\ }\textbf {\bibinfo {volume} {47}},\ \bibinfo
  {pages} {124103} (\bibinfo {year} {2023})}\BibitemShut {NoStop}%
\bibitem [{\citenamefont {Russotto}\ \emph {et~al.}(2016)\citenamefont
  {Russotto}, \citenamefont {Gannon}, \citenamefont {Kupny}, \citenamefont
  {Lasko}, \citenamefont {Acosta}, \citenamefont {Adamczyk}, \citenamefont
  {Al-Ajlan}, \citenamefont {Al-Garawi}, \citenamefont {Al-Homaidhi},
  \citenamefont {Amorini}, \citenamefont {Auditore}, \citenamefont {Aumann},
  \citenamefont {Ayyad}, \citenamefont {Basrak}, \citenamefont {Benlliure},
  \citenamefont {Boisjoli}, \citenamefont {Boretzky}, \citenamefont
  {Brzychczyk}, \citenamefont {Budzanowski}, \citenamefont {Caesar},
  \citenamefont {Cardella}, \citenamefont {Cammarata}, \citenamefont
  {Chajecki}, \citenamefont {Chartier}, \citenamefont {Chbihi}, \citenamefont
  {Colonna}, \citenamefont {Cozma}, \citenamefont {Czech}, \citenamefont
  {De~Filippo}, \citenamefont {Di~Toro}, \citenamefont {Famiano}, \citenamefont
  {Ga\ifmmode \check{s}\else \v{s}\fi{}pari\ifmmode~\acute{c}\else \'{c}\fi{}},
  \citenamefont {Grassi}, \citenamefont {Guazzoni}, \citenamefont {Guazzoni},
  \citenamefont {Heil}, \citenamefont {Heilborn}, \citenamefont {Introzzi},
  \citenamefont {Isobe}, \citenamefont {Kezzar}, \citenamefont
  {Ki\ifmmode~\check{s}\else \v{s}\fi{}}, \citenamefont {Krasznahorkay},
  \citenamefont {Kurz}, \citenamefont {La~Guidara}, \citenamefont {Lanzalone},
  \citenamefont {Le~F\`evre}, \citenamefont {Leifels}, \citenamefont {Lemmon},
  \citenamefont {Li}, \citenamefont {Lombardo}, \citenamefont {\L{}ukasik},
  \citenamefont {Lynch}, \citenamefont {Marini}, \citenamefont {Matthews},
  \citenamefont {May}, \citenamefont {Minniti}, \citenamefont {Mostazo},
  \citenamefont {Pagano}, \citenamefont {Pagano}, \citenamefont {Papa},
  \citenamefont {Paw\l{}owski}, \citenamefont {Pirrone}, \citenamefont
  {Politi}, \citenamefont {Porto}, \citenamefont {Reviol}, \citenamefont
  {Riccio}, \citenamefont {Rizzo}, \citenamefont {Rosato}, \citenamefont
  {Rossi}, \citenamefont {Santoro}, \citenamefont {Sarantites}, \citenamefont
  {Simon}, \citenamefont {Skwirczynska}, \citenamefont {Sosin}, \citenamefont
  {Stuhl}, \citenamefont {Trautmann}, \citenamefont {Trifir\`o}, \citenamefont
  {Trimarchi}, \citenamefont {Tsang}, \citenamefont {Verde}, \citenamefont
  {Veselsky}, \citenamefont {Vigilante}, \citenamefont {Wang}, \citenamefont
  {Wieloch}, \citenamefont {Wigg}, \citenamefont {Winkelbauer}, \citenamefont
  {Wolter}, \citenamefont {Wu}, \citenamefont {Yennello}, \citenamefont
  {Zambon}, \citenamefont {Zetta},\ and\ \citenamefont
  {Zoric}}]{Russotto2016_PRC94-034608}%
  \BibitemOpen
  \bibfield  {author} {\bibinfo {author} {\bibfnamefont {P.}~\bibnamefont
  {Russotto}}, \bibinfo {author} {\bibfnamefont {S.}~\bibnamefont {Gannon}},
  \bibinfo {author} {\bibfnamefont {S.}~\bibnamefont {Kupny}}, \bibinfo
  {author} {\bibfnamefont {P.}~\bibnamefont {Lasko}}, \bibinfo {author}
  {\bibfnamefont {L.}~\bibnamefont {Acosta}}, \bibinfo {author} {\bibfnamefont
  {M.}~\bibnamefont {Adamczyk}}, \bibinfo {author} {\bibfnamefont
  {A.}~\bibnamefont {Al-Ajlan}}, \bibinfo {author} {\bibfnamefont
  {M.}~\bibnamefont {Al-Garawi}}, \bibinfo {author} {\bibfnamefont
  {S.}~\bibnamefont {Al-Homaidhi}}, \bibinfo {author} {\bibfnamefont
  {F.}~\bibnamefont {Amorini}}, \bibinfo {author} {\bibfnamefont
  {L.}~\bibnamefont {Auditore}}, \bibinfo {author} {\bibfnamefont
  {T.}~\bibnamefont {Aumann}}, \bibinfo {author} {\bibfnamefont
  {Y.}~\bibnamefont {Ayyad}}, \bibinfo {author} {\bibfnamefont
  {Z.}~\bibnamefont {Basrak}}, \bibinfo {author} {\bibfnamefont
  {J.}~\bibnamefont {Benlliure}}, \bibinfo {author} {\bibfnamefont
  {M.}~\bibnamefont {Boisjoli}}, \bibinfo {author} {\bibfnamefont
  {K.}~\bibnamefont {Boretzky}}, \bibinfo {author} {\bibfnamefont
  {J.}~\bibnamefont {Brzychczyk}}, \bibinfo {author} {\bibfnamefont
  {A.}~\bibnamefont {Budzanowski}}, \bibinfo {author} {\bibfnamefont
  {C.}~\bibnamefont {Caesar}}, \bibinfo {author} {\bibfnamefont
  {G.}~\bibnamefont {Cardella}}, \bibinfo {author} {\bibfnamefont
  {P.}~\bibnamefont {Cammarata}}, \bibinfo {author} {\bibfnamefont
  {Z.}~\bibnamefont {Chajecki}}, \bibinfo {author} {\bibfnamefont
  {M.}~\bibnamefont {Chartier}}, \bibinfo {author} {\bibfnamefont
  {A.}~\bibnamefont {Chbihi}}, \bibinfo {author} {\bibfnamefont
  {M.}~\bibnamefont {Colonna}}, \bibinfo {author} {\bibfnamefont {M.~D.}\
  \bibnamefont {Cozma}}, \bibinfo {author} {\bibfnamefont {B.}~\bibnamefont
  {Czech}}, \bibinfo {author} {\bibfnamefont {E.}~\bibnamefont {De~Filippo}},
  \bibinfo {author} {\bibfnamefont {M.}~\bibnamefont {Di~Toro}}, \bibinfo
  {author} {\bibfnamefont {M.}~\bibnamefont {Famiano}}, \bibinfo {author}
  {\bibfnamefont {I.}~\bibnamefont {Ga\ifmmode \check{s}\else
  \v{s}\fi{}pari\ifmmode~\acute{c}\else \'{c}\fi{}}}, \bibinfo {author}
  {\bibfnamefont {L.}~\bibnamefont {Grassi}}, \bibinfo {author} {\bibfnamefont
  {C.}~\bibnamefont {Guazzoni}}, \bibinfo {author} {\bibfnamefont
  {P.}~\bibnamefont {Guazzoni}}, \bibinfo {author} {\bibfnamefont
  {M.}~\bibnamefont {Heil}}, \bibinfo {author} {\bibfnamefont {L.}~\bibnamefont
  {Heilborn}}, \bibinfo {author} {\bibfnamefont {R.}~\bibnamefont {Introzzi}},
  \bibinfo {author} {\bibfnamefont {T.}~\bibnamefont {Isobe}}, \bibinfo
  {author} {\bibfnamefont {K.}~\bibnamefont {Kezzar}}, \bibinfo {author}
  {\bibfnamefont {M.}~\bibnamefont {Ki\ifmmode~\check{s}\else \v{s}\fi{}}},
  \bibinfo {author} {\bibfnamefont {A.}~\bibnamefont {Krasznahorkay}}, \bibinfo
  {author} {\bibfnamefont {N.}~\bibnamefont {Kurz}}, \bibinfo {author}
  {\bibfnamefont {E.}~\bibnamefont {La~Guidara}}, \bibinfo {author}
  {\bibfnamefont {G.}~\bibnamefont {Lanzalone}}, \bibinfo {author}
  {\bibfnamefont {A.}~\bibnamefont {Le~F\`evre}}, \bibinfo {author}
  {\bibfnamefont {Y.}~\bibnamefont {Leifels}}, \bibinfo {author} {\bibfnamefont
  {R.~C.}\ \bibnamefont {Lemmon}}, \bibinfo {author} {\bibfnamefont {Q.~F.}\
  \bibnamefont {Li}}, \bibinfo {author} {\bibfnamefont {I.}~\bibnamefont
  {Lombardo}}, \bibinfo {author} {\bibfnamefont {J.}~\bibnamefont
  {\L{}ukasik}}, \bibinfo {author} {\bibfnamefont {W.~G.}\ \bibnamefont
  {Lynch}}, \bibinfo {author} {\bibfnamefont {P.}~\bibnamefont {Marini}},
  \bibinfo {author} {\bibfnamefont {Z.}~\bibnamefont {Matthews}}, \bibinfo
  {author} {\bibfnamefont {L.}~\bibnamefont {May}}, \bibinfo {author}
  {\bibfnamefont {T.}~\bibnamefont {Minniti}}, \bibinfo {author} {\bibfnamefont
  {M.}~\bibnamefont {Mostazo}}, \bibinfo {author} {\bibfnamefont
  {A.}~\bibnamefont {Pagano}}, \bibinfo {author} {\bibfnamefont {E.~V.}\
  \bibnamefont {Pagano}}, \bibinfo {author} {\bibfnamefont {M.}~\bibnamefont
  {Papa}}, \bibinfo {author} {\bibfnamefont {P.}~\bibnamefont {Paw\l{}owski}},
  \bibinfo {author} {\bibfnamefont {S.}~\bibnamefont {Pirrone}}, \bibinfo
  {author} {\bibfnamefont {G.}~\bibnamefont {Politi}}, \bibinfo {author}
  {\bibfnamefont {F.}~\bibnamefont {Porto}}, \bibinfo {author} {\bibfnamefont
  {W.}~\bibnamefont {Reviol}}, \bibinfo {author} {\bibfnamefont
  {F.}~\bibnamefont {Riccio}}, \bibinfo {author} {\bibfnamefont
  {F.}~\bibnamefont {Rizzo}}, \bibinfo {author} {\bibfnamefont
  {E.}~\bibnamefont {Rosato}}, \bibinfo {author} {\bibfnamefont
  {D.}~\bibnamefont {Rossi}}, \bibinfo {author} {\bibfnamefont
  {S.}~\bibnamefont {Santoro}}, \bibinfo {author} {\bibfnamefont {D.~G.}\
  \bibnamefont {Sarantites}}, \bibinfo {author} {\bibfnamefont
  {H.}~\bibnamefont {Simon}}, \bibinfo {author} {\bibfnamefont
  {I.}~\bibnamefont {Skwirczynska}}, \bibinfo {author} {\bibfnamefont
  {Z.}~\bibnamefont {Sosin}}, \bibinfo {author} {\bibfnamefont
  {L.}~\bibnamefont {Stuhl}}, \bibinfo {author} {\bibfnamefont
  {W.}~\bibnamefont {Trautmann}}, \bibinfo {author} {\bibfnamefont
  {A.}~\bibnamefont {Trifir\`o}}, \bibinfo {author} {\bibfnamefont
  {M.}~\bibnamefont {Trimarchi}}, \bibinfo {author} {\bibfnamefont {M.~B.}\
  \bibnamefont {Tsang}}, \bibinfo {author} {\bibfnamefont {G.}~\bibnamefont
  {Verde}}, \bibinfo {author} {\bibfnamefont {M.}~\bibnamefont {Veselsky}},
  \bibinfo {author} {\bibfnamefont {M.}~\bibnamefont {Vigilante}}, \bibinfo
  {author} {\bibfnamefont {Y.}~\bibnamefont {Wang}}, \bibinfo {author}
  {\bibfnamefont {A.}~\bibnamefont {Wieloch}}, \bibinfo {author} {\bibfnamefont
  {P.}~\bibnamefont {Wigg}}, \bibinfo {author} {\bibfnamefont {J.}~\bibnamefont
  {Winkelbauer}}, \bibinfo {author} {\bibfnamefont {H.~H.}\ \bibnamefont
  {Wolter}}, \bibinfo {author} {\bibfnamefont {P.}~\bibnamefont {Wu}}, \bibinfo
  {author} {\bibfnamefont {S.}~\bibnamefont {Yennello}}, \bibinfo {author}
  {\bibfnamefont {P.}~\bibnamefont {Zambon}}, \bibinfo {author} {\bibfnamefont
  {L.}~\bibnamefont {Zetta}}, \ and\ \bibinfo {author} {\bibfnamefont
  {M.}~\bibnamefont {Zoric}},\ }\href {\doibase 10.1103/PhysRevC.94.034608}
  {\bibfield  {journal} {\bibinfo  {journal} {Phys. Rev. C}\ }\textbf {\bibinfo
  {volume} {94}},\ \bibinfo {pages} {034608} (\bibinfo {year}
  {2016})}\BibitemShut {NoStop}%
\bibitem [{\citenamefont {{Le F\`{e}vre}}\ \emph {et~al.}(2016)\citenamefont
  {{Le F\`{e}vre}}, \citenamefont {Leifels}, \citenamefont {Reisdorf},
  \citenamefont {Aichelin},\ and\ \citenamefont
  {Hartnack}}]{LeFevre2016_NPA945-112}%
  \BibitemOpen
  \bibfield  {author} {\bibinfo {author} {\bibfnamefont {A.}~\bibnamefont {{Le
  F\`{e}vre}}}, \bibinfo {author} {\bibfnamefont {Y.}~\bibnamefont {Leifels}},
  \bibinfo {author} {\bibfnamefont {W.}~\bibnamefont {Reisdorf}}, \bibinfo
  {author} {\bibfnamefont {J.}~\bibnamefont {Aichelin}}, \ and\ \bibinfo
  {author} {\bibfnamefont {C.}~\bibnamefont {Hartnack}},\ }\href {\doibase
  https://doi.org/10.1016/j.nuclphysa.2015.09.015} {\bibfield  {journal}
  {\bibinfo  {journal} {Nucl. Phys. A}\ }\textbf {\bibinfo {volume} {945}},\
  \bibinfo {pages} {112} (\bibinfo {year} {2016})}\BibitemShut {NoStop}%
\bibitem [{\citenamefont {Zhang}\ \emph {et~al.}(2020)\citenamefont {Zhang},
  \citenamefont {Liu}, \citenamefont {Xia}, \citenamefont {Li},\ and\
  \citenamefont {Biswal}}]{Zhang2020_PRC101-034303}%
  \BibitemOpen
  \bibfield  {author} {\bibinfo {author} {\bibfnamefont {Y.}~\bibnamefont
  {Zhang}}, \bibinfo {author} {\bibfnamefont {M.}~\bibnamefont {Liu}}, \bibinfo
  {author} {\bibfnamefont {C.-J.}\ \bibnamefont {Xia}}, \bibinfo {author}
  {\bibfnamefont {Z.}~\bibnamefont {Li}}, \ and\ \bibinfo {author}
  {\bibfnamefont {S.~K.}\ \bibnamefont {Biswal}},\ }\href {\doibase
  10.1103/PhysRevC.101.034303} {\bibfield  {journal} {\bibinfo  {journal}
  {Phys. Rev. C}\ }\textbf {\bibinfo {volume} {101}},\ \bibinfo {pages}
  {034303} (\bibinfo {year} {2020})}\BibitemShut {NoStop}%
\bibitem [{\citenamefont {Essick}\ \emph {et~al.}(2021)\citenamefont {Essick},
  \citenamefont {Tews}, \citenamefont {Landry},\ and\ \citenamefont
  {Schwenk}}]{Essick2021_PRL127-192701}%
  \BibitemOpen
  \bibfield  {author} {\bibinfo {author} {\bibfnamefont {R.}~\bibnamefont
  {Essick}}, \bibinfo {author} {\bibfnamefont {I.}~\bibnamefont {Tews}},
  \bibinfo {author} {\bibfnamefont {P.}~\bibnamefont {Landry}}, \ and\ \bibinfo
  {author} {\bibfnamefont {A.}~\bibnamefont {Schwenk}},\ }\href {\doibase
  10.1103/PhysRevLett.127.192701} {\bibfield  {journal} {\bibinfo  {journal}
  {Phys. Rev. Lett.}\ }\textbf {\bibinfo {volume} {127}},\ \bibinfo {pages}
  {192701} (\bibinfo {year} {2021})}\BibitemShut {NoStop}%
\bibitem [{\citenamefont {Huth}\ \emph {et~al.}(2022)\citenamefont {Huth} \emph
  {et~al.}}]{Huth2022_Nature606-276}%
  \BibitemOpen
  \bibfield  {author} {\bibinfo {author} {\bibfnamefont {S.}~\bibnamefont
  {Huth}} \emph {et~al.},\ }\href {\doibase 10.1038/s41586-022-04750-w}
  {\bibfield  {journal} {\bibinfo  {journal} {Nature}\ }\textbf {\bibinfo
  {volume} {606}},\ \bibinfo {pages} {276} (\bibinfo {year} {2022})} \BibitemShut
  {NoStop}%
\bibitem [{\citenamefont {Zubov}\ \emph {et~al.}(2009)\citenamefont {Zubov},
  \citenamefont {Adamian},\ and\ \citenamefont
  {Antonenko}}]{Zubov2009_PPN40-847}%
  \BibitemOpen
  \bibfield  {author} {\bibinfo {author} {\bibfnamefont {A.~S.}\ \bibnamefont
  {Zubov}}, \bibinfo {author} {\bibfnamefont {G.~G.}\ \bibnamefont {Adamian}},
  \ and\ \bibinfo {author} {\bibnamefont {Antonenko}},\ }\href {\doibase
  10.1134/S1063779609060057} {\bibfield  {journal} {\bibinfo  {journal} {Phys.
  Part. Nucl.}\ }\textbf {\bibinfo {volume} {40}},\ \bibinfo {pages} {847}
  (\bibinfo {year} {2009})}\BibitemShut {NoStop}%
\bibitem [{\citenamefont {Bouyssy}\ and\ \citenamefont
  {Hufner}(1976)}]{Bouyssy1976_PLB64-276}%
  \BibitemOpen
  \bibfield  {author} {\bibinfo {author} {\bibfnamefont {A.}~\bibnamefont
  {Bouyssy}}\ and\ \bibinfo {author} {\bibfnamefont {J.}~\bibnamefont
  {Hufner}},\ }\href {\doibase 10.1016/0370-2693(76)90200-8} {\bibfield
  {journal} {\bibinfo  {journal} {Phys. Lett. B}\ }\textbf {\bibinfo {volume}
  {64}},\ \bibinfo {pages} {276} (\bibinfo {year} {1976})}\BibitemShut
  {NoStop}%
\bibitem [{\citenamefont {Millener}\ \emph {et~al.}(1988)\citenamefont
  {Millener}, \citenamefont {Dover},\ and\ \citenamefont
  {Gal}}]{Millener1988_PRC38-2700}%
  \BibitemOpen
  \bibfield  {author} {\bibinfo {author} {\bibfnamefont {D.~J.}\ \bibnamefont
  {Millener}}, \bibinfo {author} {\bibfnamefont {C.~B.}\ \bibnamefont {Dover}},
  \ and\ \bibinfo {author} {\bibfnamefont {A.}~\bibnamefont {Gal}},\ }\href
  {\doibase 10.1103/PhysRevC.38.2700} {\bibfield  {journal} {\bibinfo
  {journal} {Phys. Rev. C}\ }\textbf {\bibinfo {volume} {38}},\ \bibinfo
  {pages} {2700} (\bibinfo {year} {1988})}\BibitemShut {NoStop}%
\bibitem [{\citenamefont {Sun}\ \emph {et~al.}(2025)\citenamefont {Sun},
  \citenamefont {Tanimura}, \citenamefont {Sagawa},\ and\ \citenamefont
  {Hiyama}}]{Sun2025_PLB865-139460}%
  \BibitemOpen
  \bibfield  {author} {\bibinfo {author} {\bibfnamefont {T.-T.}\ \bibnamefont
  {Sun}}, \bibinfo {author} {\bibfnamefont {Y.}~\bibnamefont {Tanimura}},
  \bibinfo {author} {\bibfnamefont {H.}~\bibnamefont {Sagawa}}, \ and\ \bibinfo
  {author} {\bibfnamefont {E.}~\bibnamefont {Hiyama}},\ }\href {\doibase
  10.1016/j.physletb.2025.139460} {\bibfield  {journal} {\bibinfo  {journal}
  {Phys. Lett. B}\ }\textbf {\bibinfo {volume} {865}},\ \bibinfo {pages}
  {139460} (\bibinfo {year} {2025})} \BibitemShut
  {NoStop}%
\bibitem [{\citenamefont {Angeli}\ and\ \citenamefont
  {Marinova}(2013)}]{Angeli2013_ADNDT99-69}%
  \BibitemOpen
  \bibfield  {author} {\bibinfo {author} {\bibfnamefont {I.}~\bibnamefont
  {Angeli}}\ and\ \bibinfo {author} {\bibfnamefont {K.}~\bibnamefont
  {Marinova}},\ }\href {\doibase https://doi.org/10.1016/j.adt.2011.12.006}
  {\bibfield  {journal} {\bibinfo  {journal} {At. Data Nucl. Data Tables}\
  }\textbf {\bibinfo {volume} {99}},\ \bibinfo {pages} {69} (\bibinfo {year}
  {2013})}\BibitemShut {NoStop}%
\bibitem [{\citenamefont {Huang}\ \emph {et~al.}(2021)\citenamefont {Huang},
  \citenamefont {Wang}, \citenamefont {Kondev}, \citenamefont {Audi},\ and\
  \citenamefont {Naimi}}]{Huang2021_CPC45-30002}%
  \BibitemOpen
  \bibfield  {author} {\bibinfo {author} {\bibfnamefont {W.}~\bibnamefont
  {Huang}}, \bibinfo {author} {\bibfnamefont {M.}~\bibnamefont {Wang}},
  \bibinfo {author} {\bibfnamefont {F.}~\bibnamefont {Kondev}}, \bibinfo
  {author} {\bibfnamefont {G.}~\bibnamefont {Audi}}, \ and\ \bibinfo {author}
  {\bibfnamefont {S.}~\bibnamefont {Naimi}},\ }\href {\doibase
  10.1088/1674-1137/abddb0} {\bibfield  {journal} {\bibinfo  {journal} {Chin.
  Phys. C}\ }\textbf {\bibinfo {volume} {45}},\ \bibinfo {pages} {030002}
  (\bibinfo {year} {2021})}\BibitemShut {NoStop}%
\bibitem [{\citenamefont {Wang}\ \emph {et~al.}(2021)\citenamefont {Wang},
  \citenamefont {Huang}, \citenamefont {Kondev}, \citenamefont {Audi},\ and\
  \citenamefont {Naimi}}]{Wang2021_CPC45-030003}%
  \BibitemOpen
  \bibfield  {author} {\bibinfo {author} {\bibfnamefont {M.}~\bibnamefont
  {Wang}}, \bibinfo {author} {\bibfnamefont {W.}~\bibnamefont {Huang}},
  \bibinfo {author} {\bibfnamefont {F.}~\bibnamefont {Kondev}}, \bibinfo
  {author} {\bibfnamefont {G.}~\bibnamefont {Audi}}, \ and\ \bibinfo {author}
  {\bibfnamefont {S.}~\bibnamefont {Naimi}},\ }\href {\doibase
  10.1088/1674-1137/abddaf} {\bibfield  {journal} {\bibinfo  {journal} {Chin.
  Phys. C}\ }\textbf {\bibinfo {volume} {45}},\ \bibinfo {pages} {030003}
  (\bibinfo {year} {2021})}\BibitemShut {NoStop}%
\bibitem [{\citenamefont {Bally}\ \emph {et~al.}(2025)\citenamefont {Bally},
  \citenamefont {Scalesi}, \citenamefont {Som\`{a}}, \citenamefont {Zurek},\
  and\ \citenamefont {Duguet}}]{Bally2025_EPJA61-140}%
  \BibitemOpen
  \bibfield  {author} {\bibinfo {author} {\bibfnamefont {B.}~\bibnamefont
  {Bally}}, \bibinfo {author} {\bibfnamefont {A.}~\bibnamefont {Scalesi}},
  \bibinfo {author} {\bibfnamefont {V.}~\bibnamefont {Som\`{a}}}, \bibinfo
  {author} {\bibfnamefont {L.}~\bibnamefont {Zurek}}, \ and\ \bibinfo {author}
  {\bibfnamefont {T.}~\bibnamefont {Duguet}},\ }\href {\doibase
  10.1140/epja/s10050-025-01596-4} {\bibfield  {journal} {\bibinfo  {journal}
  {Eur. Phys. J. A}\ }\textbf {\bibinfo {volume} {61}},\ \bibinfo {pages} {140}
  (\bibinfo {year} {2025})}\BibitemShut {NoStop}%
\bibitem [{\citenamefont {Rong}(2023)}]{Rong2023_PRC108-054314}%
  \BibitemOpen
  \bibfield  {author} {\bibinfo {author} {\bibfnamefont {Y.-T.}\ \bibnamefont
  {Rong}},\ }\href {\doibase 10.1103/PhysRevC.108.054314} {\bibfield  {journal}
  {\bibinfo  {journal} {Phys. Rev. C}\ }\textbf {\bibinfo {volume} {108}},\
  \bibinfo {pages} {054314} (\bibinfo {year} {2023})}\BibitemShut {NoStop}%
\bibitem [{\citenamefont {Takahashi}\ \emph {et~al.}(2001)\citenamefont
  {Takahashi}, \citenamefont {Ahn}, \citenamefont {Akikawa}, \citenamefont
  {Aoki}, \citenamefont {Arai}, \citenamefont {Bahk}, \citenamefont {Baik},
  \citenamefont {Bassalleck}, \citenamefont {Chung}, \citenamefont {Chung},
  \citenamefont {Davis}, \citenamefont {Fukuda}, \citenamefont {Hoshino},
  \citenamefont {Ichikawa}, \citenamefont {Ieiri}, \citenamefont {Imai},
  \citenamefont {Iwata}, \citenamefont {Iwata}, \citenamefont {Kanda},
  \citenamefont {Kaneko}, \citenamefont {Kawai}, \citenamefont {Kawasaki},
  \citenamefont {Kim}, \citenamefont {Kim}, \citenamefont {Kim}, \citenamefont
  {Kim}, \citenamefont {Kondo}, \citenamefont {Kouketsu}, \citenamefont {Lee},
  \citenamefont {McNabb}, \citenamefont {Mitsuhara}, \citenamefont {Nagase},
  \citenamefont {Nagoshi}, \citenamefont {Nakazawa}, \citenamefont {Noumi},
  \citenamefont {Ogawa}, \citenamefont {Okabe}, \citenamefont {Oyama},
  \citenamefont {Park}, \citenamefont {Park}, \citenamefont {Parker},
  \citenamefont {Ra}, \citenamefont {Rhee}, \citenamefont {Rusek},
  \citenamefont {Shibuya}, \citenamefont {Sim}, \citenamefont {Saha},
  \citenamefont {Seki}, \citenamefont {Sekimoto}, \citenamefont {Song},
  \citenamefont {Takahashi}, \citenamefont {Takeutchi}, \citenamefont {Tanaka},
  \citenamefont {Tanida}, \citenamefont {Tojo}, \citenamefont {Torii},
  \citenamefont {Torikai}, \citenamefont {Tovee}, \citenamefont {Ushida},
  \citenamefont {Yamamoto}, \citenamefont {Yasuda}, \citenamefont {Yang},
  \citenamefont {Yoon}, \citenamefont {Yoon}, \citenamefont {Yosoi},
  \citenamefont {Yoshida},\ and\ \citenamefont
  {Zhu}}]{Takahashi2001_PRL87-212502}%
  \BibitemOpen
  \bibfield  {author} {\bibinfo {author} {\bibfnamefont {H.}~\bibnamefont
  {Takahashi}}, \bibinfo {author} {\bibfnamefont {J.~K.}\ \bibnamefont {Ahn}},
  \bibinfo {author} {\bibfnamefont {H.}~\bibnamefont {Akikawa}}, \bibinfo
  {author} {\bibfnamefont {S.}~\bibnamefont {Aoki}}, \bibinfo {author}
  {\bibfnamefont {K.}~\bibnamefont {Arai}}, \bibinfo {author} {\bibfnamefont
  {S.~Y.}\ \bibnamefont {Bahk}}, \bibinfo {author} {\bibfnamefont {K.~M.}\
  \bibnamefont {Baik}}, \bibinfo {author} {\bibfnamefont {B.}~\bibnamefont
  {Bassalleck}}, \bibinfo {author} {\bibfnamefont {J.~H.}\ \bibnamefont
  {Chung}}, \bibinfo {author} {\bibfnamefont {M.~S.}\ \bibnamefont {Chung}},
  \bibinfo {author} {\bibfnamefont {D.~H.}\ \bibnamefont {Davis}}, \bibinfo
  {author} {\bibfnamefont {T.}~\bibnamefont {Fukuda}}, \bibinfo {author}
  {\bibfnamefont {K.}~\bibnamefont {Hoshino}}, \bibinfo {author} {\bibfnamefont
  {A.}~\bibnamefont {Ichikawa}}, \bibinfo {author} {\bibfnamefont
  {M.}~\bibnamefont {Ieiri}}, \bibinfo {author} {\bibfnamefont
  {K.}~\bibnamefont {Imai}}, \bibinfo {author} {\bibfnamefont {Y.~H.}\
  \bibnamefont {Iwata}}, \bibinfo {author} {\bibfnamefont {Y.~S.}\ \bibnamefont
  {Iwata}}, \bibinfo {author} {\bibfnamefont {H.}~\bibnamefont {Kanda}},
  \bibinfo {author} {\bibfnamefont {M.}~\bibnamefont {Kaneko}}, \bibinfo
  {author} {\bibfnamefont {T.}~\bibnamefont {Kawai}}, \bibinfo {author}
  {\bibfnamefont {M.}~\bibnamefont {Kawasaki}}, \bibinfo {author}
  {\bibfnamefont {C.~O.}\ \bibnamefont {Kim}}, \bibinfo {author} {\bibfnamefont
  {J.~Y.}\ \bibnamefont {Kim}}, \bibinfo {author} {\bibfnamefont {S.~J.}\
  \bibnamefont {Kim}}, \bibinfo {author} {\bibfnamefont {S.~H.}\ \bibnamefont
  {Kim}}, \bibinfo {author} {\bibfnamefont {Y.}~\bibnamefont {Kondo}}, \bibinfo
  {author} {\bibfnamefont {T.}~\bibnamefont {Kouketsu}}, \bibinfo {author}
  {\bibfnamefont {Y.~L.}\ \bibnamefont {Lee}}, \bibinfo {author} {\bibfnamefont
  {J.~W.~C.}\ \bibnamefont {McNabb}}, \bibinfo {author} {\bibfnamefont
  {M.}~\bibnamefont {Mitsuhara}}, \bibinfo {author} {\bibfnamefont
  {Y.}~\bibnamefont {Nagase}}, \bibinfo {author} {\bibfnamefont
  {C.}~\bibnamefont {Nagoshi}}, \bibinfo {author} {\bibfnamefont
  {K.}~\bibnamefont {Nakazawa}}, \bibinfo {author} {\bibfnamefont
  {H.}~\bibnamefont {Noumi}}, \bibinfo {author} {\bibfnamefont
  {S.}~\bibnamefont {Ogawa}}, \bibinfo {author} {\bibfnamefont
  {H.}~\bibnamefont {Okabe}}, \bibinfo {author} {\bibfnamefont
  {K.}~\bibnamefont {Oyama}}, \bibinfo {author} {\bibfnamefont {H.~M.}\
  \bibnamefont {Park}}, \bibinfo {author} {\bibfnamefont {I.~G.}\ \bibnamefont
  {Park}}, \bibinfo {author} {\bibfnamefont {J.}~\bibnamefont {Parker}},
  \bibinfo {author} {\bibfnamefont {Y.~S.}\ \bibnamefont {Ra}}, \bibinfo
  {author} {\bibfnamefont {J.~T.}\ \bibnamefont {Rhee}}, \bibinfo {author}
  {\bibfnamefont {A.}~\bibnamefont {Rusek}}, \bibinfo {author} {\bibfnamefont
  {H.}~\bibnamefont {Shibuya}}, \bibinfo {author} {\bibfnamefont {K.~S.}\
  \bibnamefont {Sim}}, \bibinfo {author} {\bibfnamefont {P.~K.}\ \bibnamefont
  {Saha}}, \bibinfo {author} {\bibfnamefont {D.}~\bibnamefont {Seki}}, \bibinfo
  {author} {\bibfnamefont {M.}~\bibnamefont {Sekimoto}}, \bibinfo {author}
  {\bibfnamefont {J.~S.}\ \bibnamefont {Song}}, \bibinfo {author}
  {\bibfnamefont {T.}~\bibnamefont {Takahashi}}, \bibinfo {author}
  {\bibfnamefont {F.}~\bibnamefont {Takeutchi}}, \bibinfo {author}
  {\bibfnamefont {H.}~\bibnamefont {Tanaka}}, \bibinfo {author} {\bibfnamefont
  {K.}~\bibnamefont {Tanida}}, \bibinfo {author} {\bibfnamefont
  {J.}~\bibnamefont {Tojo}}, \bibinfo {author} {\bibfnamefont {H.}~\bibnamefont
  {Torii}}, \bibinfo {author} {\bibfnamefont {S.}~\bibnamefont {Torikai}},
  \bibinfo {author} {\bibfnamefont {D.~N.}\ \bibnamefont {Tovee}}, \bibinfo
  {author} {\bibfnamefont {N.}~\bibnamefont {Ushida}}, \bibinfo {author}
  {\bibfnamefont {K.}~\bibnamefont {Yamamoto}}, \bibinfo {author}
  {\bibfnamefont {N.}~\bibnamefont {Yasuda}}, \bibinfo {author} {\bibfnamefont
  {J.~T.}\ \bibnamefont {Yang}}, \bibinfo {author} {\bibfnamefont {C.~J.}\
  \bibnamefont {Yoon}}, \bibinfo {author} {\bibfnamefont {C.~S.}\ \bibnamefont
  {Yoon}}, \bibinfo {author} {\bibfnamefont {M.}~\bibnamefont {Yosoi}},
  \bibinfo {author} {\bibfnamefont {T.}~\bibnamefont {Yoshida}}, \ and\
  \bibinfo {author} {\bibfnamefont {L.}~\bibnamefont {Zhu}},\ }\href {\doibase
  10.1103/PhysRevLett.87.212502} {\bibfield  {journal} {\bibinfo  {journal}
  {Phys. Rev. Lett.}\ }\textbf {\bibinfo {volume} {87}},\ \bibinfo {pages}
  {212502} (\bibinfo {year} {2001})}\BibitemShut {NoStop}%
\end{thebibliography}

%

\end{document}